\documentclass[ reprint, superscriptaddress, amsmath,amssymb,aps]{revtex4-2}

\usepackage[normalem]{ulem}
\usepackage[dvipsnames]{xcolor}

\newcommand{\nic}[1]{\textcolor{black}{#1}}
\newcommand{\ric}[1]{\textcolor{black}{#1}}

\newcommand{\ricII}[1]{\textcolor{black}{#1}}
\newcommand{\rev}[1]{\textcolor{black}{#1}}

\usepackage{amsmath}
\usepackage{amsfonts}
\usepackage{pgfplots}
\DeclareUnicodeCharacter{2212}{−}
\usepgfplotslibrary{groupplots,dateplot}
\usetikzlibrary{patterns,shapes.arrows}
\usepackage{bbm}
\usepackage{cancel}
\usepackage{braket}
\pgfplotsset{compat=newest}
\usepackage{dsfont}

\begin{document}

\title{Kinetically constrained quantum dynamics
in \ric{superconducting circuits}}
\author{Riccardo J. Valencia-Tortora} \email[]{Corresponding author: rvalenci@uni-mainz.de}
\affiliation{Institut f\"{u}r Physik, Johannes Gutenberg-Universit\"{a}t Mainz, D-55099 Mainz, Germany}    
\author{Nicola Pancotti}
\thanks{Work was done prior to joining Amazon}
\affiliation{AWS Center for Quantum Computing, Pasadena, CA 91125, USA}
\author{Jamir Marino}
\affiliation{Institut f\"{u}r Physik, Johannes Gutenberg-Universit\"{a}t Mainz, D-55099 Mainz, Germany}
\date{\today}%
\begin{abstract}

We study the dynamical properties of the bosonic quantum East model at low temperature. We show that a naive generalization of the corresponding spin-1/2 quantum East model does not posses analogous slow dynamical properties. In particular, conversely to the spin case, the bosonic ground state turns out to be {\it not} localized.
We restore   localization by introducing a repulsive interaction term.  
The bosonic nature of the model allows us to construct rich families of many-body localized states, including coherent, squeezed and cat states. We formalize this finding by introducing a set of {\it superbosonic} creation-annihilation operators which satisfy the bosonic commutation relations and, when acting on the vacuum,   create excitations exponentially localized around a certain site of the lattice. 
Given the constrained nature of the model, these states retain memory of their initial conditions for  long times. Even in the presence of dissipation, we show that quantum information remains localized within decoherence times tunable with the parameters of the system. 
We propose an implementation of the bosonic quantum East model based on state-of-the-art \ric{superconducting \nic{circuits}}, which could be used in the near future to explore \nic{dynamical properties of} kinetically constrained models in \nic{modern} \ric{platforms}. 

\end{abstract}
\maketitle

\section{Introduction~\label{sec:introduction}}
Robust storage  of quantum information and decoherence induced  by  external baths are two important limiting factors that mitigate against a large-scale adoption of modern quantum technologies~\cite{preskill2018quantum}. The storage of quantum information is a challenging task, as most interacting quantum systems tend to thermalize quickly. Once equilibrium is reached, the properties of the initial configurations are hard to retrieve, as they are ergodically scattered among exponentially many degrees of freedom~\cite{polkovnikov2011colloquium}. In order to overcome this obstacle, many proposals have attempted to confine quantum information into conserved or quasi-conserved quantities~\cite{Carleo2012,PhysRevB.103.L100202, DeRoeck2014,Schiulaz2015,Papic2015,Barbiero2015,Yao2016,Smith2017,Mondaini2018,Schulz2019,Nieuwenburg2019,Shiraishi_Mori_Reply2018,kormos2017real,James2019,morong2021observation,gunawardana2021dynamical,buca2020quantum}. These proposals range from strongly disordered many-body localized~\cite{nandkishore2015many, abanin2019colloquium} or glassy systems~\cite{ritort2003glassy,Chamon2005, garrahan2018aspects,Hickey2016,Horssen2015,Lan2018,Feldmeier2019,Castelnovo2005}, in which thermalization is impeded by the presence of disordered potentials, to ``fracton'' systems, in which dynamical constraints induce fragmentation on the space of reachable configurations~\cite{Prem2017,Nandkishore2018,PhysRevB.101.174204,Sala2020,Rakovszky2020,Pretko2020,pretko2018fracton,scherg2021observing}
, and quantum scarred systems, in which certain classes of initial states show coherent oscillations for times longer than typical relaxation times~\cite{Turner2018, Turner2018b,Ho2019,Ok2019,Schecter2019,Khemani2019Signatures,Hudomal2020,Moudgalya2018Entanglement,feldmeier2020anomalous,serbyn2021quantum,PhysRevLett.126.210601,PhysRevX.11.021021,PhysRevLett.126.103002,zhao2021orthogonal}.
Most of these phenomena often rely on such delicate properties that {\it any} weak coupling with an external environment could potentially become detrimental. 

\rev{Quantum kinetically constrained models (KCMs) have recently attracted attention due to their distinctive dynamical properties.} 
Motivated by the slowness of their classical counterparts, researchers have started to investigate their quantum generalizations, such as the quantum East model, the quantum Fredricksen-Andersen model, and others~\cite{Garrahan2009, Chleboun2013, Kim2015, gopalakrishnan2018hydrodynamics,gopalakrishnan2018operator, Banuls2019,PhysRevE.102.052132}.

In this work, we explore the low-temperature dynamical properties of the bosonic quantum East model, a generalization of the spin-$1/2$ model studied in Refs.~\cite{Horssen2015,PhysRevX.10.021051}, \rev{in which} spin excitations can only be created on sites \rev{to the ``east''} of a previously occupied one. Our contributions can be summarized as follows. ({\it i}) We show that repulsive density-density interactions are necessary to entail localization in  the ground state, in contrast to East models with \rev{a finite-dimensional} local Hilbert space. ({\it ii}) We exploit the properties of the localized phase and the bosonic nature of the model, to    construct families  of non-Gaussian many-body  states \rev{that are useful for quantum-information processing}. ({\it iii}) We illustrate how localization   enhances the robustness of these states    against decoherence. ({\it iv}) Finally, we propose an implementation of the bosonic quantum East model based on chains of superconducting qubits.

In the spin-$1/2$ case, evidence has been provided in support of 
a dynamical transition from a fast thermalizing regime to a slow, non-ergodic one~\cite{Horssen2015,PhysRevX.10.021051}. In particular, in Ref.~\cite{PhysRevX.10.021051}, it \rev{has been} argued that the slow dynamics is a byproduct of the localized nature of the low-energy eigenstates of the model. Namely, the corresponding wavefunctions contain nontrivial excitations only on a small compact region of the lattice and they are in the vacuum state everywhere else. 
This has direct consequences \rev{for} the dynamical properties of the system\rev{,} as the localized states can be used as building blocks to construct exponentially many ``slow'' states in the size of the system.

The dynamical transition observed in Ref.~\cite{PhysRevX.10.021051} is not guaranteed to survive in the bosonic case. In fact, we provide strong numerical evidence that this is not the case for the most naive bosonic generalization of the spin-$1/2$ model. In order to restore localization at low temperature, we consider a modified model \rev{in which} density-density interactions - absent in the bare spin case - play a crucial role.  More precisely, we show that the ground state remains localized as we increase the finite cutoff of the local Fock-space dimension only in the presence of repulsive interactions. We support our findings by combining numerical and analytical approaches. Within the localized phase, the ground state is well approximated by a product state for any value of interaction. It is therefore well approximated by a  matrix product state, making large system size and local Fock space dimension numerically accessible (cf. Secs. \ref{section_bosonic_quantum_east_model} and \ref{section_localization}).

The bosonic generalization of the spin-1/2 East model opens \rev{up} a number of directions including the construction of many-body versions of archetypal states \rev{that are} relevant for quantum information applications such as coherent states, squeezed states, and cat states~\cite{walls2007quantum}. These states possess the same properties \rev{as} their single-mode counterparts, although they are   supported on a few neighboring sites. We provide a formal description of these objects by proposing a simple adiabatic protocol \rev{that} defines a set of {\it superbosonic} creation-annihilation operators (Sec.~\ref{section_state_preparation}). These operators fulfill the canonical bosonic commutation relations and they are exponentially localized in the neighborhood of a given site on the lattice. This allows us to construct an effective, non-interacting, theory at low temperature in terms of these operators\rev{, in which} the Hamiltonian is reminiscent of the l-bit construction in \rev{many-body localization (MBL)}~\cite{chandran2015constructing,ros2015integrals,imbrie2017local,huse2014phenomenology}.  

\begin{figure}[t!]
    \centering
    \includegraphics[width=\linewidth]{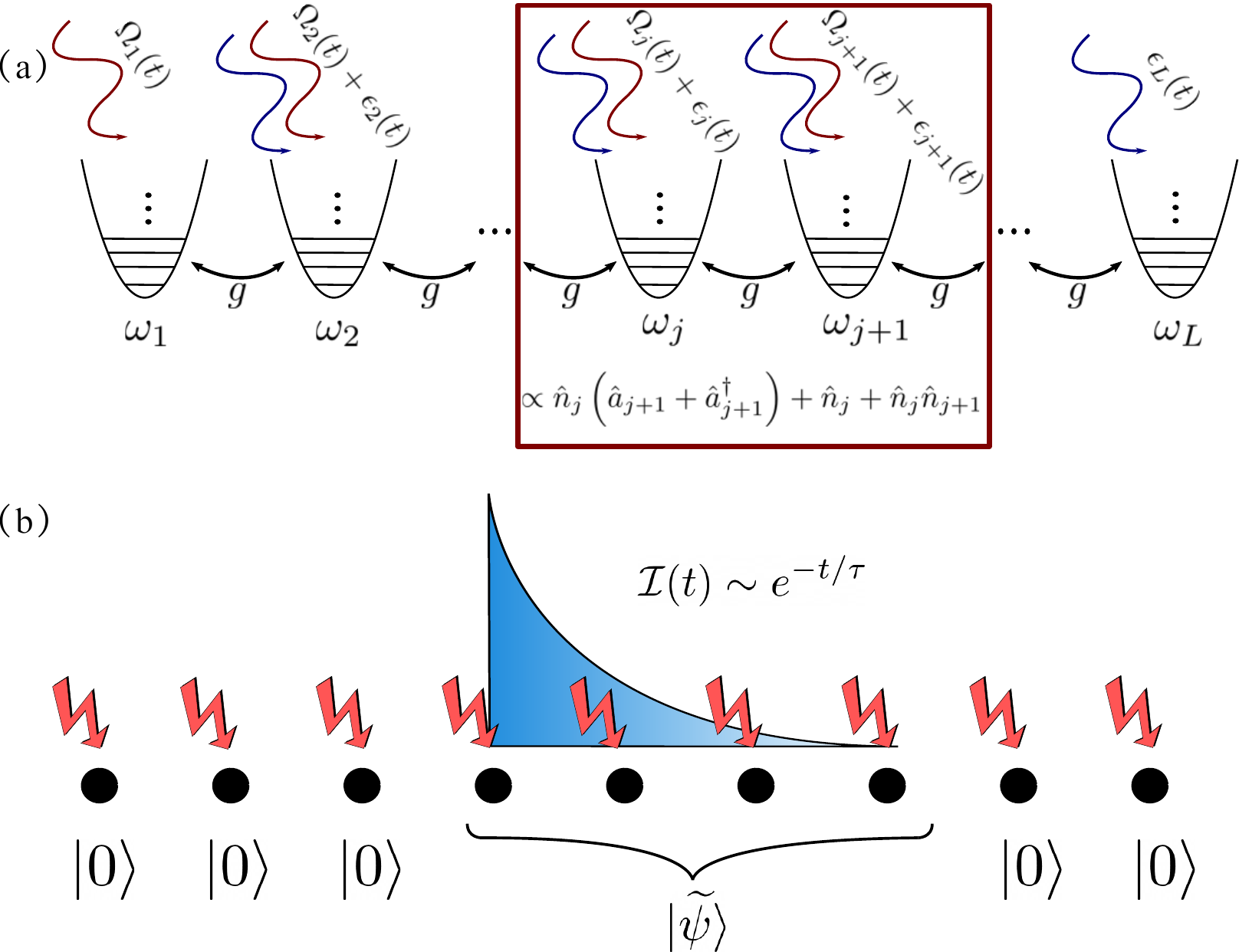}
    \caption{(a): A chain of driven superconducting qubits coupled via exchange interaction $g$. In the red box we write the low-energy effective interaction between the $j$-th and $(j+1)$-th superconducting qubits. (b): A sketch of a localized state subject to external noise (arrows). The visibility of the initial peak with respect to the rest of the system (measured by the imbalance $\mathcal{I}(t)$) decays exponentially with a   time $\tau$ much larger than the characteristic operational  timescales of state-of-the-art superconducting circuits.
    }\label{figintro}
\end{figure}

  In Sec.~\ref{section_noise_robustness_super_cat_state}, we couple the system to different noise sources and, via a detailed numerical analysis, we show that localized states retain some memory of their initial condition even in the presence of strong dissipation (see Fig.~\ref{figintro}). First, we \ric{consider the effects of dephasing     noise coupled to bosonic occupations}, which preserves the ``East symmetry'' (see the definition in Sec.~\ref{section_bosonic_quantum_east_model}). In this scenario, the localized states are barely altered by the environment. We show that the fidelity between the time-evolved state and the initial state decays exponentially   with a long decoherence time, controlled by the parameters of the Hamiltonian, the initial state, and the strength of the noise. 
  Second, we consider \ric{the effects of particle losses} \rev{that} break the ``East symmetry.'' As expected in this situation, the magnitude of the fidelity decays exponentially fast in time, with a decoherence time \rev{that is} parametrically small in the \ric{loss rate}. 
  It is important to stress that as the localized states have non-trivial structure only on a small support, any external noise that does not act in their immediate vicinity leaves them essentially invariant. 
This set of noise-resilient properties renders the many-body states studied in this work qualitatively different from localization induced by disorder, which is inherently fragile to decoherence (\rev{for studies on MBL systems coupled to a bath or external noise} see Refs.~\cite{PhysRevX.7.011034,PhysRevLett.125.116601,PhysRevB.93.094205,PhysRevB.90.064203,nandkishore2017many,fischer2016dynamics}). In particular, in Sec.~\ref{section_cQED_implementation} we argue that our localized states can be manipulated on timescales shorter than the characteristic relaxation and decoherence times of superconducting qubit wires.

In fact, our proposal for an implementation of the bosonic quantum East model based on superconducting qubits is one of the key findings of this work. 
In recent years, unprecedented quantum control of interacting superconducting qubits with microwave photons has been reached in circuit-QED platforms~\cite{RevModPhys.93.025005,Blais2020,Joshi_2021,eickbusch2021fast,Ma2021,Wang2021,PhysRevX.10.021060,Wallraff2004,Houck2012}. 
These circuits allow quantum-information-processing tasks and the quantum simulation of paradigmatic light-matter interfaces. Superconducting Josephson junctions allow us to introduce nonlinearity
in quantum electrical circuits, which is a key factor in protecting quantum resources, by making these platforms resilient to noise and errors.  \ric{This is a key factor of merit for any superconducting qubit, ranging from the established transmon to, for instance, the more recently developed \rev{superconducting nonlinear asymmetric inductive element (SNAIL)}~\cite{PhysRevA.76.042319,Frattini2017}}. Here, we consider a chain of superconducting qubits (see Refs.~\cite{Carusotto2020,PhysRevA.76.042319,Yanay2020,PhysRevB.103.L220202,Schmidt2013,Devoret2013,Roushan2017,PhysRevLett.123.050502,chiaro2020direct}), which can be described as anharmonic oscillators,  coupled via a hopping term 
(cf. Fig.~\ref{figintro}). 
In the limit of weak coupling and low anharmonicity, we find an effective description of such superconducting qubits array in terms of  the bosonic quantum East chain.

The paper is organized as follows. In Sec.~\ref{section_bosonic_quantum_east_model}, we introduce the Hamiltonian of the model, enumerate its symmetries, and compare it to previous works on similar models. In Sec.~\ref{section_localization}, we explore the localization properties of the ground state of the model. In particular, we show when the transition point is independent of the size of the cutoff of the local \rev{Fock-space} dimension and how the localization length behaves in the proximity of the transition. On the localized side of the transition, we quantitatively compare results extracted with tensor-network methods and mean field, and we show that they are in excellent agreement. In Sec.~\ref{section_state_preparation}, we introduce a description in terms of superbosonic operators, which allows us to generalize coherent, squeezed, and cat states. In Sec.~\ref{section_noise_robustness_super_cat_state}, we study the robustness of these localized states
against noise source. In Sec.~\ref{section_cQED_implementation}, we present the implementation of the Hamiltonian for
the bosonic quantum East model, based on a chain of superconducting qubits.

\section{Bosonic Quantum East Model \label{section_bosonic_quantum_east_model}}
We investigate the following Hamiltonian with open boundary conditions
\begin{equation}
\label{eq_H_full_model}
H=- \frac{1}{2}\sum_{j=0}^L \hat{n}_j \left[e^{-s} \left(\hat{a}_{j+1}+\hat{a}_{j+1}^\dagger\right) - \epsilon \hat{n}_{j} - U \hat{n}_{j+1}- 1\right],
\end{equation}
where $\hat{a}_j$ and $\hat{a}_j^\dagger$ are bosonic annihilation and creation operators acting on site $j$ respectively; $e^{-s}$ controls the constrained creation and annihilation of bosons; $\epsilon$ is the on-site density-density interaction; and $U$ is the nearest-neighbor density-density interaction. 

As discussed in Sec.~\ref{sec:introduction}, Eq.~\eqref{eq_H_full_model} is a kinetically constrained ``East'' model. The unidirectional constrained feature has consequences \rev{for} the accessible portion of the Hilbert space by the dynamics. Namely, any initial state with a product of vacua from the left edge up to a given site in the bulk will exhibit nontrivial dynamics only on the right side of the lattice after the first occupied site. For sake of concreteness, let us consider the state $|00100\dots 0\rangle$. Via subsequent application of the 
Hamiltonian \rev{given in Eq.}~\eqref{eq_H_full_model} we have,
\begin{equation}
\begin{split}
& |00120 \dots 0\rangle \dots\\
& \nearrow\\
|00100\dots 0\rangle \rightarrow |00110\dots 0\rangle & \rightarrow |001110\dots 0\rangle \dots\\
& \searrow \\
& |00100\dots 0\rangle \dots
\end{split}
\end{equation}
where $\rightarrow$ represents the action of the constrained creation and annihilation of bosons at each step of perturbation theory.
The  occupation of the first nonvacant site and of those at its left cannot change as a consequence of the ``East'' constraint. 
More formally, the Hamiltonian commutes with the projectors
\begin{equation}
\label{eq_operator_commuting_with_H}
P(n_0,k) = \mathcal{P}_{0,j}^{\otimes_{j=0}^{k-1}} \otimes \mathcal{P}_{n_0,k}\otimes \mathds{1}_j^{\otimes_{j > k}},
\end{equation}
where $\mathcal{P}_{s,j}=|s\rangle_j {}_j\langle s|$ is the projector on the Fock state with $s$ particles on site $j$, $\mathds{1}_j$ is the identity acting on site $j$, and $k$ and $n_0$ are, respectively, the position and occupation of the first nonvacant site. We can split the Hilbert space into dynamically disconnected sectors $\mathcal{H}_{n_0,k}$, such that the action of $P(n_0,k)$ is equivalent to the identity, while the action of the other projectors gives zero. For example, the state $|00100\dots 0\rangle \in \mathcal{H}_{1,2}$ (note that the first site index is $0$). Furthermore, since $\sum_{k=0}^L\sum_{n_0=1}^\infty P(n_0,k)=\mathds{1}$ these sectors $\{\mathcal{H}_{k,n_0}\}$ constitute a complete and orthogonal basis of the whole Hilbert space $\mathcal{H}$, namely $\mathcal{H}=\bigoplus_{k=0}^L \bigoplus_{n_0=1}^\infty \mathcal{H}_{n_0,k}$.\\

In the following, we focus on a certain block specified by $k$, $n_0$, and the number of ``active'' sites $L$ right next to the $k$-th one. Since the action of $H$ on sites to the left of the $k$-th one is trivial, the index $k$ is physically irrelevant for our purpose and we therefore choose $k=0$ without any loss generality.
Exploiting this property, we   write the Hamiltonian \rev{given in Eq.}~\eqref{eq_H_full_model} as $H_{L+1} = \sum_{n_0}H_{L+1}(n_0)$, where $H_{L+1}(n_0)$ is
\begin{equation}
\label{eq_H_symmetry_sector}
\begin{split}
&H_{L+1}(n_0) = \hat{h}_1+\\
&-\frac{1}{2}\sum_{j=\ric{1}}^{L}\hat{n}_j \left[e^{-s} \left(\hat{a}_{j+1}+\hat{a}_{j+1}^\dagger\right) - \epsilon \hat{n}_{j} - U \hat{n}_{j+1}- 1\right],
\end{split}
\end{equation}
{with $\hat{h}_1 \equiv -\frac{1}{2}n_0 \left[e^{-s} \left(\hat{a}_{1}+\hat{a}_{1}^\dagger\right) - \epsilon n_0 - U \hat{n}_{1}- 1\right]$ and $n_0 \in \mathbb{N}^+$}. Furthermore, since $H_{L+1}(n_0)$ commutes with the operators acting on the $(L+1)$-th site, we can represent it as the sum of an infinite number of commuting terms $H_{L+1}(n_0)=\sum_{\beta_r}H_{L}^{\beta_r}(n_0)\otimes \Pi_{L+1}^{\beta_r}$, where $\Pi_L^\beta$ is the projector over the eigenstate $|\beta_r\rangle$ with eigenvalue $\beta_r=rU-e^{-2s}/U$ of the operator $\left( U \hat{n}_{L+1}-e^{-s} \left(\hat{a}_{L+1}+\hat{a}_{L+1}^\dagger\right) \right)$, where $r \in \mathbb{N}$, and,
\begin{equation}
\label{eq_H_specific_symmetry_sector}
\begin{split}
&H_{L}^{\beta_r}(n_0) = \hat{h}_1+\\
&-\frac{1}{2}\sum_{j=\ric{1}}^{L-1}\hat{n}_j \left[e^{-s} \left(\hat{a}_{j+1}+\hat{a}_{j+1}^\dagger\right) - \epsilon \hat{n}_{j} - U \hat{n}_{j+1}- 1\right]+\\
&+\frac{1}{2}\hat{n}_{L} \left[\beta_r + \epsilon \hat{n}_{L} + 1\right].
\end{split}
\end{equation} 
In Sec.~\ref{section_localization}, we focus on the properties of the ground state of the Hamiltonian \rev{given in Eq.}~\eqref{eq_H_specific_symmetry_sector} within a certain symmetry sector.\\

The Hamiltonian \rev{given in Eq.}~\eqref{eq_H_full_model} can be linked to its spin-$1/2$ version~\cite{PhysRevX.10.021051} by setting $U=\epsilon=0$ and replacing the bosons with hard-core ones. Since the Hilbert space of each spin is finite, the ``East'' symmetry is largely reduced with respect to the bosonic case. Each symmetry sector $\mathcal{H}_{k,n_0=1}$ is specified only by the position of the first excitation, since $n_0$ is bound to be zero or one. The ground state properties within a symmetry sector $\mathcal{H}_{k,n_0=1}$, where the position $k$ of the first nonempty is again irrelevant, \rev{have been investigated in Ref.~\cite{PhysRevX.10.021051}}. It \rev{has been} observed that the probability of finding an occupied site in the ground state decays exponentially fast around the first occupied site when $s>0$, namely
\begin{equation}
\label{eq_average_occupation_number}
\langle \hat{n}_j\rangle \sim \exp (-j/\xi (s)),
\end{equation}
where the expectation value is taken on the ground state and we introduce the localization length $\xi>0$. The localization length $\xi$ is the typical distance from the first occupied site such that the state becomes a trivial product state \rev{that is} well approximated by the vacuum.\\
In Sec.~\ref{section_localization}, we investigate the conditions for localization of the ground state at finite values of $s$ upon trading spins (hard-core bosons) for bosons. Such generalization is not granted. The amplitude for \ric{``}eastern'' particle creation can now be enhanced by the prefactor $n_0$, suggesting that the transition may be qualitatively established when $(n_0e^{-s})\sim1$. This would imply a critical value $s_c\propto \log n_0$, which is parametrically large in $n_0$, pushing the extension of the localized phase up to $s\to\infty$. 
Nonetheless, we show in Sec.~\ref{section_localization} that a localized phase  still occurs for $s>0$ whenever  repulsive interactions are included in Eq.~\eqref{eq_H_full_model}. 


\section{Localization transition \label{section_localization} }
\begin{figure}[t]
\centering
\includegraphics[width=\linewidth]{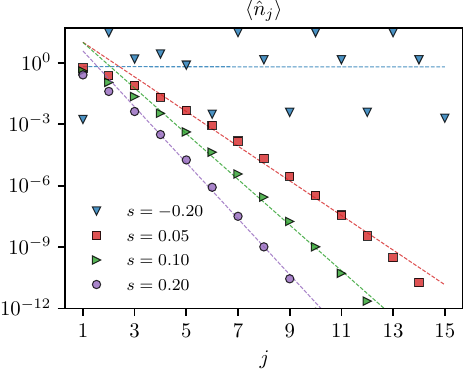}
\caption{The average occupation number of the ground state for different values of $s$ at fixed nearest-neighbor density-density interaction $U=1$. We fix $L=15$, a cutoff $\Lambda=30$ to the maximal occupation number, and $n_0=1$. \ric{In the plot, we do not display the occupation $n_0$ of the zeroth site that fixes the ``East symmetry'' sector.} The dashed lines are the exponential fit, the slope of which is $-1/\xi$, where $\xi$ is the localization length (cf. Eq.~\eqref{eq_average_occupation_number}).}
\label{fig_occupation_number_localization}
\end{figure}

\begin{figure}[t]
\centering
\includegraphics[width=\linewidth]{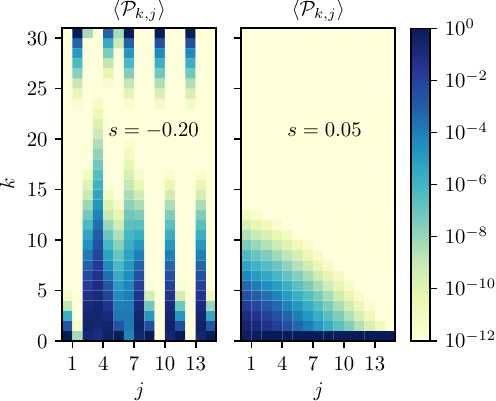}
\caption{{The probability of having $k \in [0,\Lambda]$ bosons on site $j \in[1,L]$ in the ground state. \ric{In the plot, we do not display the occupation $n_0$ of the zeroth site that fixes the ``East symmetry'' sector.} We fix $L=15$, $\Lambda=30$, $n_0=1$ and $U=1$. \rev{In the left panel,} we consider a typical configuration in the delocalized phase ($s=-0.20$). The cutoff is saturated over many sites. The staggered feature is due to the repulsive nearest-neighbor interaction. \rev{In the right panel,} we consider a typical localized ground state ($s=0.05$). Along each site \ric{$j$}, the probability of having $k$ bosons, $\langle \mathcal{P}_{k,j}\rangle$, drops exponentially fast with $k$. The light color means that the value is smaller than $10^{-12}$.}}
\label{fig_xi_F_xi}
\end{figure}

In this section we show that the Hamiltonian in Eq.~\eqref{eq_H_specific_symmetry_sector} displays a localization-delocalization transition at finite $s$ and $U>0$. We give numerical evidence corroborated by analytical observations that  repulsive  interactions are   necessary  to observe such a transition at finite $s$.
We use the inverse localization length $\xi^{-1}$ controlling the decay of the average occupation number in space (cf. Eq.~\eqref{eq_average_occupation_number}), as proxy for the transition.

In the following, we fix $\epsilon=0$ and the symmetry sector $\beta_{r=0}$ in Eq.~\eqref{eq_H_specific_symmetry_sector}, unless mentioned otherwise.  \ric{The additional nonlinear term proportional to $  \epsilon$   would complicate the   analysis from a technical standpoint   without altering the main contents of the paper. For the sake of clarity,  Appendix~\ref{appendix_on_site_density_density_interaction}   shows that, for   $U=0$ and $\epsilon>0$,   the localization properties of  the ground state remain qualitatively similar to those discussed in the main text.}

In order to investigate the properties of the ground state, we resort to a combination of mean-field arguments, exact diagonalization (ED),  and density matrix renormalization group (DMRG) methods~\cite{SCHOLLWOCK201196}.  
Since we aim to explore large system sizes, we mainly resort to the DMRG and we use ED as a benchmark when both methods can be used. 
Interestingly, we find  that mean field is able to analytically predict the location of the transition point obtained via the DMRG.

We compute the ground state $|\psi_0(n_0)\rangle$ at fixed $n_0$, $s$, and $U$.
We fix the system size at $L = 15$. This value is sufficiently large   to   capture the localized tail of the ground state, without relevant finite-size effects.  
Although the   local Fock space   is infinite, in order to treat the model numerically, we need to fix a finite cutoff $\Lambda$.
We work with Fock states $|0\rangle$ through $|\Lambda\rangle$, such that the spin-$1/2$ case of Ref.~\cite{PhysRevX.10.021051} is recovered at $\Lambda=1$. In Appendix~\ref{section_appendix_properties_localized_state}, we show how localization is only mildly dependent on the   sector selected by the occupation $n_0$ of the zeroth site. Accordingly, in the following, we set $n_0=1$.

The Hamiltonian is one dimensional, local, and gapped at finite $\Lambda$; therefore, its ground state can be  efficiently accessed via a matrix product state (MPS) formulation of the  DMRG~\cite{SCHOLLWOCK201196}.
The main source of error is given by the finite cutoff $\Lambda$. Indeed, the properties of $|\psi_0(n_0)\rangle$ can change nontrivially as a function of $\Lambda$. More precisely, for any finite cutoff $\Lambda$, the model falls into the class of localized systems studied in Ref.~\cite{PhysRevX.10.021051}. As a result, $|\psi_0(n_0)\rangle$ is always localized for a large enough $s$ at finite $\Lambda$ but this does not imply localization for $\Lambda\to \infty$. Indeed, although  $U>0$ makes the spectrum of the Hamiltonian in Eq.~\eqref{eq_H_full_model} bounded from below, it does not ensure that its ground state is still localized in space when $s$ is finite. In the following, we extract the $\Lambda \to \infty$ limit via a scaling analysis.
\begin{figure}[t]
\centering
\includegraphics[width=\linewidth]{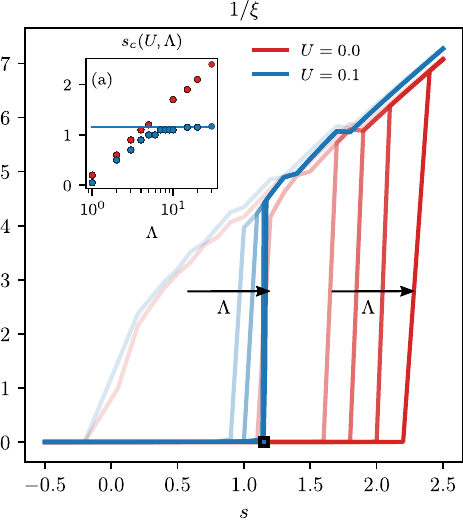}
\caption{
The inverse of the localization length $\xi$ in a system of $L=15$ ``active'' sites in the   symmetry sector $n_0=1$ and $\beta_{r=0}$. The main plot shows the inverse of the localization length $\xi^{-1}$ as a function of $s$ for different values of $\Lambda \in [1,30]$ and $U$. The darker lines correspond to larger values of $\Lambda$. The square is the mean-field estimate of $s_c$ in the bosonic case ($\Lambda=\infty$). The inset (a) shows the behavior of $s_c(U,\Lambda)$ as a function of $\Lambda$ for $U=0$ (red) and $U=0.1$ (blue). The circles correspond to  numerically extracted values from the DMRG results, while the continuous lines are the mean-field estimate $s_c \approx \log(1/\sqrt{U})$, which matches the numerics at large $\Lambda$.}
\label{fig_localization_length}
\end{figure}

In Fig.~\ref{fig_occupation_number_localization}, we show the average occupation number $\langle \hat{n}_j\rangle$ as a function of site $j$ for some values of $s$ at fixed $U=1$. For $s$ not large enough, the average occupation does not change smoothly with the site $j$ and it saturates the cutoff $\Lambda$, meaning that there are strong finite-cutoff effects. In contrast, for $s$ large enough, the occupation decays exponentially in $j$, matches Eq.~\eqref{eq_average_occupation_number} well, and does not change upon increasing the cutoff $\Lambda$. The value of $s$ at which this change of behavior occurs depends on $U$, as we discuss in more detail in this section.

{In order to check \rev{the effects of a} finite $\Lambda$ cutoff, we compute the probability of having $k$ bosons on site $j$, namely the expectation value of the projector $\mathcal{P}_{k,j} = |k\rangle_j {}_j\langle k|$, where $|k\rangle_j$ is the Fock state with $k$ particles on site $j$. In Fig.~\ref{fig_xi_F_xi}, we show $\langle \mathcal{P}_{k,j} \rangle$ as a function of $k$ and $j$ for typical localized and delocalized ground states, respectively.
The results in the delocalized phase are not reliable, since the observable suffers finite-cutoff effects. Instead, in the localized phase,
\begin{equation}
\label{eq_decay_projector_localized_phase}
\langle \mathcal{P}_{k,j}\rangle \sim e^{-k/\xi_{F,j}},
\end{equation}
with $\xi_{F,j}>0$ for any site $j$. The exponential decay in the localized phase sheds additional light on the fact that the system is well described by a finite effective cutoff (for additional details, see Appendix~\ref{scaling_analysis_in_lambda}).}\\

\begin{figure}[t]
\centering
\includegraphics[width=\linewidth]{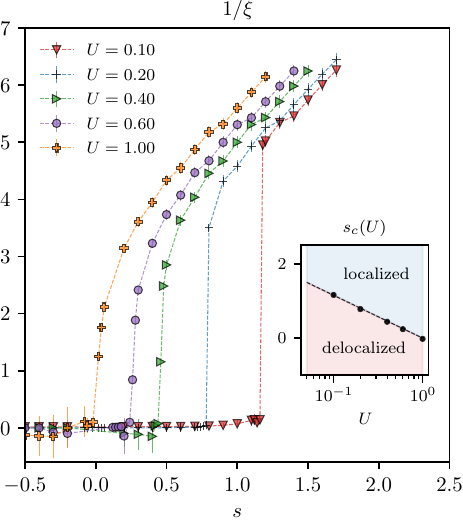}
\caption{
The inverse of the localization length $\xi$ in a system of $L=15$ ``active'' sites in the symmetry sector $n_0=1$ and $\beta_{r=0}$. We fix the cutoff $\Lambda=30$. The main plot shows the inverse of the localization length $\xi^{-1}$ as a function of $s$ for different values of $U$. We plot the error bars on top of each point. In the inset we plot the transition point $s_c(U)$ as a function of $U$. \rev{The dots} represent the extracted $s_c(U)\equiv \lim_{\Lambda \to \infty} s_c(U,\Lambda)$. The dashed line is the mean-field estimate for the transition point $s_c^\text{MF}(U) = \log(1/\sqrt{U})$.
}
\label{fig_sc_vs_U}
\end{figure}
For each value of $U$ and $\Lambda$, the inverse of the localization length goes from values smaller than or equal to zero to positive values as $s$ increases. We identify the region where $1/\xi \leq 0$ as the delocalized phase, while the region where $1/\xi>0$ \rev{is identified} as the localized phase. \ric{In the delocalized phase, strong finite cutoff effects \ricII{can lead to a positive} localization length $\xi$. In order not to mistakenly identify these points as belonging to the localized phase, we fix a threshold $\lambda>0$ and for each $\Lambda$ and $U$  we identify the transition point $s_c(U,\Lambda)$ as the value of $s$ such that $1/\xi \leq \ric{\lambda}$ and $1/\xi > \ric{\lambda}$ for $s$ smaller and greater than $s_c(U,\Lambda)$, respectively.  We choose $\lambda \approx 10^{-1}$. The results are weakly affected by this choice of $\ric{\lambda}$.  Furthermore, the precise location of the transition point $s_c(U,\Lambda)$ is beyond the scope of this work, since we are interested in engineering states deep in the localized phase, as we discuss extensively in Sec.~\ref{section_state_preparation}}.

As discussed above, in the delocalized phase,  results are strongly dependent on the cutoff, since the average occupations  always saturate their artificial upper bound. This circumstance allows us to draw only qualitative conclusions on the physics at $s<s_c$ in the case of the bosonic East model ($\Lambda\to\infty$).

In Fig.~\ref{fig_localization_length}, we show the inverse of the localization length $\xi$ swiping $s$ for different values of $\Lambda$ at fixed $U$. For $U=0$, the transition point $s_c(U=0,\Lambda)$ always increases with $\Lambda$. Instead, when $U>0$, the transition point converges to a finite value independent of $\Lambda$ for $\Lambda\to\infty$.
In Fig.~\ref{fig_localization_length}.$(a)$, we show the numerically extracted transition point $s_c(U,\Lambda)$  as a function of $\Lambda$ and $U$. For $U>0$, it is possible to extract a finite value of $s_c(U) \equiv \lim_{\Lambda \to \infty}s_c(U,\Lambda)$. Instead, for $U=0$, the transition point scales as $s_c(U=0,\Lambda) \propto \log(\Lambda)$, suggesting that in the actual bosonic system we have $s_c(U=0)=\infty$, meaning \rev{that there} is no transition. Therefore, whenever $U>0$, the system undergoes a delocalized-localized transition at finite $s_c(U)$. In Fig.~\ref{fig_sc_vs_U}, we show the inverse of \rev{the} localization length $\xi$ as a function of $s$ for different values of $U$ at fixed $\Lambda$. The transition point $s_c$ depends on the competition between the dynamical term, controlled by $e^{-s}$, and the nearest-neighbor density term, proportional to $U$. The former favors the delocalization of the state, while the latter \rev{favors} its localization. Indeed, in the $U\to 0$ limit, \rev{we provide evidence} that the bosonic system is always delocalized if $s<\infty$. Instead, in the large $U$ limit, the Hamiltonian is approximated by $U \sum_i \hat{n}_i \hat{n}_{i+1} + \hat{n}_i$, \rev{the} ground state of which in a specific symmetry sector at given total particle number is simply $|n_0\rangle |00\dots 0\rangle$.

The role of the interaction term $U$ in the localization of the bosonic system can be appreciated in a mean-field treatment.
We project the Hamiltonian into the manifold of coherent product states $|\phi\rangle = \bigotimes_{j=1}^L |\alpha_j\rangle_j$, with $\hat{a}_j|\alpha_j\rangle_j = \alpha_j |\alpha_j\rangle_j$. We evaluate the Hamiltonian \rev{given} in Eq.~\eqref{eq_H_symmetry_sector} in this basis:
\begin{equation}
\langle \phi | H(n_0)|\phi\rangle = - \frac{1}{2} \sum_{j=0}^L |\alpha_j|^2 \left(2 e^{-s}\alpha_{j+1} - U |\alpha_{j+1}|^2 - 1\right),
\end{equation}
where $|\alpha_j|^2$ is the average number of particles in the coherent state at site $j$. From unidirectionality of the interaction, we can write $\langle \phi|H(n_0)|\phi\rangle = -\frac{1}{2}\sum_{j}|\alpha_j|^2 h_j(\alpha_{j+1},s,U)$, where $h_j(\alpha_{j+1},s,U)=\left(2 e^{-s}\alpha_{j+1} - U |\alpha_{j+1}|^2 - 1\right)$. For energetic stability the effective field $h_j(\alpha_{j+1},s,U)$ on site $j$ should be negative:  
\begin{equation}
\label{eq_mean_field_condition_localization}
\begin{split}
&\left(2 e^{-s}\alpha_{j+1} - U |\alpha_{j+1}|^2 - 1\right) < 0 \Rightarrow\\
& \Rightarrow s > \log \left( \frac{2\alpha_{j+1}}{1+U|\alpha_{j+1}|^2}\right) \equiv s_c(\alpha_{j+1}).
\end{split}
\end{equation}
Since the system does not conserve the number of particles there can be an unbounded number of excitations in the ground state within a fixed symmetry sector. Therefore, in order to have localization at a mean-field level it is necessary that Eq.~\eqref{eq_mean_field_condition_localization} holds for any value of $\alpha_{j+1} \in [0,\infty)$, namely $s>\max_{\alpha_{j+1}}s_c(\alpha_{j+1})$, and for all sites. For $U>0$, such condition is satisfied if $s>\log(1/\sqrt{U})$, which turns to be in very good agreement with the DMRG numerical findings (see Fig.~\ref{fig_sc_vs_U}). Instead, for $U \leq 0$, there is no finite value of $s$ \rev{that} fulfills Eq.~\eqref{eq_mean_field_condition_localization} for all $\alpha_{j+1}$. 

The excellent agreement between the DMRG and the mean-field analysis can be explained by observing that the ground state $|\psi_0\rangle$ (excluding the \rev{zeroth} site, which fixes the symmetry sector) obtained via the DMRG is well approximated via a product state, namely $|\psi_0\rangle \approx \bigotimes_{j =1}^L |\phi_j\rangle$. To further investigate the nature of the state $|\psi_0\rangle$, we consider the correlator $\Delta_{j} \equiv (\langle \hat{n}_j \hat{n}_{j+1}\rangle - \langle \hat{n}_j\rangle \langle \hat{n}_{j+1}\rangle)$. We use this operator as a proxy for non-Gaussian correlations. We compare $\Delta_j$ computed on the ground state obtained via the DMRG and the one computed assuming \rev{that} the same state is Gaussian in the operators $\{\hat{a}_j^{(\dagger)}\}_{j=1}^L$,   using   Wick's theorem. As shown in Appendix~\ref{appendix_signatures_gaussianity_ground_state}, the closer we are to the transition point $s_c$, the more the state develops non-Gaussian features at distances $ j \lesssim \xi$. On the contrary, deep in the localized phase, the Gaussian \textit{ansatz} captures the actual correlations at all sites well. Indeed, in the large $s$ limit, the Hamiltonian turns out to be diagonal in the number basis, namely $H(s\gg 1)\sim \sum_j (\hat{n}_j \hat{n}_{j+1}+\hat{n}_j)$, the ground state of which is $|n_0\rangle |00\dots 0\rangle$, which is a product state of Gaussian states (excluding the zeroth site, which fixes the symmetry sector). 

The localized tail can be explained in a more intuitive way via the adiabatic theorem. Indeed, the Hamiltonian is gapped in the localized phase when $U>0$; therefore, we can adiabatically connect two ground states within it. In particular, we can link any localized ground state to the one   at $s=\infty$. This choice is particularly convenient since the Hamiltonian is diagonal in the number basis at $s=\infty$,   $H(s\to\infty)=\sum_{j=1}(U\hat{n}_j \hat{n}_{j+1} + \hat{n}_j)/2$ and its ground state at fixed symmetry sector is simply $|n_0\rangle \bigotimes_{j=1}^L|0\rangle_j$. Then, the evolution with the adiabatically changing Hamiltonian will dress the initial site with an exponentially localized tail. In Sec.~\ref{section_state_preparation}, we further exploit the adiabatic theorem to design the many-body version of a variety of states \rev{that are} relevant in quantum-information setups, such as coherent states, cat states, and squeezed states.


\section{Localized states engineering \label{section_state_preparation}}

In Sec.~\ref{section_localization}, we have discussed the localization properties of the ground state of the bosonic quantum East model within each symmetry sector specified by the occupation $n_0$ of the first nonvacant site. In this section, we show that the ground states of different symmetry sectors are connected via bosonic creation and annihilation operators. We use this infinite set of localized states to construct \ric{the localized} versions of cat, coherent, and squeezed states that are relevant for quantum-information purposes. \ric{These states share the same properties as their single-mode counterparts, although they are supported on a few neighboring sites toward the East as the ground states.} 

Starting with a given symmetry sector fixed by $n_0$, our aim is to find operators $\mathcal{A}$ and $\mathcal{A}^{\dag}$ that obey the bosonic canonical commutation relations $\left[ \mathcal{A}, \mathcal{A}^{\dag} \right] = 1$, with the defining property  
\begin{equation}
\label{eq_definition_superbosons}
\left(\mathcal{A}^{\dag}\right)^{n_0} | 0 \rangle = \mathcal{N}|n_0\rangle \otimes |\psi_0(n_0)\rangle :=\mathcal{N}|\widetilde{n}_0\rangle,
\end{equation}
where \ric{$|\psi_0(n_0)\rangle$ is the localized tail of the ground state at fixed symmetry sector $n_0$ and} $\mathcal{N}$ is a constant. In other words, by acting $n_0$ times on the bosonic vacuum state with the operator $\mathcal{A}^\dagger$, we aim \rev{to retrieve}  the localized ground state {of the Hamiltonian in Eq.}~\eqref{eq_H_full_model} in the symmetry sector with $n_0$ particles on the first nonvacant site. 
From now on, we refer to these operators as \textit{superbosonic} creation and annihilation operators since, in contrast to single site annihilation and creation operators, they act on a localized region of the system, by creating or destroying a bosonic localized tail along the chain. Likewise, we refer to the localized ground states $|\widetilde{n}_0\rangle$ as \textit{superbosons}.

In order to find an explicit form for such  operators, we employ the adiabatic theorem. 
From numerical evidence our Hamiltonian, is gapped within the whole localized phase \ric{(see Fig.~\ref{fig_gap_adiabatic_protocol})}. Therefore, there exists a slow  tuning of $s$ that enables us to connect two localized ground states at fixed values of $U$ and $n_0$. We consider such a unitary transformation $\mathcal{U}(s,U)$ linking the ground state for $s=\infty$ with the target one at $s>s_c(U)$ in a fixed symmetry sector specified by the occupation $n_0$ of the first nonvacant site. We fix $s=\infty$ as our starting point since the Hamiltonian is diagonal in the number operator when $s\to\infty$ and its ground state is simply the tensor product $|n_0\rangle \otimes_{j\geq 1}|0\rangle_j$. By the adiabatic theorem, the unitary operator takes the following form~\cite{sakurai2017modern,messiah2014quantum}:
\begin{equation}
\label{eq_adiabatic_transformation}
\mathcal{U}(s,U) = \mathcal{T}\exp\left[-i \int_0^T dt H\left(s(t)\right)\right],
\end{equation}
where $\mathcal{T}$ indicates the time-ordering operator and $s(t)$ is a function that interpolates from $s(t=0)=\infty$ and $s(t=T)=s$. The function  $s(t)$ has to be chosen such that it satisfies~\cite{sakurai2017modern,messiah2014quantum},
\begin{equation}
\label{eq_condition_adiabatic_theorem}
\frac{1}{\Delta(t)^2} \max_{n \neq 0}\left|\langle \Psi_{n}(t)| \dot{H}(t) |\Psi_0(t)\rangle\right| \ll 1,
\end{equation}
at all times $t$. In Eq.~\eqref{eq_condition_adiabatic_theorem}, the state $|\Psi_n(t)\rangle$ is the $n$-th excited eigenstate of the Hamiltonian computed at time $t$; $\dot{H}(t)$ is the time derivative of the Hamiltonian, which encodes the information about the specific protocol; finally, $\Delta(t)\equiv E_1(t)-E_0(t)$ is the gap at time $t$. 
\nic{For a  reasonably fast protocol,  we require  $\Delta(s) \sim \mathcal{O}(1)$ in the parameter regime of interest}. \ric{We write $H(s(t))=H(s=\infty)+J(t)V$, where $H(s=\infty)=\sum_j (\hat{n}_j + U \hat{n}_j \hat{n}_{j+1})/2$,  and $V=\sum_j \hat{n}_j(\hat{a}_{j+1}+\hat{a}_{j+1}^\dagger)$ is the coupling that we adiabatically switch on through the time-dependent protocol $J(t)=-e^{-s(t)}/2$. 
The time derivative of the Hamiltonian then reads $\dot{H}(t)=\dot{J}(t) V$. 
Let us focus on the perturbation $V$ and the gap $\Delta$ at first and then on the specific protocol $J(t)$.
In Fig.~\ref{fig_gap_adiabatic_protocol}, we show the gap of the Hamiltonian and the maximum matrix element $\max_n V_{n}(s) \equiv \langle \psi_n(s)|V|\psi_0(s)\rangle$ connecting the ground to the $n$-th excited state as a function of $s$ at fixed $U$. 
Within the localized phase, the gap is $\mathcal{O}(1)$ and the maximum matrix element $\max_n V_{n}(s) \sim n_0$, where $n_0$ is the occupation of the first nonempty site fixing the symmetry sector. \nic{Due to the kinetic constraint,} the largest matrix element $\max_n V_{n}(s)$ is between the localized ground state and \nic{the second localized state perturbatively close to the product states} $|n_0 1 0 0 \dots \rangle$ \nic{(\ricII{note that this} is not necessarily the first excited state)}. Therefore, the leading contribution comes from the first few sites, since the other terms are exponentially suppressed in the localization length of $|\Psi_{0}\rangle$. Let us consider, as a possible adiabatic protocol, the linear ramping $J(t)=-e^{-s}t/(2T)$, where $t \in [0,T]$, with $T$ as the total duration time. 
\begin{figure}
\centering
\includegraphics[width=\linewidth]{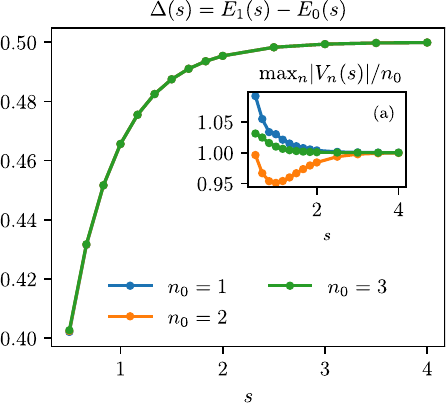}
\caption{\ric{The gap of the Hamiltonian in Eq.~\eqref{eq_H_specific_symmetry_sector} as a function of $s\in[0.5,4]$ for different values of the occupation $n_0$ of the first nonempty site. The inset (a) shows the maximum matrix element $\max_n V_{n}(s)/n_0 \equiv \max_n \langle \psi_n(s)|V|\psi_0(s)\rangle/n_0$ of the perturbation $V=\sum_j \hat{n}_j(\hat{a}_{j+1}+\hat{a}_{j+1}^\dagger)$  between the $n$-th excited state and the ground state at fixed $s$. We fix a system size $L=6$, cutoff $\Lambda=3$ and nearest-neighbor density-density interaction $U=1$. The transition point is at $s_c(U=1)\approx 0$. The results are weakly affected (of the order of few percent) by the finite cutoff $\Lambda$ for $s \lesssim 2$.} 
}
\label{fig_gap_adiabatic_protocol}
\end{figure}
From Eq.~\eqref{eq_condition_adiabatic_theorem}, the total time $T$ has to satisfy $T \gg n_0 e^{-s}$. \nic{Recall} that we set the on-site bare frequency of the bosons as our energy scale and therefore the time $T$ is expressed in that unit as well. In Sec.~\ref{section_cQED_implementation}, we propose a possible experimental implementation of the bosonic quantum East model based on superconducting \ricII{qubits}. The typical on-site bare frequency of superconducting \ricII{qubits} is $\mathcal{O}(\text{GHz})$, leading to $T \gg (n_0 e^{-s})\text{ns} \sim 1\text{ns}$, which is within the typical coherence time of $\mathcal{O}(1\mu$s$)$ of state-of-the-art superconducting \ricII{qubits}~\cite{RevModPhys.93.025005}.}\\
For $s(t)$   that   satisfies Eq.~\eqref{eq_condition_adiabatic_theorem}, we obtain
\begin{equation}
\mathcal{U}(s,U)|n_0\rangle_0 \bigotimes_{j=1}^L |0\rangle = e^{i\theta}|\widetilde{n}_0\rangle,
\end{equation}
where $\theta$ is a phase acquired during the adiabatic time evolution~\cite{messiah2014quantum,sakurai2017modern}. Using $|n_0\rangle_0 =\left(\hat{a}_0^\dagger\right)^{n_0}|0\rangle /\sqrt{n_0!}$ and $\mathcal{U}(s,U)|00\dots 0\rangle = |00\dots 0\rangle$, we obtain
\begin{equation}
\label{eq_superbosonic_operator}
\left(\mathcal{A}(s,U)^\dagger\right)^{n_0}|\widetilde{0}\rangle =e^{i\theta} \sqrt{n_0!}|\widetilde{n}_0\rangle,
\end{equation}
where $|\widetilde{0}\rangle \equiv |00\dots 0\rangle$ and $\mathcal{A}(s,U)^\dagger = \mathcal{U}(s,U) \hat{a}_0^\dagger \mathcal{U}(s,U)^\dagger$. 
We can straightforwardly generalize Eq.~\eqref{eq_superbosonic_operator} taking into account the position $j$ starting from which we want to embed the state $|\widetilde{n}_0\rangle$. We define $\mathcal{A}_j(s,U)^{\dagger}= \mathcal{U}(s,U) \hat{a}_j^{\dagger} \mathcal{U}(s,U)^\dagger$, the action $n_0$ times of which on the bosonic vacuum generates the state $|0\rangle_\ell^{\otimes_{\ell<j}}\otimes|\widetilde{n}_0\rangle$. Differently from the generic interacting case, the dressed operator $\mathcal{A}_j^{(\dagger)}(s,U)$ acts nontrivially in a region exponentially localized around $j$. The operator $\mathcal{A}_j(s,U)^{(\dagger)}$ satisfies the bosonic commutation relations, since  they are connected via a unitary transform to the bare bosonic operators $\hat{a}_j^{(\dagger)}$. Therefore, they are bosonic operators. As anticipated, we call the operators $\mathcal{A}_j(s,U)^{(\dagger)}$ \textit{superbosonic} annihilation(creation) operators.

\begin{figure}
\centering
\includegraphics[width=\linewidth]{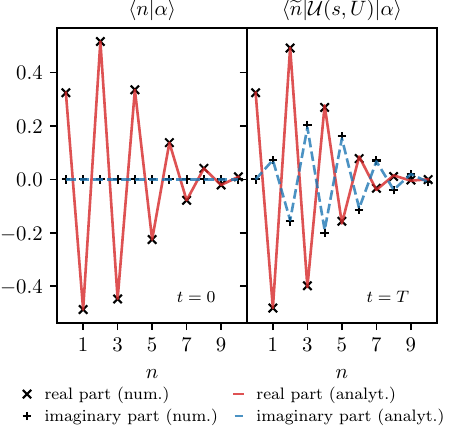}
\caption{
At $t=0$, the system is prepared in a single-body coherent state $|\alpha\rangle=|\alpha\rangle_0 \bigotimes_{j=1}^L |0\rangle_j$, where $|\alpha\rangle_0$ is a coherent state on the first site with $\alpha=1.5$. At time $t\geq 0$, we apply the adiabatic protocol defined in Eq.~\eqref{eq_adiabatic_transformation} to the state $|\alpha\rangle$ up to time $t=T$, obtaining $|\widetilde{\alpha}'\rangle$. In the left panel, we compute the probability amplitudes $\langle n |\alpha\rangle$, where $|n\rangle = |n\rangle_0 \bigotimes_{j=1}^L |0\rangle_j$ is an eigenstate of the number operator $\hat{n}_0$. The data (symbols) match the amplitudes of a single-site coherent state with $\alpha=1.5$ (continuous and dashed line). In the right panel, we compute the probability amplitudes $\langle \widetilde{n}|\mathcal{U}(s,U)|\alpha\rangle$, where $|\widetilde{n}\rangle$ is a \textit{superboson} (cf. Eq.~\eqref{eq_definition_superbosons}) with $n$ excitations on the first site. The data (symbols) match the amplitudes of the localized version of a coherent state defined in Eq.~\eqref{eq_dressed_coherent_state} well, with $\alpha'=1.5 e^{1.42i}$ (continuous and dashed line).
}
\label{fig_overlap_dressed_coherent_state}
\end{figure}
Since the transition point $s_c$ is essentially independent of the value of $n_0$ (see Appendix~\ref{section_appendix_properties_localized_state}), we can design a protocol that obeys the adiabatic theorem for any initial state $|n_0\rangle \otimes |0 \ldots 0\rangle $. { Furthermore, since these states belong to dynamically disconnected symmetry sectors, $\mathcal{H}_{k=0,n_0}$, for any values of $s$ and $U$, it is possible to adiabatically evolve them independently of each other.} Therefore, any linear combination of initial states turns under the adiabatic protocol   into   
\begin{equation}
\label{eq_generic_dressed_state}
\begin{split}
\mathcal{U}(s,U)\sum_{n_0}c_{n_0}|n_0\rangle \otimes |0 \ldots 0\rangle
&=\sum_{n_0}c_{n_0}\left(\mathcal{A}(s,u)^\dagger\right)^{n_0}|\widetilde{0}\rangle\\
&=\sum_{n_0}c_{n_0}e^{i\theta(n_0,s,U,T)}|\widetilde{n}_0\rangle,
\end{split}
\end{equation}  
where $\theta(n_0,s,U,T)$ is the phase acquired during the adiabatic time evolution. 
As discussed in Appendix~\ref{section_appendix_properties_localized_state}, deep in the localized phase the spectrum depends linearly on $n_0$, with small corrections. Since the phase acquired during the adiabatic evolution depends on the energy of the given state during the protocol, we have $\theta(n_0,s,U,T) \sim n_0 f(s,U,T)$, where $f(s,U,T)$ is a function that is dependent on the specific protocol. This has important consequences for the state engineering we discuss in the following. 
As an example, let us consider as initial state of the adiabatic preparation the coherent state 
$|\alpha\rangle \equiv |\alpha\rangle_0 \bigotimes_{j\geq 1} |0\rangle_j$, where 
\ric{
\begin{equation}
|\alpha\rangle_0 = \sum_{n=0}^\infty \frac{e^{-|\alpha|^2/2}\alpha^n}{\sqrt{n!}} |n\rangle_0.
\end{equation}
}
Using Eq.~\eqref{eq_generic_dressed_state}, the state $|\alpha\rangle$ turns into
\begin{equation}
\label{eq_dressed_coherent_state}
\begin{split}
\mathcal{U}(s,U)|\alpha\rangle &= \sum_{n=0}^\infty \frac{e^{-|\alpha|^2/2}\alpha^n}{\sqrt{n!}} e^{i\theta(n,s,U,T)}|\widetilde{n}\rangle\\
&= \sum_{n=0}^\infty \frac{e^{-|\alpha'|^2/2}\alpha'^n}{\sqrt{n!}}|\widetilde{n}\rangle,
\end{split}
\end{equation}
where $\alpha' = \alpha e^{if(s,U,T)}$. In Fig.~\ref{fig_overlap_dressed_coherent_state}, we compute the overlap between $\mathcal{U}(s(t),U)|\alpha\rangle$ and the \textit{superbosons} $|\widetilde{n}(s(t),U)\rangle$ for different values of $\alpha$ at the initial time $t=0$ and at the final time $t=T$ of the adiabatic transformation. At the initial time, we have $\mathcal{U}(s(0),U)|\alpha\rangle=|\alpha\rangle$ and $|\widetilde{n}(s(0),U)\rangle = |n\rangle \otimes |00\dots 0\rangle$. At the final time we have $|\widetilde{n}(s(T),U)\rangle = |\widetilde{n}\rangle$. In Fig.~\ref{fig_overlap_dressed_coherent_state}, the overlaps are in very good agreement with Eq.~\eqref{eq_dressed_coherent_state} and we obtain the desired state in Eq.~\eqref{eq_dressed_coherent_state} with a fidelity $\approx 0.9994$ for $\alpha=1.5$. 
We expect that when $\alpha$ is large, the fidelity achieved by the protocol becomes small, since corrections to the linear dependence of $\theta(n,s,U,T)$ from $n$ become important. \ric{We call the localized version of a coherent state $|\widetilde{\alpha}\rangle \equiv \mathcal{U}(s,U)|\alpha\rangle$ a \textit{supercoherent} state.}\\
Analogously, we perform the same analysis considering as initial state a cat state \ric{$|C\rangle$ on site} $j=0$.
Indeed, since the phase factor $e^{if(s,U,T)}$ does not depend on $\alpha$, given a cat state
\begin{equation}
|C\rangle \bigotimes_{j>1} |0\rangle_j = \frac{1}{\mathcal{N}}\left( |\alpha\rangle_0 + e^{i\phi}|-\alpha\rangle_0 \right) \bigotimes_{j>1} |0\rangle_j,
\end{equation}
where $\mathcal{N}$ is a normalization constant, its localized version is
\begin{equation}
\label{eq_super_cat_state}
|\widetilde{\mathcal{C}}\rangle = \frac{1}{\mathcal{N}}\left( |\widetilde{\alpha'}\rangle + e^{i\phi}|-\widetilde{\alpha'}\rangle \right) 
\end{equation}
where $|\widetilde{\mathcal{C}}\rangle \equiv \mathcal{U}(s,U)|C\rangle$, 
and $\alpha' = \alpha e^{if(s,U,T)}$. We call $|\widetilde{\mathcal{C}}\rangle$ a \textit{supercat} state.\\ 
We can extend {Eq.}~\eqref{eq_dressed_coherent_state} to states of the form 
\begin{equation}
\label{eq_general_state_valid_adiabatic_protocol}
|\psi\rangle = |00\dots 0\rangle \otimes \left( \sum_{n=0}^\infty \rho_n \beta^{\theta n} |n\rangle_j\right)\otimes|00\dots 0\rangle,
\end{equation}
where $\rho_n \in \mathbb{R}$ and $\beta,\theta \in \mathbb{C}$. Indeed, if we apply the adiabatic protocol to the state defined in Eq.~\eqref{eq_general_state_valid_adiabatic_protocol}, the phase acquired can be absorbed into $\beta$. Coherent states, cat states, and squeezed states all fall into the class described in Eq.~\eqref{eq_general_state_valid_adiabatic_protocol}. In other words, using the adiabatic protocol, not only can we engineer the localized versions of states such as coherent and squeezed states but we can do so preserving their single-mode properties.

For instance, the localized versions of coherent and squeezed states can be implemented either via the adiabatic time evolution or the application of an operator $\mathcal{M}$ that is linear or quadratic in the superbosonic operators $\mathcal{A}$. The operator $\mathcal{M}$ can be obtained applying the adiabatic protocol to its single-site counterpart $M$, namely $\mathcal{M}=\mathcal{U}(s,U) M \mathcal{U}(s,U)^\dagger$. For instance, we define the dressed displacement operator,
\begin{equation}
\mathcal{D}(\alpha) = \exp\left(\alpha \mathcal{A}^\dagger - \alpha^* \mathcal{A}\right),
\end{equation}
where $\alpha\in \mathbb{C}$ is the displacement parameter, and the dressed  squeezed operator,
\begin{equation}
\mathcal{S}(\xi) = \exp\left[ \frac{1}{2}\left(\zeta^* \mathcal{A}^2 - h.c.\right)\right],
\end{equation}
where $\zeta\in \mathbb{C}$ is the squeezing parameter, the action of which on the vacuum creates a \textit{supercoherent} and \textit{supersqueezed} state, respectively. 
However, the most natural way to prepare such states is by starting from their single-mode version and then adiabatically turning on the off-diagonal term $\propto e^{-s}$ in the Hamiltonian. Note that these states are Gaussian with respect to the \textit{superbosonic} operators $\mathcal{A}^{(\dagger)}$ and not with respect to the bare operators $\hat{a}^{(\dagger)}$. We call these states \textit{super-Gaussian}.\\

We find that \textit{superbosons} $|\widetilde{n}_0\rangle$, with different $n_0$ and the same position $j$ of the first nonvacant site, are connected via the operators $\mathcal{A}_j^{(\dagger)}$. We see that their localized feature makes their energies approximately evenly spaced as a function of $n_0$ (cf. Appendix~\ref{section_appendix_properties_localized_state}). The evenly spaced energies of different ground states and the fact that the different ground states are connected via a bosonic operator $\mathcal{A}_j(s,U)^{(\dagger)}$ resemble the features of a quadratic Hamiltonian, such as the one-dimensional harmonic oscillator. Adding up    these properties, the action of the interacting Hamiltonian $H(s,U)$ in Eq.~\eqref{eq_H_full_model} in the manifold of the ground states is approximatively equivalent to a free theory in the \textit{superbosonic} operators $\mathcal{A}_j(s,U)^{(\dagger)}$, namely
\begin{equation}
\label{eq_H_quadratic}
H(s,U) \approx \sum_{j=-\infty}^{+\infty}\epsilon_0 \mathcal{A}_j(s,U)^\dagger \mathcal{A}_j(s,U).
\end{equation}
the eigenstates of which are $\bigotimes_{j=-\infty}^{+\infty}(\mathcal{A}_j(s,U)^\dagger)^{k_j}|0\rangle$, where $k_j \in [0,\infty)$. 
The effective Hamiltonian in Eq.~\eqref{eq_H_quadratic} well captures the action of the full Hamiltonian Eq.~\eqref{eq_H_full_model} on a \textit{superboson} $|\widetilde{n}\rangle$ well up to a certain $n$ that is parametrically large in $s$ and $U$, since corrections to the evenly spaced feature of the ground states energies become important as $n$ increases. Moreover, the effective Hamiltonian in Eq.~\eqref{eq_H_quadratic} neglects the interaction between neighboring \textit{superbosons}. Therefore, in the infinite set of eigenstates of {Eq.}~\eqref{eq_H_quadratic}, only those given by \textit{superbosons} separated by a large number of empty sites with respect to the typical localization length $\xi$ approximate eigenstates of the original model well (up to corrections that are exponentially small with the distance of   two \textit{superbosons}). For instance, the state $\mathcal{A}_1(s,U)^\dagger \mathcal{A}_{j \gg \xi}(s,U)^\dagger |0\rangle$, which describes two far localized excitations, is an eigenstate of the effective theory in {Eq.}~\eqref{eq_H_quadratic} and, approximately, of the original Hamiltonian in {Eq.}~\eqref{eq_H_full_model}. Instead, the state $\mathcal{A}_1(s,U)^\dagger \mathcal{A}_{2}(s,U)^\dagger |0\rangle$, which describes two nearly localized excitations, is an eigenstate of {Eq.}~\eqref{eq_H_quadratic} with energy $2\epsilon_0$, while it is not an eigenstate of the original model {Eq.}~\eqref{eq_H_full_model}, since we are neglecting the   contribution coming from the interacting part of the Hamiltonian. Despite these limitations, the effective Hamiltonian {in Eq.}~\eqref{eq_H_quadratic} captures the equilibrium properties in the localized phase and the dynamical features of states such as the \textit{supercat} state and \textit{supersqueezed} state well when the interacting part bewteen \textit{superbosons} can be neglected.
In this regard, the properties of the localized phase of quantum East models are reminiscent of the $l$-bits construction in MBL~\cite{chandran2015constructing,ros2015integrals,imbrie2017local,huse2014phenomenology}.\\
Let us consider a \textit{supercat} state $|\psi(t=0)\rangle=|\widetilde{\mathcal{C}}\rangle$ defined in Eq.~\eqref{eq_super_cat_state} as initial state \ricII{in order to test the effective quadratic theory in Eq.~\eqref{eq_H_quadratic}}. We evolve it and compute the fidelity 
\begin{figure}[t!]
\centering
\includegraphics[width=\linewidth]{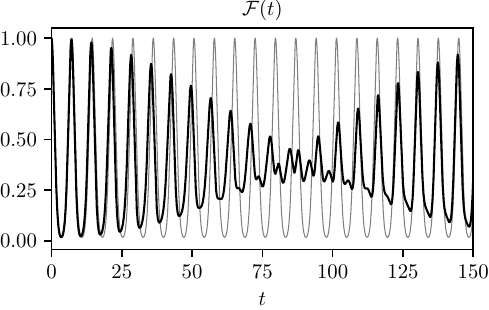}
\caption{The coherent dynamics of a \textit{supercat} state with $\alpha=1.5$. We simulate a system of size $L=15$. We fix $s=1$ and $U=1$. We show the dynamics of the fidelity $\mathcal{F}$ (dark black line). The light black line is the expected value from the effective quadratic theory in Eq.~\eqref{eq_H_quadratic} with a numerically extracted $\epsilon_0 \approx 0.43$.}
\label{figure_fidelity_super_cat_state}
\end{figure}
\begin{equation}
\label{eq_fidelity}
\mathcal{F}(t)=|\langle \psi(t)|\psi(t=0)\rangle|^2.
\end{equation}
As shown in Fig.~\ref{figure_fidelity_super_cat_state}, the fidelity displays almost perfect oscillations at short times, followed by a drop and almost perfect revivals. The short-time behavior is compatible with a rotation of the \textit{supercat} state in the \textit{dressed} phase-space $\tilde{X}_0=(\mathcal{A}_0+\mathcal{A}_0^\dagger)$ and $\tilde{P}_0=-i(\mathcal{A}_0-\mathcal{A}_0^\dagger)$, as expected from the effective Hamiltonian in Eq.~\eqref{eq_H_quadratic}.
We can approximately compute the dynamics of the \textit{supercat} state $|\widetilde{\mathcal{C}}\rangle$ \ric{generated by Eq.~\eqref{eq_H_quadratic}} as
\begin{equation}
\label{eq_psi_super_cat_evolved_gamma_0}
e^{-iHt}|\widetilde{\mathcal{C}}\rangle \approx \frac{1}{\mathcal{N}}\ric{\Big(}|\widetilde{\alpha}(t)\rangle + e^{i\phi}|-\widetilde{\alpha}(t)\rangle\ric{\Big)},
\end{equation}
where $\alpha(t)=\alpha(t=0)e^{-i\epsilon_0 t}$. The state {in Eq.}~\eqref{eq_psi_super_cat_evolved_gamma_0} is a rotating \textit{supercat} state in the \textit{dressed} space. From Eq.~\eqref{eq_psi_super_cat_evolved_gamma_0} we can estimate the expected fidelity. In Fig.~\ref{figure_fidelity_super_cat_state}, we compare the expected value and the numerical results. The former matches the numerical results up to times parametrically large in $s$ and $1/\alpha$. On the one hand,   nonlinear corrections are suppressed the more the system is localized. On the other, corrections to the linear dependence of the energies $\langle \widetilde{n}|\hat{H}|\widetilde{n}\rangle$ become important the larger $n$ is or, equivalently, $\alpha$, leading to dephasing processes~\cite{PhysRevLett.100.100601}. 
The revivals can be explained considering nonlinear effects; indeed, perfect revivals are observed for  single-mode cat states with self-Kerr interaction~\cite{yurke1988dynamic} (for a circuit-QED implementation, see Ref.~\cite{2013}). \ric{Differently from the latter case, we have an extended state and  nearest-neighbor density-density interactions. As a consequence, pushing the simulations to longer times we observe no perfect revivals \nic{as in the case of} single bosonic modes with Kerr nonlinearities. Such behavior might be captured by improving the effective theory introduced in Eq.~\eqref{eq_H_quadratic},     adding nonlinearities in the basis of \textit{superbosonic} degrees of freedom. This is beyond our current scope and therefore left as a potential interesting follow-up.}\\
We can extend these dynamical properties to any state prepared via the adiabatic protocol starting from a state of the form given in {Eq.}~\eqref{eq_general_state_valid_adiabatic_protocol}. Indeed, these states evolve analogously to the \textit{supercat} state under the effective quadratic theory defined in Eq.~\eqref{eq_H_quadratic}.
The \textit{super-Gaussian} states fall into this class. Once again, we highlight that these states are Gaussian with respect to the \textit{superbosonic} operators $\mathcal{A}^{(\dagger)}$ but not with respect to the bare operators $\hat{a}^{(\dagger)}$. \\

%
We have discussed the application of the adiabatic protocol to a single-site state embedded in the vacuum; however, this extends directly to more general initial states. For instance, we could have started from a product state made of single-body states separated by a large number of empty sites, with respect to the localization length $\xi$, or from a superposition of those. At the end of the protocol, each one will be dressed independently from the others. Therefore, the final state will be made of localized states concatenated one after the other. 
\ric{
\section{Effects of dephasing and losses} \label{section_noise_robustness_super_cat_state}}
In this section, we investigate the dynamical properties of the localized states introduced in Sec.~\ref{section_state_preparation} when coupled to \ric{the environment}. Here, we study the effects of two different couplings with an external bath, namely a global \ric{dephasing due to a noise} coupled to the local densities, which commutes with the ``East'' symmetry, and \ric{global losses}, which break the ``East'' symmetry. \ric{Both of these couplings are experimentally relevant in superconducting circuits setups}~\cite{RevModPhys.93.025005}, which are at the core of the experimental implementation we propose in Sec.~\ref{section_cQED_implementation}. We provide numerical evidence that local information is erased very slowly  when the \ric{environment} is coupled \ric{via} densities \ric{to the system}. We show how the characteristic time scales depend on the parameters of the Hamiltonian, the initial state, and the strength of the coupling to the \ricII{environment}. On the contrary, we show that \ric{losses are} highly disruptive and that the time scales are dependent on the strength of the \ric{coupling to the} \ricII{environment} and the initial state, while the underlying coherent dynamics does not play a substantial role. \ricII{At the end of the section, we show that the typical couplings to the environment currently achieved in superconducting circuits are small enough to make the effects of the coherent dynamics appreciable and observable in the presence of losses.}\\

We consider the following Linbland master equation:
\begin{equation}
\label{eq_quantum_master_equation}
\begin{split}
&\ric{\dot{\hat{\rho}}=-i[\hat{H},\hat{\rho}] + \gamma\sum_{j}\left(\hat{L}_j\hat{\rho}\hat{L}_j^\dagger - \frac{1}{2}\left\{\hat{L}_j^\dagger \hat{L}_j,\hat{\rho}\right\}\right),}
\end{split}
\end{equation}
where $\hat{\rho}$ is the state of the system, $\hat{H}$ is the Hamiltonian in Eq.~\eqref{eq_H_full_model} with $\epsilon=0$, \ric{$\hat{L}_j$ is the quantum jump operator acting on site $j$ and $\gamma$ is the corresponding rate. In order to efficiently simulate the Lindbland master equation in Eq.~\eqref{eq_quantum_master_equation}, we resort to the quantum trajectories algorithm, which is based on defining the effective non-Hermitian Hamiltonian
\begin{equation}
\label{eq_effective_Hamiltonian}
\hat{H}_\text{eff} = \hat{H} - i\frac{\gamma}{2}\sum_j \hat{L}_j^\dagger \hat{L}_j,
\end{equation}
and alternating the action of the Hamiltonian given in Eq.~\eqref{eq_effective_Hamiltonian} with the jump operators $\{\hat{L}_j\}$ based on a stochastic process (for the details, we refer to Refs.~\cite{Jaschke2018,Daley2014}). The dynamics of any observable $\hat{\mathcal{O}}$ result from averaging over $N$ different uncorrelated stochastic trajectories labeled by $\eta\in[1,N]$,
\begin{equation}
\label{eq_average}
\ric{\langle\hat{\mathcal{O}}(t)\rangle = \left\langle \mathcal{O}_\eta(t)\right\rangle_\eta,\quad  \mathcal{O}_\eta(t) = {}_\eta\langle \psi(t)| \hat{\mathcal{O}}|\psi(t)\rangle_\eta,}
\end{equation}
\ric{where $|\psi(t)\rangle_\eta$ is the state for a given stochastic trajectory $\eta \in [1,N]$ at time $t$ and $\langle \: \cdot \:\rangle_\eta$ denotes the average over the different trajectories.}
We resort to tensor-network methods for performing the simulations (see Appendix~\ref{appendix_numerical_methods}). 
We consider two different jump operators, namely $\hat{L}_j=\hat{n}_j$ and $\hat{L}_j=\hat{a}_j$. The former corresponds to dephasing, while the latter corresponds to losses}. We choose such \ric{jump operators} in order to investigate the effects of \ric{the environment} when it preserves the ``East'' symmetry, as for the \ric{dephasing process}, or when it does not, as for the \ric{global losses}. \ric{Both situations are relevant in superconducing-circuit setups~\cite{RevModPhys.93.025005}. We compute the observables averaging over $1000$ to $3000$ stochastic realizations} depending on the value of $\gamma$ and the \ric{jump operator}.\\
We study the dynamical properties of \textit{superbosons} $|\widetilde{n}\rangle$ defined in Eq.~\eqref{eq_definition_superbosons}, since they constitute the building blocks of any localized state that we can engineer. Then, we turn our attention to a paradigmatic superposition of \textit{superbosons}, namely the \textit{supercat} state, providing arguments to extend our findings to a class of states to which \textit{supersqueezed} and \textit{supercoherent} states belong. We consider as initial state $|\psi_k(t=0)\rangle=\bigotimes_{j = -\infty}^{k-1}|0\rangle_j \otimes |\widetilde{n}\rangle$, where the subscript $k$ in $|\psi_k(t=0)\rangle$ refers to the position of the first site of the embedded \textit{superboson}.
Since $|\widetilde{n}\rangle$ is localized with localization length $\xi$  (cf. Eq.~\eqref{eq_average_occupation_number}), we can truncate its support to $L' \gg \xi$ sites. Thus, our initial state is
\begin{equation}
\label{eq_psit0}
|\psi_k(t=0)\rangle = |0\rangle_j^{\otimes_{j=-\infty}^{k-1}} \otimes |\widetilde{n}\rangle_{L'}  \otimes |0\rangle_j^{\otimes_{j=k+L'}^{+\infty}},
\end{equation}
where $L'$ is the size of the \textit{superboson} support. 

In a generic non-integrable system, we expect information about   initial states encoded in local observables to be washed out fast. Here, we want to study how localization   and   slow dynamics instead protect the information encoded in local quantities. We compute    the fidelity and the imbalance.
The fidelity (cf. Eq.~\eqref{eq_fidelity})
provides global information about the state and sets an upper bound on the time dependence of the expectation value of any local observable.
Nonetheless, the fidelity is highly sensitive to any local perturbation of the state. Indeed, it is enough to have even a single occupied site far from the superbosons $|\widetilde{n}\rangle$ to make {Eq.}~\eqref{eq_fidelity} negligibly small. Among all the possible local observables, we want to investigate if the initial localized peak remains well resolved. We therefore compute the imbalance between the occupation of the initial peak and the second highest peak in the system, namely
\begin{equation}
\label{eq_imbalance}
\mathcal{I} = \frac{n_{k}-\max_{j\neq k}n_j}{n_{k}+\max_{j\neq k}n_j},
\end{equation} 
where $k$ is the position of the first site of the embedded state (cf. Eq.~\eqref{eq_psit0}).
The imbalance $\mathcal{I} \in [-1,1]$ and for $\mathcal{I}>0$ the initial peak is the largest one in the system.

\begin{figure}[t!]
\centering
\includegraphics[width=\linewidth]{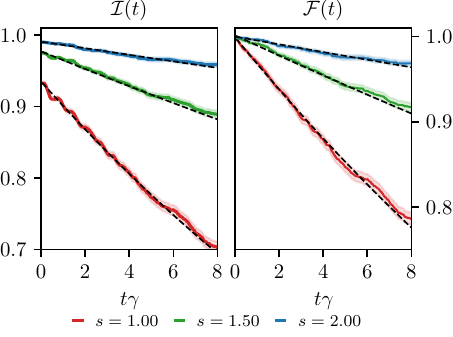}
\caption{The time evolution of the imbalance (cf. {Eq.}~\eqref{eq_imbalance}) and fidelity (cf. Eq.~\eqref{eq_fidelity})  starting from the state {in Eq.}~\eqref{eq_initial_state_ni} and \ric{subjected to the dissipative dynamics given by Eq.~\eqref{eq_quantum_master_equation} with quantum jump operator $\hat{L}_j=\hat{n}_j$}. We fix $n_0=1$, $U=1$\ric{, $\gamma=0.1$} and we swipe across different values of $s\in[1,2]$. The initial value $\mathcal{I}(0)$ ranges from $\approx  \ric{0.93}$ to $\approx 0.99$ as $s$ increases. We show   results for $s = \{1,1.5,2\}$, on top of which we plot the exponential fit (dashed black line). Both plots are in \ric{linear}-linear scale.
The light area surrounding the curves represents the statistical error due to the finite number of sampled trajectories.}
\label{fig_dynamics_noise_n}
\end{figure}

%
%
When dissipation enters in the form of a dephasing noise coupled to the bosonic densities, the Lindbland equation respects the ``East'' symmetry. The jump operators commute with the operator in Eq.~\eqref{eq_operator_commuting_with_H}. Thus, the $n$ excitations on the first site of the \textit{superbosons} $|\widetilde{n}\rangle$ and the empty sites to its left are conserved. Furthermore, since the \ric{the jump operators are} not able to generate excitations out of the vacuum and the state is exponentially localized, we can keep only a few empty sites to the left of $|\widetilde{n}\rangle_{L'}$ without introducing relevant size effects. 
For the set of parameters that we choose, restricting the \textit{superboson} support to $L' \approx 10$ sites and keeping only one empty site to its right turns out to be sufficient. Thus, our initial state is
\begin{equation}
\label{eq_initial_state_ni}
|\psi(t=0)\rangle = |\widetilde{n}_0\rangle_{L'} \otimes |0\rangle.
\end{equation}
In Fig.~\ref{fig_dynamics_noise_n} we show the dynamics of the fidelity and imbalance for different values of $s$ and noise strength $\gamma$ keeping $U=1$, starting from the state in {Eq.}~\eqref{eq_initial_state_ni} with $n_0=1$.
The imbalance displays an exponential decay $\mathcal{I}(t) \sim \mathcal{I}(0)e^{-t/\tau}$, with $\tau$ dependent on the initial state, the parameters of the Hamiltonian, and the coupling strength $\gamma$ with the external bath. 
The decay time $\tau$ increases the more the system is in the localized phase and the larger is the initial occupation $n_0$, while it decreases with the noise strength $\gamma$ as $\tau \propto 1/\gamma$.
Therefore, the time decay $\tau$ can be enhanced by either tuning the parameters of the Hamiltonian or embedding a \textit{superboson} with $n_0$ large (cf. Eq.~\eqref{eq_initial_state_ni}). On the one hand, increasing $s$ or $U$ helps to protect the local memory at longer times, at the cost of making the initial state less entangled. Indeed, in the $s,U\to\infty$ limit, the Hamiltonian tends to $\propto \sum_i (Un_in_{i+1}+n_i)$, the ground state of which is a product state of eigenstates of number operators. 
On the other hand, we can exploit the bosonic nature of the system and embed a \textit{superboson} with a larger initial $n_0$, keeping $s$ small and enhancing the initial state entanglement.
It is important to stress that despite the exponential feature of the decay, the time scale $\tau$ is generally very large with respect to the time scales of the coherent dynamics of the system. 
From Eq.~\eqref{eq_imbalance}, and inspecting the late times average occupation number, the initial peak remains still well resolved and so does the information encoded within it.\\
The fidelity decays exponentially fast in time $\mathcal{F}(t) \sim e^{-t/\tau'}$,
with a decoherence time $\tau'$ dependent on the parameters of the Hamiltonian, the initial state, and the strength of the noise. Analogously to the decay time $\tau$ of the imbalance, the decoherence time $\tau'$ increases the more the system is in the localized phase and decreases with the noise strength $\gamma$ as $\tau' \propto 1/\gamma$.
Contrary to the imbalance, the fidelity drops faster the larger is $n_0$. Indeed, the conserved initial occupation $n_0$ pumps excitations on the next site, reducing the typical coherent time scales by approximatively $1/(n_0 e^{-s})$ and effectively enhancing the effects of the environment.\\
\begin{figure}[t!]
\centering
\includegraphics[width=\linewidth]{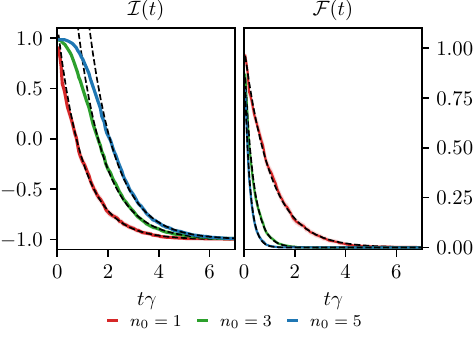}
\caption{The time evolution of the imbalance (cf. {Eq.}~\eqref{eq_imbalance}) and fidelity (cf. Eq.~\eqref{eq_fidelity})  starting from the state \ric{in} \ric{Eq.~\eqref{eq_initial_state_ni}}, with $n_0 \in \ric{\{1,3,5\}}$, and \ric{subjected to the dissipative dynamics given by Eq.~\eqref{eq_quantum_master_equation} with quantum jump operator $\hat{L}_j=\hat{a}_j$}. We fix $U=1$, $s=1.50$ ($e^{-s}\approx 0.22$), and $\gamma=0.1$. The imbalance initial value is $\mathcal{I}(0)\approx 0.99$. In the main figures we show   results for three different values of $n_0 = \ric{\{1,3,5\}}$, on top of which we plot the exponential fit (dashed black line).  The imbalance and fidelity decay as an exponential (the dashed lines are the associated fits).  \ric{Both plots are in linear-linear scale.}
The light area surrounding the curves represents the statistical error due to the finite number of sampled trajectories.}
\label{fig_dynamics_noise_x}
\end{figure}
\ric{Under the action of single-body losses, the dynamics no longer preserve the ``East'' symmetry. Indeed, losses can deplete the occupation of the first site, which fixes the ``East'' symmetry sector.\\ 
Since the vacuum is invariant under the action of losses and coherent dynamics cannot create excitations to the West of the initial embedded \textit{superboson}, we can still consider Eq.~\eqref{eq_initial_state_ni} as our initial state.}
In Fig.~\ref{fig_dynamics_noise_x}, we show the dynamics of the fidelity and imbalance for different values of $n_0$, keeping $U=1$, $s=1.5$ and $\gamma=0.1$ fixed.
\ric{ Losses turn out to be detrimental to the initial state independent of the parameters of the Hamiltonian. Instead, the height of the initial peak plays a substantial role in enhancing the conservation of the imbalance. 
Intuitively, \nic{if the first site $n_0$ is highly occupied at time $t=0$, it will require longer times to drain all the particles. This leads} to an initial \textit{plateaux} in the imbalance, followed by an exponential decay toward the minimum value $\mathcal{I}(t\to\infty)=-1$. 
The decay is well fitted by $\mathcal{I}(t)=\left(A e^{-t/\tau}-1\right)$ at long times, \ricII{where $\tau \propto 1/\gamma$ is the relaxation time} and $A$ is a constant.}
\ric{The insensitivity of the time decay with respect to the parameters of the Hamiltonian indicates that the slow dynamics do not provide additional protection against this type of coupling \ricII{to the environment}. Indeed, the decay of the imbalance is due to the emission of particles from the first occupied site, which fixes the symmetry sector, and since the coherent dynamics cannot create excitations on top of it the initial peak is depleted in time $\propto 1/\gamma$.\\
The fidelity drops to zero exponentially fast, as expected, with a decay time that is parametrically small in the occupation of the initial peak. Indeed, the higher the peak is, the larger is the probability that the emission occurs, which immediately produces a state orthogonal to the initial one.}\\
\ricII{Despite losses being more detrimental with respect to dephasing, we show at the end of the section that the coherent dynamics takes place on time scales that are small with respect to the relaxation time in typical superconducting circuits (cf. Sec.~\ref{section_cQED_implementation} for the experimental implementation of the bosonic quantum East model).}

Note that we can immediately extend our analysis to a large variety of states. For instance, we can consider states given by the superposition of \textit{superbosons} embedded in different regions of the systems, namely
\begin{equation}
|\Psi\rangle \propto |\psi_k(t=0)\rangle + e^{i\theta}|\psi_s(t=0)\rangle_,
\end{equation}
where $|\psi_k(t=0)\rangle$ is defined in {Eq.}~\eqref{eq_psit0}, $\theta$ is a phase, and $|s-k|\gg \xi$. These two states are weakly coupled by the coherent and dissipative dynamics. In a first approximation, we can apply our analysis to each of them separately, and therefore predict their dynamics easily.

The extension of these results to superposition of \textit{superbosons} embedded in the same support (cf. Eq.~\eqref{eq_generic_dressed_state}) is less trivial and depends on the specific \ric{coupling to the environment}. For instance, a coupling that does not preserve the ``East'' symmetry makes the different states dynamically connected, likely leading to different results from the ones observed for the single \textit{superbosons}. On the other hand, a coupling which preserves the ``East'' symmetry can also lead to additional phenomena such as dephasing processes between the superimposed states. Indeed, we observe that \ric{coupling to the densities} is also detrimental. We give further details in Sec.~\ref{section_local_impurities}, exploring the effects of \ric{local dephasing} in the system.
\vspace{10pt}\\



\subsection{\ric{Local dephasing} \label{section_local_impurities}}

We now investigate the effects of \ric{local dephasing} in the dynamical properties of a state given by the superposition of \textit{superbosons} embedded in the same support. Among the possible choices, we consider a paradigmatic \textit{super-Gaussian} state, namely the \textit{supercat} state, and then we generalize.


We consider local \ric{dephasing due to noise coupled to the densities} (see e.g.~\cite{dolgirev2020non}). \ric{In the case of local dephasing acting on a compact support $\mathcal{S}$, the effective theory in Eq.~\eqref{eq_effective_Hamiltonian} turns into
\begin{equation}
\label{eq_effective_H_local}
\hat{H}_\text{eff} = \hat{H} - i\frac{\gamma}{2}\sum_{j\in\mathcal{S}} \hat{L}_j^\dagger \hat{L}_j,
\end{equation}
where the summation is along the support $\mathcal{S}$. We consider $\hat{L}_j=\hat{n}_j$ as jump operator.}

\begin{figure}[t!]
\centering
\includegraphics[width=\linewidth]{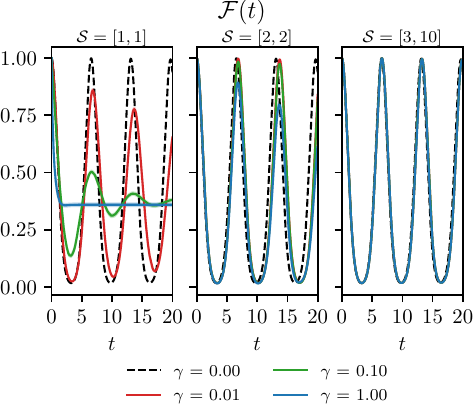}
\caption{
The dynamics of the fidelity ({cf. Eq.}~\eqref{eq_fidelity}) of a \textit{supercat} state with $\alpha=1.5$ upon changing the noise strength $\gamma$ and its support $\mathcal{S}=[a,b]$, starting from site $a$ and ending at site $b$. We fix $U=1$ and $s=1.5$. The initial state is exponentially localized around the site $j=1$. 
The \ric{dephasing} is highly disruptive only in an exponentially localized region around the peak (see the first two columns). Instead, if it acts on a region far from the localized peak, \ric{it} does not produce any appreciable effect \ric{at the scale shown in the plots}. The light area surrounding the curves represents the statistical error due to the finite number of \ric{sampled trajectories}.}
\label{fig_dynamics_super_cat}
\end{figure}

We study the impact of the \ric{dephasing} as a function of the strength $\gamma$ and the extension of its support $\mathcal{S}$. Since the \ric{dephasing} preserves the ``East'' symmetry, we can once again focus on system comprising a few sites without introducing relevant finite-size effects. We initialize our system in the state
\begin{equation}
|\psi(t=0)\rangle = |\widetilde{\mathcal{C}}\rangle_{L},
\end{equation}
where $|\widetilde{\mathcal{C}}\rangle_L$ is a \textit{supercat} state (cf. Eq.~\eqref{eq_super_cat_state}) with support $L$ and average number of particles $|\alpha|^2$. A support of $L=10$ turns out to be large enough for the parameters explored ($\alpha=1.50$, $s=1.5$ and $U=1$). 
In Fig.~\ref{fig_dynamics_super_cat} we show the dynamics of the fidelity as a function of the coupling strength $\gamma$ and support $\mathcal{S}$. The \textit{supercat} state is still localized in space for any $\gamma$ and $\mathcal{S}$. Nonetheless, the coherence of the state is highly dependent on $\gamma$ and $\mathcal{S}$.
Indeed, \ric{local dephasing} is highly disruptive in an exponentially localized region around the peak, where the state is mostly located. If, instead, the \ric{local dephasing}  acts on a region far from the localized peak it does not produce any appreciable effect. More precisely, we estimate that the typical time $\tau$ at which the embedded state is appreciably affected by the noise scales as $\tau \sim \min_{|k-j|\in \mathcal{S}} 1/(\gamma \langle n_j\rangle) \sim   \min_{|k-j|\in \mathcal{S}} e^{|k-j|/\xi}/\gamma$, where $k$ is the site where the peak is located. We numerically verify the polynomial dependence of $\tau$ on $\gamma$. On the contrary, it is not possible to extract the dependence on the support $\mathcal{S}$ with high enough accuracy from the times explored, because of the slowness of the decay.

%
Our findings can be extended to \ric{other} channels that do not \ric{necessarily} preserve the ``East'' symmetry. For instance, \ric{losses} acting far from the localized peak will not affect local information encoded in the localized state.
Furthermore, we expect that the observed dynamical properties can be easily extended to any state prepared via the adiabatic protocol from a state of the form given in {Eq.}~\eqref{eq_general_state_valid_adiabatic_protocol}, to which \textit{super-Gaussian} states belong.\\ 
\\
\ricII{In this section we have discussed the effects of dephasing and losses, without much emphasis on the actual value of the coupling strength $\gamma$ to the environment in typical superconducting circuits (cf. Sec.~\ref{section_cQED_implementation} for the   implementation). As previously mentioned, we set the on-site bare frequency of bosons as our energy scale, which is $\mathcal{O}(\text{GHz})$ in typical superconducting circuits~\cite{RevModPhys.93.025005}. The typical strength of the coupling to the environment $\gamma$ is $\mathcal{O}(\text{MHz})$~\cite{RevModPhys.93.025005}. Therefore, $\gamma \approx 10^{-3}$ in our nondimensional units. As a consequence, 
  coherent dynamics take place on smaller  scales  with respect to the operational times of typical superconducting platforms of $\mathcal{O}(1 \mu\text{s})$, hinting  that the physics of localized states is potentially observable in state-of-the-art experiments. Corroboration of this statement with more quantitative calculations would require an \emph{ab initio} study of the dynamics of the architecture introduced in Sec.~\ref{section_cQED_implementation}, which constitutes   an interesting follow-up project per se.   }
  
\ric{
\section{Superconducting circuit implementation \label{section_cQED_implementation}}}
In this section, we propose an experimental implementation of the Hamiltonian in Eq.~\eqref{eq_H_full_model} in terms of a simple superconducting-circuit \ric{setup}. 
\ric{We consider a chain of driven superconducting qubits. A superconducting \ricII{qubit} is basically a quantized $LC$ oscillator with capacitance $C$ and nonlinear inductance $L$~\cite{RevModPhys.93.025005}}. This nonlinear dependence can be achieved via a Josephson junction working in the superconducting regime without introducing undesired dissipative effects~\cite{JOSEPHSON1962251,tinkham2004introduction,RevModPhys.93.025005}.
\ricII{In particular, we consider here the SNAIL introduced in Ref.~\cite{Frattini2017} as our building block. We consider specifically the SNAIL parameters in Ref.~\cite{PhysRevA.102.062408}, where   kinetically constrained terms (at just two sites) are obtained using  the second-order nonlinearity $\propto (\hat{a}^\dagger \hat{a}^\dagger a + h.c.)$ of the SNAILs. Differently from Ref.~\cite{PhysRevA.102.062408}, we do not use the second-order nonlinearity of SNAILs. Indeed, any superconducting \ricII{qubit} that can be approximated as an anharmonic oscillator with positive anharmonicity could be a suitable candidate for our setup (e.g., the C-shunt flux qubit~\cite{Yan2016}).}\\

We consider an array of $L$ driven superconducting (SC) qubits coupled via an exchange interaction as our starting point. We retain all the energy levels of each SC qubit.  The Hamiltonian can be decomposed as a sum of three terms, $H = H_0 + H_\text{drive} + V$, where
\begin{equation}
\label{eq_H_start_experiment}
\begin{split}
H_0 =& \sum_{j=1}^L \omega_j \hat{a}_j^\dagger \hat{a}_j \ric{+} \frac{E_C}{2} \hat{a}_j^\dagger \hat{a}_j^\dagger \hat{a}_j \hat{a}_j,\\
\ric{H_\text{drive} =}&\ric{ \sum_{j=1}^{L-1} e^{-i \alpha_j t}\left(\Omega_j \hat{a}_j^\dagger + \epsilon_{j+1} \hat{a}_{j+1}^\dagger\right) + h.c.},\\
\ric{V=}&\ric{ \sum_{j=1}^{L-1}g \left(\hat{a}_j \hat{a}_{j+1}^\dagger + h.c.\right)},
\end{split}
\end{equation}
where $\hat{a}_j^\dagger$ ($\hat{a}_j$) creates (destroys) an excitation in the $j$-th SC qubit; $H_0$ is the bare Hamiltonian of the SC qubits with qubit frequencies $\{\omega_j\}_{j=1}^L$, and anharmonicity $E_C>0$~\cite{RevModPhys.93.025005}; $H_\text{drive}$ describes the action of classical drive fields on the bare SC qubits; and $V$ describes hopping processes that can be engineered by a common bus resonator~\cite{DiCarlo2009} or a direct capacitance~\cite{Barends2014}. An illustration of the scheme of {Eq.}~\eqref{eq_H_start_experiment} is given in Fig.~\ref{figintro}(a).\\
We work in the weak-coupling regime $g \ll |\omega_j - \omega_{j+1}|$ and in the low-anharmonicity limit $E_C\ll |\omega_j-\omega_{j+1}|$ for all $j$. The former condition is necessary in order to have far-detuned processes connected by $V$, and therefore to treat $V$ in perturbation theory~\cite{auerbach2012interacting}. The low-anharmonicity limit is necessary to retrieve a bosonic model in the effective perturbative  Hamiltonian achieved after treating $V$ with a Schrieffer-Wolf (SW) transformation in the small $g$ limit. 
Each SC qubit $j\in [1,L-1]$ is driven by a classical drive field of amplitude $\Omega_j$ and frequency $\alpha_j$.
These classical drive fields give rise to the desired interaction together with undesired single-site fields in the low-energy effective Hamiltonian~\cite{PhysRevA.101.052308}. 
In order to get rid of them, we add another drive field on each SC qubit $j\in [2,L]$ of amplitude $\epsilon_j$ and frequency $\alpha_{j-1}$~\cite{PhysRevA.93.060302,PhysRevA.87.030301}.\\

%
We are interested in exploiting the multilevel (bosonic) structure of SC qubits. We do not reduce each component of the system to a qubit. 
We therefore introduce the ladder operators
\begin{equation}
\label{eq_definition_ladder_operators}
\hat{a}_j = \sum_{\ell=0}^\infty \sqrt{\ell+1}|\ell,j\rangle \langle \ell+1,j| \equiv \sum_{\ell=0}^\infty \hat{c}_{\ell,j},
\end{equation}
where $\hat{c}_{\ell,j}$ is the ladder operator which destroys an excitation in the $(\ell+1)$-th level and creates an excitation in the $\ell$-th level on the $j$-th SC qubit. Analogously, we can define its Hermitian conjugate, $\hat{c}^\dagger_{\ell,j}$.\\

We work in the dispersive regime, $g \ll \Delta_{j,j+1}$, where $\Delta_{i,j}=\omega_i - \omega_j$. We perturbatively diagonalize the Hamiltonian $H_0+V$ to second order in $g$ via a SW transformation $S$~\cite{PhysRevA.75.032329}. The drive field terms in $H_\text{drive}$ are modified by the same SW transformation. From now on, we neglect terms of order $\mathcal{O}(g^2\Omega_j/\Delta_{j,j+1}^2)$ and higher. 
We move to the frame that rotates at the frequencies of the drives and we  neglect the fast oscillating terms by employing the rotating-wave approximation (RWA).
Before detailing the calculations, we discuss the physics of each term in the Hamiltonian defined in Eq.~\eqref{eq_H_start_experiment}. The bare Hamiltonian $H_0$ provides the necessary anharmonicity that we desire. The perturbation $V$ gives rise to the nearest-neighbor interaction, a renormalization of the bare energies of the SC qubits, and some additional two-excitation processes. The drive field yields the constrained terms $\hat{n}_j(\hat{a}_{j\pm 1}+\hat{a}_{j\pm 1}^\dagger)$ toward ``East'' and ``West''. The time dependence of the drive fields in the laboratory frame enables us to get rid of the undesired processes, such as the two-excitation processes and the ``West'' terms, passing in the rotating frame of the drive fields and employing the RWA.

In order to find the explicit form of the SW transformation, we follow the prescription in Ref.~\cite{haq2019explicit}. First, we compute $\eta = [H_0,V]$; we consider $\eta$ with arbitrary coefficients as an \textit{ansatz} for $S$. Finally, we fix these coefficients, imposing the condition $[S,H_0]=-V$. We obtain (cf. Appendix~\ref{section_SW_transformation})
\begin{equation}
\label{eq_SW}
\ric{S = \sum_{j=1}^{L-1} \sum_{\ell,s=0}^\infty \frac{g}{\tilde{\Delta}_{\ell,j+1} -\tilde{\Delta}_{s,j}} \left(\hat{c}_{s,j} \hat{c}_{\ell,j+1}^\dagger - \hat{c}_{s,j}^\dagger \hat{c}_{\ell,j+1}\right)},
\end{equation}
where $\tilde{\Delta}_{\ell,j} = (\omega_j \ric{+} E_C \ell)$, the first summation is along the system, while the second summation is along all the levels of the SC qubits. 
Using the Baker-Campbell-Hausdorff expansion, the Hamiltonian in Eq.~\eqref{eq_H_full_model} after the SW transformation reads
\begin{equation}
\begin{split}
\tilde{H} \equiv& e^{S}He^{-S}\\ 
\approx& H_0 + H_\text{drive} +[S,H_\text{drive}] + \frac{1}{2}[S,V] + \mathcal{O}\left(\frac{g^2\Omega}{\Delta^2}\right).
\end{split}
\end{equation}
After lengthy yet standard calculations, we obtain $\tilde{H}$ explicitly dependent on the ladder operators $\hat{c}_{\ell,j}^{(\dagger)}$ introduced in Eq.~\eqref{eq_definition_ladder_operators} and with coefficients dependent on the site and internal levels (see Appendix~\ref{section_after_S}).
Our aim is to write $\tilde{H}$ as a function of the bosonic operators $\hat{a}_j^{(\dagger)}$. We need to find a regime in which the coefficients in $\tilde{H}$ are approximately independent of the specific level, so that we can use Eq.~\eqref{eq_definition_ladder_operators}. These coefficients are similar to the one appearing in Eq.~\eqref{eq_SW}. In order to make them level independent, we need 
\begin{equation}
\label{eq_condition_bosons}
\tilde{\Delta}_{\ell,j+1} -\tilde{\Delta}_{s,j} \approx \omega_{j+1}-\omega_j \equiv \Delta_{j+1,j},
\end{equation}
which holds if $|\ell-s| \ll |\Delta_{j+1,j}|/E_C$. Since the SC qubit can have an infinite number of excitations, we have $(\ell-s) \in (-\infty,+\infty)$. This means that Eq.~\eqref{eq_condition_bosons} cannot be satisfied for all possible $\ell$ and $s$ if $E_C \neq 0$. Nonetheless, it can be achieved up to a certain value $N$ of $\ell$ and $s$, such that $N \ll |\Delta_{j+1,j}/E_C|$. Therefore, the coefficients in $\tilde{H}$ satisfy Eq.~\eqref{eq_condition_bosons} up to the $N$-th energy level, leading to a bosonic Hamiltonian that approximates \ric{the action of the full Hamiltonian to} states with an occupation that is small with respect to $N$ (cf. Appendix~\ref{sez_small_anharmonicity_limit}).
The bosonic $\tilde{H}$ still displays undesired processes, such as hopping and local fields. We move to a rotating frame of reference via the unitary transformation
\begin{equation}
U= \exp\left(it\sum_{j=1}^{L-1}\alpha_j \hat{n}_{j+1}\right)
\end{equation}
and we neglect all the oscillating terms by employing the RWA (cf. Appendix~\ref{section_appendix_RWA}). 
In doing so, we get rid of almost all the undesired processes except for some local fields at the sites $j\geq 2$.
These fields can be eliminated via the additional drive fields of amplitudes $\{\epsilon_j\}$\ric{, analogously to what has been done in similar scenarios (see, e.g., Refs.}~\cite{PhysRevA.93.060302,PhysRevA.87.030301}). We tune their amplitudes such that they cancel the undesired local terms. We obtain the matching condition $\epsilon_j =   g\Omega_{j-1}/\Delta_{j-1,j}$, with $j \geq 2$.
This leads to the effective Hamiltonian
\begin{equation}
\label{eq_H_implemented}
\begin{split}
\tilde{H}=&\sum_{j=1}^L \tilde{\omega}_j \hat{n}_j \ric{+ \frac{E_C}{2}\hat{n}_j\hat{n}_j } +\\
&+ \frac{2g^2 E_C}{\Delta_{j,j+1}^2}\hat{n}_j \hat{n}_{j+1} \ric{+} \frac{g\Omega_j E_C}{\Delta_{j,j+1}^2}\hat{n}_j \left(\hat{a}_{j+1}^\dagger + \hat{a}_{j+1}\right)
\end{split}
\end{equation}
where $\tilde{\omega}_1 = \omega_1 \ric{- E_C/2} + \mathcal{O}(g^2/\Delta_{12})$ and $\tilde{\omega}_{j \neq 1} = \omega_j \ric{-E_C/2} -\alpha_{j-1} + \mathcal{O}(g^2/\Delta_{j,j+1})$.\\
We now evaluate the couplings in Eq.~\eqref{eq_H_implemented}, \ric{considering the SNAIL as our SC \ricII{qubit} and using the parameters of Ref.~\cite{PhysRevA.102.062408}. We work in the parameter regime in which the SNAILs Hamiltonian is given by $H_0$ in Eq.~\eqref{eq_H_start_experiment}. We fix $E_C \approx 150$MHz, $g=75$MHz and $\omega_j \approx 3$GHz. We consider the classical drive fields with amplitude $\Omega_j=-100$MHz  (the amplitude  has to be negative to have the correct sign for the constrained hopping), which can be achieved by adding a $\pi$ phase to the external drive fields. Any real system is inevitably coupled to the environment and SC circuits are no exception. In the context of SC circuits, two different time scales are defined, namely $T_1$ and $T_2$~\cite{RevModPhys.93.025005}. The time scale $T_1$ is the typical time at which the coupling with the environment leads excited states to decay to lower-energy states. The time scale $T_2$ quantifies \nic{the coherence time of the system}. For consistency with the chosen parameters (taken from Ref.~\cite{PhysRevA.102.062408}), we also consider, as $T_1$ and $T_2$, the values from Ref.~\cite{PhysRevA.102.062408}, which are $T_{1,2}\approx 1\mu$s}. We fix the qubit frequencies $\omega_j$ and the drive field frequencies $\alpha_j$ in order to satisfy: ({\it i}) the dispersive regime, valid for $g/\Delta_{j,j+1}\ll 1$; ({\it ii}) the low-anharmonicity limit, $E_C \ll \Delta_{j,j+1}$; ({\it iii}) the validity of the RWA, namely $|\alpha_j|\gg \Omega_j$\ric{,} $|\alpha_{j+1}-\alpha_j|\gg \Omega_j$ \ric{and $|\alpha_{j+2}-\alpha_{j}|>g\Omega_{j+2}/\Delta_{j+1,j+2}$}; ({\it iv}) $\tilde{\omega}_j\approx \omega_j-\alpha_{j-1}>0$ for $j>1$, necessary in order to have localization; ({\it v}) $1/T_{1,2}$ small with respect to the typical energies in the effective Hamiltonian in Eq.~\eqref{eq_H_implemented}\ric{; and ({\it vi}) the system is in the localized phase}.\\
The more stringent conditions are given by ({\it ii}) and ({\it v}). 
A good trade-off between ({\it ii}) and ({\it v}) is obtained  at $|\Delta_{j,j+1}|\equiv \Delta \approx 5 E_C \approx \ric{750\text{MHz}}$, for which the typical time scale of the kinetically constrained term is approximately $T_{1,2}/2$. 
\begin{table}[t!]
\centering
\begin{tabular}{|c|c|c|c|c|c|}
\hline 
 & $j=1$ & $j=2$ & $j=3$ & $j=4$ & $j=5$ \\ 
\hline 
\ric{$\alpha_j$(GHz)} & \ric{0.75} & \ric{1.6} & \ric{0.65} & \ric{1.7} & \ric{0.75 }\\ 
\hline 
\ric{$\omega_j$(GHz)} & \ric{3} & \ric{3.75} & \ric{4.5} & \ric{3.75} & \ric{4.5} \\ 
\hline 
\end{tabular} 
\caption{\ric{A possible configuration for the external classical drive field frequencies $\{\alpha_j\}$ and bare frequency $\{\omega_j\}$ of SNAILs for the experimental implementation of the bosonic quantum East model in a system of size $L=5$. For bigger system sizes, it is enough to periodically repeat the configuration from site $j=2$ to $j=5$. The other parameters are as follows: anharmonicity $E_C=150$MHz, bare capacitive coupling $g=75$MHz, and classical drive field amplitude $\Omega=-100$MHz.}}
\label{table_parameters}
\end{table}
We have $g/\Delta_{j,j+1}\approx \ric{0.1}$, meaning that ({\it i}) is reasonably satisfied. Condition ({\it iii}) is satisfied by a staggered configuration of the drive field frequencies \ric{with an additional dishomogeneity between next-neighbor drive field frequencies, for instance: 
\ric{$\alpha_{j}=\alpha_{j-1}+(-1)^{\ric{j}} (\delta + (j-1) \zeta)$ for $j \in [2,4]$} and boundary condition $\alpha_1 \gg \Omega$ \ric{(for larger systems, it is enough to periodically repeat the configuration of the frequencies)}, with $\delta \gg \Omega$, $\alpha_1 \gg \Omega$, and $\zeta \gg g\Omega/\Delta \approx 10$MHz. Condition ({\it iv}) is satisfied by a staggered configuration of the qubit frequencies as well: $\omega_{j+1}=\omega_{j}+(-1)^{j}\Delta$ for $j\geq 2$, $\omega_2 = \omega_1 +\Delta$}, with boundary condition $\omega_1>  \alpha_1$. For instance, we can consider $\alpha_1=$\ric{$750$MHz, $\delta=750$MHz, $\zeta=100$MHz}, and $\omega_1=\ric{3}$GHz. 
These conditions lead to Eq.~\eqref{eq_H_implemented} being \ric{almost} translationally invariant \ric{(except for dishomogeneities in the frequencies $\tilde{\omega}_j$ of the order of approximately $5\%$, which can be eliminated via a more fine-tuned choice of $\{\omega_j\}$)}. Moreover, condition ({\it vi}) is satisfied for these set of parameters. In Table~\ref{table_parameters}, we summarize a possible set of parameters available in state-of-the-art superconducting circuits for implementing the bosonic quantum East model.
\vspace{20pt}
\section{Perspectives}

The implementation of a kinetically constrained East model using \ric{superconducting circuits} represents a   bridge between the two   communities of circuit-QED and nonergodic quantum dynamics. 
It has the potential to attract the former toward fundamental questions regarding dynamical phase transitions and to stimulate the latter toward the  search  for quantum-information and metrological applications of constrained dynamics. Our explicit construction of localized analogs of squeezed and cat states relying on the East constraint represents a first stepping stone in this direction.

A fruitful prosecution of this work is the study of an analog of the mobility edge  separating localized from delocalized states in the spectrum of East models (for the mobility edge in  MBL see Refs.~\cite{nandkishore2015many,abanin2019colloquium}). An understanding of how such a mobility edge scales with $\Lambda$, is essential for predicting the onset of dynamical   transitions in platforms with unidirectional constraints, as well as of practical interest. For instance, a mobility edge at finite energy density is a feature of  direct relevance for experimental realizations, since it would yield the conditions for performing efficient   quantum manipulations deep in the localized phase when    finite-temperature  or heating effects are present. A related interesting question is the survival of the effective integrable description of the   localized phase discussed in Sec.~\ref{section_state_preparation} upon increasing the density of  energy above the ground state. This   would   have implications for    heat and particle transport features of the   East model  in the nonergodic phase, which would be governed by the effective integrable description in~\eqref{eq_H_quadratic}, as it happens for      MBL systems~\cite{vznidarivc2016diffusive}. 
 
The insensitivity to noise acting away from  localized peaks could open up a path toward the study of the protection of spatially separated  macroscopic superpositions of superbosonic states.   Given the slow decay of localized wave packets in the presence of noise, one could conceive the storage and noise resilience of long-lived many-body entangled states in faraway regions, with applications to  quantum communication.   

To conclude, we observe that the implementation discussed in Sec.~\ref{section_state_preparation} may be easily adapted to retain kinetic terms with both East and West symmetries. This could, for instance, lead to the formation of localized modes at edges of the wire, with exciting perspectives for novel forms of topological states in kinetically constrained models that are realizable with circuit QED. We are currently focusing our research efforts in this direction.  \\

\emph{Acknowledgements} We are indebted to S. M. Girvin for careful proof-reading of the manuscript and for providing valuable comments. We thank Mari Carmen Ba\~nuls and Juan P. Garrahan for insightful discussions.  
This project has been supported by the Deutsche Forschungsgemeinschaft (DFG, German Research Foundation) through the Project-ID 429529648 -- TRR 306 QuCoLiMa (``Quantum Cooperativity of Light and Matter''), and the  grant HADEQUAM- MA7003/3-1; and by the Dynamics and Topology Center funded by the State of Rhineland Palatinate. Parts of this research were conducted using the Mogon supercomputer and/or advisory services offered by Johannes Gutenberg University Mainz (hpc.uni-mainz.de), which is a member of the AHRP (Alliance for High Performance Computing in Rhineland Palatinate,  www.ahrp.info) and the Gauss Alliance e.V. We gratefully acknowledge the computing time granted on the Mogon supercomputer at Johannes Gutenberg University Mainz (hpc.uni-mainz.de) through the project ``DysQCorr''.
\bibliography{bosonic_quantum_east_model_draft}

\begin{thebibliography}{112}%
\makeatletter
\providecommand \@ifxundefined [1]{%
 \@ifx{#1\undefined}
}%
\providecommand \@ifnum [1]{%
 \ifnum #1\expandafter \@firstoftwo
 \else \expandafter \@secondoftwo
 \fi
}%
\providecommand \@ifx [1]{%
 \ifx #1\expandafter \@firstoftwo
 \else \expandafter \@secondoftwo
 \fi
}%
\providecommand \natexlab [1]{#1}%
\providecommand \enquote  [1]{``#1''}%
\providecommand \bibnamefont  [1]{#1}%
\providecommand \bibfnamefont [1]{#1}%
\providecommand \citenamefont [1]{#1}%
\providecommand \href@noop [0]{\@secondoftwo}%
\providecommand \href [0]{\begingroup \@sanitize@url \@href}%
\providecommand \@href[1]{\@@startlink{#1}\@@href}%
\providecommand \@@href[1]{\endgroup#1\@@endlink}%
\providecommand \@sanitize@url [0]{\catcode `\\12\catcode `\$12\catcode
  `\&12\catcode `\#12\catcode `\^12\catcode `\_12\catcode `\%12\relax}%
\providecommand \@@startlink[1]{}%
\providecommand \@@endlink[0]{}%
\providecommand \url  [0]{\begingroup\@sanitize@url \@url }%
\providecommand \@url [1]{\endgroup\@href {#1}{\urlprefix }}%
\providecommand \urlprefix  [0]{URL }%
\providecommand \Eprint [0]{\href }%
\providecommand \doibase [0]{https://doi.org/}%
\providecommand \selectlanguage [0]{\@gobble}%
\providecommand \bibinfo  [0]{\@secondoftwo}%
\providecommand \bibfield  [0]{\@secondoftwo}%
\providecommand \translation [1]{[#1]}%
\providecommand \BibitemOpen [0]{}%
\providecommand \bibitemStop [0]{}%
\providecommand \bibitemNoStop [0]{.\EOS\space}%
\providecommand \EOS [0]{\spacefactor3000\relax}%
\providecommand \BibitemShut  [1]{\csname bibitem#1\endcsname}%
\let\auto@bib@innerbib\@empty
\bibitem [{\citenamefont {Preskill}(2018)}]{preskill2018quantum}%
  \BibitemOpen
  \bibfield  {author} {\bibinfo {author} {\bibfnamefont {J.}~\bibnamefont
  {Preskill}},\ }\bibfield  {title} {\bibinfo {title} {Quantum computing in the
  nisq era and beyond},\ }\href@noop {} {\bibfield  {journal} {\bibinfo
  {journal} {Quantum}\ }\textbf {\bibinfo {volume} {2}},\ \bibinfo {pages} {79}
  (\bibinfo {year} {2018})}\BibitemShut {NoStop}%
\bibitem [{\citenamefont {Polkovnikov}\ \emph {et~al.}(2011)\citenamefont
  {Polkovnikov}, \citenamefont {Sengupta}, \citenamefont {Silva},\ and\
  \citenamefont {Vengalattore}}]{polkovnikov2011colloquium}%
  \BibitemOpen
  \bibfield  {author} {\bibinfo {author} {\bibfnamefont {A.}~\bibnamefont
  {Polkovnikov}}, \bibinfo {author} {\bibfnamefont {K.}~\bibnamefont
  {Sengupta}}, \bibinfo {author} {\bibfnamefont {A.}~\bibnamefont {Silva}},\
  and\ \bibinfo {author} {\bibfnamefont {M.}~\bibnamefont {Vengalattore}},\
  }\bibfield  {title} {\bibinfo {title} {Colloquium: Nonequilibrium dynamics of
  closed interacting quantum systems},\ }\href@noop {} {\bibfield  {journal}
  {\bibinfo  {journal} {Reviews of Modern Physics}\ }\textbf {\bibinfo {volume}
  {83}},\ \bibinfo {pages} {863} (\bibinfo {year} {2011})}\BibitemShut
  {NoStop}%
\bibitem [{\citenamefont {Carleo}\ \emph {et~al.}(2012)\citenamefont {Carleo},
  \citenamefont {Becca}, \citenamefont {Schir{\'o}},\ and\ \citenamefont
  {Fabrizio}}]{Carleo2012}%
  \BibitemOpen
  \bibfield  {author} {\bibinfo {author} {\bibfnamefont {G.}~\bibnamefont
  {Carleo}}, \bibinfo {author} {\bibfnamefont {F.}~\bibnamefont {Becca}},
  \bibinfo {author} {\bibfnamefont {M.}~\bibnamefont {Schir{\'o}}},\ and\
  \bibinfo {author} {\bibfnamefont {M.}~\bibnamefont {Fabrizio}},\ }\bibfield
  {title} {\bibinfo {title} {Localization and glassy dynamics of many-body
  quantum systems},\ }\href {https://doi.org/10.1038/srep00243} {\bibfield
  {journal} {\bibinfo  {journal} {Scientific Reports}\ }\textbf {\bibinfo
  {volume} {2}},\ \bibinfo {pages} {243} (\bibinfo {year} {2012})}\BibitemShut
  {NoStop}%
\bibitem [{\citenamefont {Doggen}\ \emph {et~al.}(2021)\citenamefont {Doggen},
  \citenamefont {Gornyi},\ and\ \citenamefont
  {Polyakov}}]{PhysRevB.103.L100202}%
  \BibitemOpen
  \bibfield  {author} {\bibinfo {author} {\bibfnamefont {E.~V.~H.}\
  \bibnamefont {Doggen}}, \bibinfo {author} {\bibfnamefont {I.~V.}\
  \bibnamefont {Gornyi}},\ and\ \bibinfo {author} {\bibfnamefont {D.~G.}\
  \bibnamefont {Polyakov}},\ }\bibfield  {title} {\bibinfo {title} {Stark
  many-body localization: Evidence for hilbert-space shattering},\ }\href
  {https://doi.org/10.1103/PhysRevB.103.L100202} {\bibfield  {journal}
  {\bibinfo  {journal} {Phys. Rev. B}\ }\textbf {\bibinfo {volume} {103}},\
  \bibinfo {pages} {L100202} (\bibinfo {year} {2021})}\BibitemShut {NoStop}%
\bibitem [{\citenamefont {{De Roeck}}\ and\ \citenamefont
  {{Huveneers}}(2014)}]{DeRoeck2014}%
  \BibitemOpen
  \bibfield  {author} {\bibinfo {author} {\bibfnamefont {W.}~\bibnamefont {{De
  Roeck}}}\ and\ \bibinfo {author} {\bibfnamefont {F.}~\bibnamefont
  {{Huveneers}}},\ }\bibfield  {title} {\bibinfo {title} {Scenario for
  delocalization in translation-invariant systems},\ }\href
  {https://doi.org/10.1103/PhysRevB.90.165137} {\bibfield  {journal} {\bibinfo
  {journal} {Phys. Rev. B}\ }\textbf {\bibinfo {volume} {90}},\ \bibinfo
  {pages} {165137} (\bibinfo {year} {2014})}\BibitemShut {NoStop}%
\bibitem [{\citenamefont {Schiulaz}\ \emph {et~al.}(2015)\citenamefont
  {Schiulaz}, \citenamefont {Silva},\ and\ \citenamefont
  {M\"uller}}]{Schiulaz2015}%
  \BibitemOpen
  \bibfield  {author} {\bibinfo {author} {\bibfnamefont {M.}~\bibnamefont
  {Schiulaz}}, \bibinfo {author} {\bibfnamefont {A.}~\bibnamefont {Silva}},\
  and\ \bibinfo {author} {\bibfnamefont {M.}~\bibnamefont {M\"uller}},\
  }\bibfield  {title} {\bibinfo {title} {Dynamics in many-body localized
  quantum systems without disorder},\ }\href
  {https://doi.org/10.1103/PhysRevB.91.184202} {\bibfield  {journal} {\bibinfo
  {journal} {Phys. Rev. B}\ }\textbf {\bibinfo {volume} {91}},\ \bibinfo
  {pages} {184202} (\bibinfo {year} {2015})}\BibitemShut {NoStop}%
\bibitem [{\citenamefont {Papi{\'c}}\ \emph {et~al.}(2015)\citenamefont
  {Papi{\'c}}, \citenamefont {Stoudenmire},\ and\ \citenamefont
  {Abanin}}]{Papic2015}%
  \BibitemOpen
  \bibfield  {author} {\bibinfo {author} {\bibfnamefont {Z.}~\bibnamefont
  {Papi{\'c}}}, \bibinfo {author} {\bibfnamefont {E.~M.}\ \bibnamefont
  {Stoudenmire}},\ and\ \bibinfo {author} {\bibfnamefont {D.~A.}\ \bibnamefont
  {Abanin}},\ }\bibfield  {title} {\bibinfo {title} {Many-body localization in
  disorder-free systems: The importance of finite-size constraints},\ }\href
  {https://doi.org/https://doi.org/10.1016/j.aop.2015.08.024} {\bibfield
  {journal} {\bibinfo  {journal} {Ann. of Phys.}\ }\textbf {\bibinfo {volume}
  {362}},\ \bibinfo {pages} {714 } (\bibinfo {year} {2015})}\BibitemShut
  {NoStop}%
\bibitem [{\citenamefont {Barbiero}\ \emph {et~al.}(2015)\citenamefont
  {Barbiero}, \citenamefont {Menotti}, \citenamefont {Recati},\ and\
  \citenamefont {Santos}}]{Barbiero2015}%
  \BibitemOpen
  \bibfield  {author} {\bibinfo {author} {\bibfnamefont {L.}~\bibnamefont
  {Barbiero}}, \bibinfo {author} {\bibfnamefont {C.}~\bibnamefont {Menotti}},
  \bibinfo {author} {\bibfnamefont {A.}~\bibnamefont {Recati}},\ and\ \bibinfo
  {author} {\bibfnamefont {L.}~\bibnamefont {Santos}},\ }\bibfield  {title}
  {\bibinfo {title} {Out-of-equilibrium states and quasi-many-body localization
  in polar lattice gases},\ }\href {https://doi.org/10.1103/PhysRevB.92.180406}
  {\bibfield  {journal} {\bibinfo  {journal} {Phys. Rev. B}\ }\textbf {\bibinfo
  {volume} {92}},\ \bibinfo {pages} {180406(R)} (\bibinfo {year}
  {2015})}\BibitemShut {NoStop}%
\bibitem [{\citenamefont {Yao}\ \emph {et~al.}(2016)\citenamefont {Yao},
  \citenamefont {Laumann}, \citenamefont {Cirac}, \citenamefont {Lukin},\ and\
  \citenamefont {Moore}}]{Yao2016}%
  \BibitemOpen
  \bibfield  {author} {\bibinfo {author} {\bibfnamefont {N.~Y.}\ \bibnamefont
  {Yao}}, \bibinfo {author} {\bibfnamefont {C.~R.}\ \bibnamefont {Laumann}},
  \bibinfo {author} {\bibfnamefont {J.~I.}\ \bibnamefont {Cirac}}, \bibinfo
  {author} {\bibfnamefont {M.~D.}\ \bibnamefont {Lukin}},\ and\ \bibinfo
  {author} {\bibfnamefont {J.~E.}\ \bibnamefont {Moore}},\ }\bibfield  {title}
  {\bibinfo {title} {Quasi-many-body localization in translation-invariant
  systems},\ }\href {https://doi.org/10.1103/PhysRevLett.117.240601} {\bibfield
   {journal} {\bibinfo  {journal} {Phys. Rev. Lett.}\ }\textbf {\bibinfo
  {volume} {117}},\ \bibinfo {pages} {240601} (\bibinfo {year}
  {2016})}\BibitemShut {NoStop}%
\bibitem [{\citenamefont {Smith}\ \emph {et~al.}(2017)\citenamefont {Smith},
  \citenamefont {Knolle}, \citenamefont {Kovrizhin},\ and\ \citenamefont
  {Moessner}}]{Smith2017}%
  \BibitemOpen
  \bibfield  {author} {\bibinfo {author} {\bibfnamefont {A.}~\bibnamefont
  {Smith}}, \bibinfo {author} {\bibfnamefont {J.}~\bibnamefont {Knolle}},
  \bibinfo {author} {\bibfnamefont {D.~L.}\ \bibnamefont {Kovrizhin}},\ and\
  \bibinfo {author} {\bibfnamefont {R.}~\bibnamefont {Moessner}},\ }\bibfield
  {title} {\bibinfo {title} {Disorder-free localization},\ }\href
  {https://doi.org/10.1103/PhysRevLett.118.266601} {\bibfield  {journal}
  {\bibinfo  {journal} {Phys. Rev. Lett.}\ }\textbf {\bibinfo {volume} {118}},\
  \bibinfo {pages} {266601} (\bibinfo {year} {2017})}\BibitemShut {NoStop}%
\bibitem [{\citenamefont {Mondaini}\ \emph {et~al.}(2018)\citenamefont
  {Mondaini}, \citenamefont {Mallayya}, \citenamefont {Santos},\ and\
  \citenamefont {Rigol}}]{Mondaini2018}%
  \BibitemOpen
  \bibfield  {author} {\bibinfo {author} {\bibfnamefont {R.}~\bibnamefont
  {Mondaini}}, \bibinfo {author} {\bibfnamefont {K.}~\bibnamefont {Mallayya}},
  \bibinfo {author} {\bibfnamefont {L.~F.}\ \bibnamefont {Santos}},\ and\
  \bibinfo {author} {\bibfnamefont {M.}~\bibnamefont {Rigol}},\ }\bibfield
  {title} {\bibinfo {title} {Comment on ``systematic construction of
  counterexamples to the eigenstate thermalization hypothesis''},\ }\href
  {https://doi.org/10.1103/PhysRevLett.121.038901} {\bibfield  {journal}
  {\bibinfo  {journal} {Phys. Rev. Lett.}\ }\textbf {\bibinfo {volume} {121}},\
  \bibinfo {pages} {038901} (\bibinfo {year} {2018})}\BibitemShut {NoStop}%
\bibitem [{\citenamefont {Schulz}\ \emph {et~al.}(2019)\citenamefont {Schulz},
  \citenamefont {Hooley}, \citenamefont {Moessner},\ and\ \citenamefont
  {Pollmann}}]{Schulz2019}%
  \BibitemOpen
  \bibfield  {author} {\bibinfo {author} {\bibfnamefont {M.}~\bibnamefont
  {Schulz}}, \bibinfo {author} {\bibfnamefont {C.~A.}\ \bibnamefont {Hooley}},
  \bibinfo {author} {\bibfnamefont {R.}~\bibnamefont {Moessner}},\ and\
  \bibinfo {author} {\bibfnamefont {F.}~\bibnamefont {Pollmann}},\ }\bibfield
  {title} {\bibinfo {title} {Stark many-body localization},\ }\href
  {https://doi.org/10.1103/PhysRevLett.122.040606} {\bibfield  {journal}
  {\bibinfo  {journal} {Phys. Rev. Lett.}\ }\textbf {\bibinfo {volume} {122}},\
  \bibinfo {pages} {040606} (\bibinfo {year} {2019})}\BibitemShut {NoStop}%
\bibitem [{\citenamefont {{van Nieuwenburg}}\ \emph {et~al.}(2019)\citenamefont
  {{van Nieuwenburg}}, \citenamefont {{Baum}},\ and\ \citenamefont
  {{Refael}}}]{Nieuwenburg2019}%
  \BibitemOpen
  \bibfield  {author} {\bibinfo {author} {\bibfnamefont {E.}~\bibnamefont {{van
  Nieuwenburg}}}, \bibinfo {author} {\bibfnamefont {Y.}~\bibnamefont
  {{Baum}}},\ and\ \bibinfo {author} {\bibfnamefont {G.}~\bibnamefont
  {{Refael}}},\ }\bibfield  {title} {\bibinfo {title} {From bloch oscillations
  to many-body localization in clean interacting systems},\ }\href
  {https://doi.org/10.1073/pnas.1819316116} {\bibfield  {journal} {\bibinfo
  {journal} {Proc. Natl. Acad. Sci. USA}\ }\textbf {\bibinfo {volume} {116}},\
  \bibinfo {pages} {9269} (\bibinfo {year} {2019})}\BibitemShut {NoStop}%
\bibitem [{\citenamefont {Shiraishi}\ and\ \citenamefont
  {Mori}(2018)}]{Shiraishi_Mori_Reply2018}%
  \BibitemOpen
  \bibfield  {author} {\bibinfo {author} {\bibfnamefont {N.}~\bibnamefont
  {Shiraishi}}\ and\ \bibinfo {author} {\bibfnamefont {T.}~\bibnamefont
  {Mori}},\ }\bibfield  {title} {\bibinfo {title} {Shiraishi and mori reply},\
  }\href {https://doi.org/10.1103/PhysRevLett.121.038902} {\bibfield  {journal}
  {\bibinfo  {journal} {Phys. Rev. Lett.}\ }\textbf {\bibinfo {volume} {121}},\
  \bibinfo {pages} {038902} (\bibinfo {year} {2018})}\BibitemShut {NoStop}%
\bibitem [{\citenamefont {Kormos}\ \emph {et~al.}(2017)\citenamefont {Kormos},
  \citenamefont {Collura}, \citenamefont {Tak{\'a}cs},\ and\ \citenamefont
  {Calabrese}}]{kormos2017real}%
  \BibitemOpen
  \bibfield  {author} {\bibinfo {author} {\bibfnamefont {M.}~\bibnamefont
  {Kormos}}, \bibinfo {author} {\bibfnamefont {M.}~\bibnamefont {Collura}},
  \bibinfo {author} {\bibfnamefont {G.}~\bibnamefont {Tak{\'a}cs}},\ and\
  \bibinfo {author} {\bibfnamefont {P.}~\bibnamefont {Calabrese}},\ }\bibfield
  {title} {\bibinfo {title} {Real-time confinement following a quantum quench
  to a non-integrable model},\ }\href@noop {} {\bibfield  {journal} {\bibinfo
  {journal} {Nature Physics}\ }\textbf {\bibinfo {volume} {13}},\ \bibinfo
  {pages} {246} (\bibinfo {year} {2017})}\BibitemShut {NoStop}%
\bibitem [{\citenamefont {James}\ \emph {et~al.}(2019)\citenamefont {James},
  \citenamefont {Konik},\ and\ \citenamefont {Robinson}}]{James2019}%
  \BibitemOpen
  \bibfield  {author} {\bibinfo {author} {\bibfnamefont {A.~J.~A.}\
  \bibnamefont {James}}, \bibinfo {author} {\bibfnamefont {R.~M.}\ \bibnamefont
  {Konik}},\ and\ \bibinfo {author} {\bibfnamefont {N.~J.}\ \bibnamefont
  {Robinson}},\ }\bibfield  {title} {\bibinfo {title} {Nonthermal states
  arising from confinement in one and two dimensions},\ }\href
  {https://doi.org/10.1103/PhysRevLett.122.130603} {\bibfield  {journal}
  {\bibinfo  {journal} {Phys. Rev. Lett.}\ }\textbf {\bibinfo {volume} {122}},\
  \bibinfo {pages} {130603} (\bibinfo {year} {2019})}\BibitemShut {NoStop}%
\bibitem [{\citenamefont {Morong}\ \emph {et~al.}(2021)\citenamefont {Morong},
  \citenamefont {Liu}, \citenamefont {Becker}, \citenamefont {Collins},
  \citenamefont {Feng}, \citenamefont {Kyprianidis}, \citenamefont {Pagano},
  \citenamefont {You}, \citenamefont {Gorshkov},\ and\ \citenamefont
  {Monroe}}]{morong2021observation}%
  \BibitemOpen
  \bibfield  {author} {\bibinfo {author} {\bibfnamefont {W.}~\bibnamefont
  {Morong}}, \bibinfo {author} {\bibfnamefont {F.}~\bibnamefont {Liu}},
  \bibinfo {author} {\bibfnamefont {P.}~\bibnamefont {Becker}}, \bibinfo
  {author} {\bibfnamefont {K.}~\bibnamefont {Collins}}, \bibinfo {author}
  {\bibfnamefont {L.}~\bibnamefont {Feng}}, \bibinfo {author} {\bibfnamefont
  {A.}~\bibnamefont {Kyprianidis}}, \bibinfo {author} {\bibfnamefont
  {G.}~\bibnamefont {Pagano}}, \bibinfo {author} {\bibfnamefont
  {T.}~\bibnamefont {You}}, \bibinfo {author} {\bibfnamefont {A.}~\bibnamefont
  {Gorshkov}},\ and\ \bibinfo {author} {\bibfnamefont {C.}~\bibnamefont
  {Monroe}},\ }\bibfield  {title} {\bibinfo {title} {Observation of stark
  many-body localization without disorder},\ }\href@noop {} {\bibfield
  {journal} {\bibinfo  {journal} {arXiv preprint arXiv:2102.07250}\ } (\bibinfo
  {year} {2021})}\BibitemShut {NoStop}%
\bibitem [{\citenamefont {Gunawardana}\ and\ \citenamefont
  {Bu{\v{c}}a}(2021)}]{gunawardana2021dynamical}%
  \BibitemOpen
  \bibfield  {author} {\bibinfo {author} {\bibfnamefont {T.}~\bibnamefont
  {Gunawardana}}\ and\ \bibinfo {author} {\bibfnamefont {B.}~\bibnamefont
  {Bu{\v{c}}a}},\ }\bibfield  {title} {\bibinfo {title} {Dynamical l-bits in
  stark many-body localization},\ }\href@noop {} {\bibfield  {journal}
  {\bibinfo  {journal} {arXiv preprint arXiv:2110.13135}\ } (\bibinfo {year}
  {2021})}\BibitemShut {NoStop}%
\bibitem [{\citenamefont {Buca}\ \emph {et~al.}(2020)\citenamefont {Buca},
  \citenamefont {Purkayastha}, \citenamefont {Guarnieri}, \citenamefont
  {Mitchison}, \citenamefont {Jaksch},\ and\ \citenamefont
  {Goold}}]{buca2020quantum}%
  \BibitemOpen
  \bibfield  {author} {\bibinfo {author} {\bibfnamefont {B.}~\bibnamefont
  {Buca}}, \bibinfo {author} {\bibfnamefont {A.}~\bibnamefont {Purkayastha}},
  \bibinfo {author} {\bibfnamefont {G.}~\bibnamefont {Guarnieri}}, \bibinfo
  {author} {\bibfnamefont {M.~T.}\ \bibnamefont {Mitchison}}, \bibinfo {author}
  {\bibfnamefont {D.}~\bibnamefont {Jaksch}},\ and\ \bibinfo {author}
  {\bibfnamefont {J.}~\bibnamefont {Goold}},\ }\bibfield  {title} {\bibinfo
  {title} {Quantum many-body attractors},\ }\href@noop {} {\bibfield  {journal}
  {\bibinfo  {journal} {arXiv preprint arXiv:2008.11166}\ } (\bibinfo {year}
  {2020})}\BibitemShut {NoStop}%
\bibitem [{\citenamefont {Nandkishore}\ and\ \citenamefont
  {Huse}(2015)}]{nandkishore2015many}%
  \BibitemOpen
  \bibfield  {author} {\bibinfo {author} {\bibfnamefont {R.}~\bibnamefont
  {Nandkishore}}\ and\ \bibinfo {author} {\bibfnamefont {D.~A.}\ \bibnamefont
  {Huse}},\ }\bibfield  {title} {\bibinfo {title} {Many-body localization and
  thermalization in quantum statistical mechanics},\ }\href@noop {} {\bibfield
  {journal} {\bibinfo  {journal} {Annu. Rev. Condens. Matter Phys.}\ }\textbf
  {\bibinfo {volume} {6}},\ \bibinfo {pages} {15} (\bibinfo {year}
  {2015})}\BibitemShut {NoStop}%
\bibitem [{\citenamefont {Abanin}\ \emph {et~al.}(2019)\citenamefont {Abanin},
  \citenamefont {Altman}, \citenamefont {Bloch},\ and\ \citenamefont
  {Serbyn}}]{abanin2019colloquium}%
  \BibitemOpen
  \bibfield  {author} {\bibinfo {author} {\bibfnamefont {D.~A.}\ \bibnamefont
  {Abanin}}, \bibinfo {author} {\bibfnamefont {E.}~\bibnamefont {Altman}},
  \bibinfo {author} {\bibfnamefont {I.}~\bibnamefont {Bloch}},\ and\ \bibinfo
  {author} {\bibfnamefont {M.}~\bibnamefont {Serbyn}},\ }\bibfield  {title}
  {\bibinfo {title} {Colloquium: Many-body localization, thermalization, and
  entanglement},\ }\href@noop {} {\bibfield  {journal} {\bibinfo  {journal}
  {Reviews of Modern Physics}\ }\textbf {\bibinfo {volume} {91}},\ \bibinfo
  {pages} {021001} (\bibinfo {year} {2019})}\BibitemShut {NoStop}%
\bibitem [{\citenamefont {Ritort}\ and\ \citenamefont
  {Sollich}(2003)}]{ritort2003glassy}%
  \BibitemOpen
  \bibfield  {author} {\bibinfo {author} {\bibfnamefont {F.}~\bibnamefont
  {Ritort}}\ and\ \bibinfo {author} {\bibfnamefont {P.}~\bibnamefont
  {Sollich}},\ }\bibfield  {title} {\bibinfo {title} {Glassy dynamics of
  kinetically constrained models},\ }\href@noop {} {\bibfield  {journal}
  {\bibinfo  {journal} {Advances in physics}\ }\textbf {\bibinfo {volume}
  {52}},\ \bibinfo {pages} {219} (\bibinfo {year} {2003})}\BibitemShut
  {NoStop}%
\bibitem [{\citenamefont {Chamon}(2005)}]{Chamon2005}%
  \BibitemOpen
  \bibfield  {author} {\bibinfo {author} {\bibfnamefont {C.}~\bibnamefont
  {Chamon}},\ }\bibfield  {title} {\bibinfo {title} {Quantum glassiness in
  strongly correlated clean systems: An example of topological
  overprotection},\ }\href {https://doi.org/10.1103/PhysRevLett.94.040402}
  {\bibfield  {journal} {\bibinfo  {journal} {Phys. Rev. Lett.}\ }\textbf
  {\bibinfo {volume} {94}},\ \bibinfo {pages} {040402} (\bibinfo {year}
  {2005})}\BibitemShut {NoStop}%
\bibitem [{\citenamefont {Garrahan}(2018)}]{garrahan2018aspects}%
  \BibitemOpen
  \bibfield  {author} {\bibinfo {author} {\bibfnamefont {J.~P.}\ \bibnamefont
  {Garrahan}},\ }\bibfield  {title} {\bibinfo {title} {Aspects of
  non-equilibrium in classical and quantum systems: Slow relaxation and
  glasses, dynamical large deviations, quantum non-ergodicity, and open quantum
  dynamics},\ }\href@noop {} {\bibfield  {journal} {\bibinfo  {journal}
  {Physica A: Statistical Mechanics and its Applications}\ }\textbf {\bibinfo
  {volume} {504}},\ \bibinfo {pages} {130} (\bibinfo {year}
  {2018})}\BibitemShut {NoStop}%
\bibitem [{\citenamefont {Hickey}\ \emph {et~al.}(2016)\citenamefont {Hickey},
  \citenamefont {Genway},\ and\ \citenamefont {Garrahan}}]{Hickey2016}%
  \BibitemOpen
  \bibfield  {author} {\bibinfo {author} {\bibfnamefont {J.~M.}\ \bibnamefont
  {Hickey}}, \bibinfo {author} {\bibfnamefont {S.}~\bibnamefont {Genway}},\
  and\ \bibinfo {author} {\bibfnamefont {J.~P.}\ \bibnamefont {Garrahan}},\
  }\bibfield  {title} {\bibinfo {title} {Signatures of many-body localisation
  in a system without disorder and the relation to a glass transition},\ }\href
  {https://doi.org/10.1088/1742-5468/2016/05/054047} {\bibfield  {journal}
  {\bibinfo  {journal} {J. Stat. Mech.}\ }\textbf {\bibinfo {volume} {2016}},\
  \bibinfo {pages} {054047} (\bibinfo {year} {2016})}\BibitemShut {NoStop}%
\bibitem [{\citenamefont {van Horssen}\ \emph {et~al.}(2015)\citenamefont {van
  Horssen}, \citenamefont {Levi},\ and\ \citenamefont
  {Garrahan}}]{Horssen2015}%
  \BibitemOpen
  \bibfield  {author} {\bibinfo {author} {\bibfnamefont {M.}~\bibnamefont {van
  Horssen}}, \bibinfo {author} {\bibfnamefont {E.}~\bibnamefont {Levi}},\ and\
  \bibinfo {author} {\bibfnamefont {J.~P.}\ \bibnamefont {Garrahan}},\
  }\bibfield  {title} {\bibinfo {title} {Dynamics of many-body localization in
  a translation-invariant quantum glass model},\ }\href
  {https://doi.org/10.1103/PhysRevB.92.100305} {\bibfield  {journal} {\bibinfo
  {journal} {Phys. Rev. B}\ }\textbf {\bibinfo {volume} {92}},\ \bibinfo
  {pages} {100305(R)} (\bibinfo {year} {2015})}\BibitemShut {NoStop}%
\bibitem [{\citenamefont {Lan}\ \emph {et~al.}(2018)\citenamefont {Lan},
  \citenamefont {van Horssen}, \citenamefont {Powell},\ and\ \citenamefont
  {Garrahan}}]{Lan2018}%
  \BibitemOpen
  \bibfield  {author} {\bibinfo {author} {\bibfnamefont {Z.}~\bibnamefont
  {Lan}}, \bibinfo {author} {\bibfnamefont {M.}~\bibnamefont {van Horssen}},
  \bibinfo {author} {\bibfnamefont {S.}~\bibnamefont {Powell}},\ and\ \bibinfo
  {author} {\bibfnamefont {J.~P.}\ \bibnamefont {Garrahan}},\ }\bibfield
  {title} {\bibinfo {title} {Quantum slow relaxation and metastability due to
  dynamical constraints},\ }\href
  {https://doi.org/10.1103/PhysRevLett.121.040603} {\bibfield  {journal}
  {\bibinfo  {journal} {Phys. Rev. Lett.}\ }\textbf {\bibinfo {volume} {121}},\
  \bibinfo {pages} {040603} (\bibinfo {year} {2018})}\BibitemShut {NoStop}%
\bibitem [{\citenamefont {Feldmeier}\ \emph {et~al.}(2019)\citenamefont
  {Feldmeier}, \citenamefont {Pollmann},\ and\ \citenamefont
  {Knap}}]{Feldmeier2019}%
  \BibitemOpen
  \bibfield  {author} {\bibinfo {author} {\bibfnamefont {J.}~\bibnamefont
  {Feldmeier}}, \bibinfo {author} {\bibfnamefont {F.}~\bibnamefont
  {Pollmann}},\ and\ \bibinfo {author} {\bibfnamefont {M.}~\bibnamefont
  {Knap}},\ }\bibfield  {title} {\bibinfo {title} {Emergent glassy dynamics in
  a quantum dimer model},\ }\href
  {https://doi.org/10.1103/PhysRevLett.123.040601} {\bibfield  {journal}
  {\bibinfo  {journal} {Phys. Rev. Lett.}\ }\textbf {\bibinfo {volume} {123}},\
  \bibinfo {pages} {040601} (\bibinfo {year} {2019})}\BibitemShut {NoStop}%
\bibitem [{\citenamefont {Castelnovo}\ \emph {et~al.}(2005)\citenamefont
  {Castelnovo}, \citenamefont {Chamon}, \citenamefont {Mudry},\ and\
  \citenamefont {Pujol}}]{Castelnovo2005}%
  \BibitemOpen
  \bibfield  {author} {\bibinfo {author} {\bibfnamefont {C.}~\bibnamefont
  {Castelnovo}}, \bibinfo {author} {\bibfnamefont {C.}~\bibnamefont {Chamon}},
  \bibinfo {author} {\bibfnamefont {C.}~\bibnamefont {Mudry}},\ and\ \bibinfo
  {author} {\bibfnamefont {P.}~\bibnamefont {Pujol}},\ }\bibfield  {title}
  {\bibinfo {title} {From quantum mechanics to classical statistical physics:
  Generalized rokhsar--kivelson hamiltonians and the ``stochastic matrix form''
  decomposition},\ }\href
  {https://doi.org/https://doi.org/10.1016/j.aop.2005.01.006} {\bibfield
  {journal} {\bibinfo  {journal} {Ann. of Phys.}\ }\textbf {\bibinfo {volume}
  {318}},\ \bibinfo {pages} {316 } (\bibinfo {year} {2005})}\BibitemShut
  {NoStop}%
\bibitem [{\citenamefont {Prem}\ \emph {et~al.}(2017)\citenamefont {Prem},
  \citenamefont {Haah},\ and\ \citenamefont {Nandkishore}}]{Prem2017}%
  \BibitemOpen
  \bibfield  {author} {\bibinfo {author} {\bibfnamefont {A.}~\bibnamefont
  {Prem}}, \bibinfo {author} {\bibfnamefont {J.}~\bibnamefont {Haah}},\ and\
  \bibinfo {author} {\bibfnamefont {R.}~\bibnamefont {Nandkishore}},\
  }\bibfield  {title} {\bibinfo {title} {Glassy quantum dynamics in translation
  invariant fracton models},\ }\href
  {https://doi.org/10.1103/PhysRevB.95.155133} {\bibfield  {journal} {\bibinfo
  {journal} {Phys. Rev. B}\ }\textbf {\bibinfo {volume} {95}},\ \bibinfo
  {pages} {155133} (\bibinfo {year} {2017})}\BibitemShut {NoStop}%
\bibitem [{\citenamefont {{Nandkishore}}\ and\ \citenamefont
  {{Hermele}}(2019)}]{Nandkishore2018}%
  \BibitemOpen
  \bibfield  {author} {\bibinfo {author} {\bibfnamefont {R.~M.}\ \bibnamefont
  {{Nandkishore}}}\ and\ \bibinfo {author} {\bibfnamefont {M.}~\bibnamefont
  {{Hermele}}},\ }\bibfield  {title} {\bibinfo {title} {{Fractons}},\ }\href
  {https://doi.org/10.1146/annurev-conmatphys-031218-013604} {\bibfield
  {journal} {\bibinfo  {journal} {Annu. Rev. Condens. Matter Phys.}\ }\textbf
  {\bibinfo {volume} {10}},\ \bibinfo {pages} {295} (\bibinfo {year}
  {2019})}\BibitemShut {NoStop}%
\bibitem [{\citenamefont {Khemani}\ \emph {et~al.}(2020)\citenamefont
  {Khemani}, \citenamefont {Hermele},\ and\ \citenamefont
  {Nandkishore}}]{PhysRevB.101.174204}%
  \BibitemOpen
  \bibfield  {author} {\bibinfo {author} {\bibfnamefont {V.}~\bibnamefont
  {Khemani}}, \bibinfo {author} {\bibfnamefont {M.}~\bibnamefont {Hermele}},\
  and\ \bibinfo {author} {\bibfnamefont {R.}~\bibnamefont {Nandkishore}},\
  }\bibfield  {title} {\bibinfo {title} {Localization from hilbert space
  shattering: From theory to physical realizations},\ }\href
  {https://doi.org/10.1103/PhysRevB.101.174204} {\bibfield  {journal} {\bibinfo
   {journal} {Phys. Rev. B}\ }\textbf {\bibinfo {volume} {101}},\ \bibinfo
  {pages} {174204} (\bibinfo {year} {2020})}\BibitemShut {NoStop}%
\bibitem [{\citenamefont {Sala}\ \emph {et~al.}(2020)\citenamefont {Sala},
  \citenamefont {Rakovszky}, \citenamefont {Verresen}, \citenamefont {Knap},\
  and\ \citenamefont {Pollmann}}]{Sala2020}%
  \BibitemOpen
  \bibfield  {author} {\bibinfo {author} {\bibfnamefont {P.}~\bibnamefont
  {Sala}}, \bibinfo {author} {\bibfnamefont {T.}~\bibnamefont {Rakovszky}},
  \bibinfo {author} {\bibfnamefont {R.}~\bibnamefont {Verresen}}, \bibinfo
  {author} {\bibfnamefont {M.}~\bibnamefont {Knap}},\ and\ \bibinfo {author}
  {\bibfnamefont {F.}~\bibnamefont {Pollmann}},\ }\bibfield  {title} {\bibinfo
  {title} {Ergodicity breaking arising from hilbert space fragmentation in
  dipole-conserving hamiltonians},\ }\href
  {https://doi.org/10.1103/PhysRevX.10.011047} {\bibfield  {journal} {\bibinfo
  {journal} {Phys. Rev. X}\ }\textbf {\bibinfo {volume} {10}},\ \bibinfo
  {pages} {011047} (\bibinfo {year} {2020})}\BibitemShut {NoStop}%
\bibitem [{\citenamefont {Rakovszky}\ \emph {et~al.}(2020)\citenamefont
  {Rakovszky}, \citenamefont {Sala}, \citenamefont {Verresen}, \citenamefont
  {Knap},\ and\ \citenamefont {Pollmann}}]{Rakovszky2020}%
  \BibitemOpen
  \bibfield  {author} {\bibinfo {author} {\bibfnamefont {T.}~\bibnamefont
  {Rakovszky}}, \bibinfo {author} {\bibfnamefont {P.}~\bibnamefont {Sala}},
  \bibinfo {author} {\bibfnamefont {R.}~\bibnamefont {Verresen}}, \bibinfo
  {author} {\bibfnamefont {M.}~\bibnamefont {Knap}},\ and\ \bibinfo {author}
  {\bibfnamefont {F.}~\bibnamefont {Pollmann}},\ }\bibfield  {title} {\bibinfo
  {title} {Statistical localization: From strong fragmentation to strong edge
  modes},\ }\href {https://doi.org/10.1103/PhysRevB.101.125126} {\bibfield
  {journal} {\bibinfo  {journal} {Phys. Rev. B}\ }\textbf {\bibinfo {volume}
  {101}},\ \bibinfo {pages} {125126} (\bibinfo {year} {2020})}\BibitemShut
  {NoStop}%
\bibitem [{\citenamefont {Pretko}\ \emph {et~al.}(2020)\citenamefont {Pretko},
  \citenamefont {Chen},\ and\ \citenamefont {You}}]{Pretko2020}%
  \BibitemOpen
  \bibfield  {author} {\bibinfo {author} {\bibfnamefont {M.}~\bibnamefont
  {Pretko}}, \bibinfo {author} {\bibfnamefont {X.}~\bibnamefont {Chen}},\ and\
  \bibinfo {author} {\bibfnamefont {Y.}~\bibnamefont {You}},\ }\href@noop {}
  {\bibinfo {title} {Fracton phases of matter}} (\bibinfo {year} {2020}),\
  \Eprint {https://arxiv.org/abs/2001.01722} {arXiv:2001.01722
  [cond-mat.str-el]} \BibitemShut {NoStop}%
\bibitem [{\citenamefont {Pretko}\ and\ \citenamefont
  {Radzihovsky}(2018)}]{pretko2018fracton}%
  \BibitemOpen
  \bibfield  {author} {\bibinfo {author} {\bibfnamefont {M.}~\bibnamefont
  {Pretko}}\ and\ \bibinfo {author} {\bibfnamefont {L.}~\bibnamefont
  {Radzihovsky}},\ }\bibfield  {title} {\bibinfo {title} {Fracton-elasticity
  duality},\ }\href@noop {} {\bibfield  {journal} {\bibinfo  {journal}
  {Physical review letters}\ }\textbf {\bibinfo {volume} {120}},\ \bibinfo
  {pages} {195301} (\bibinfo {year} {2018})}\BibitemShut {NoStop}%
\bibitem [{\citenamefont {Scherg}\ \emph {et~al.}(2021)\citenamefont {Scherg},
  \citenamefont {Kohlert}, \citenamefont {Sala}, \citenamefont {Pollmann},
  \citenamefont {Madhusudhana}, \citenamefont {Bloch},\ and\ \citenamefont
  {Aidelsburger}}]{scherg2021observing}%
  \BibitemOpen
  \bibfield  {author} {\bibinfo {author} {\bibfnamefont {S.}~\bibnamefont
  {Scherg}}, \bibinfo {author} {\bibfnamefont {T.}~\bibnamefont {Kohlert}},
  \bibinfo {author} {\bibfnamefont {P.}~\bibnamefont {Sala}}, \bibinfo {author}
  {\bibfnamefont {F.}~\bibnamefont {Pollmann}}, \bibinfo {author}
  {\bibfnamefont {B.~H.}\ \bibnamefont {Madhusudhana}}, \bibinfo {author}
  {\bibfnamefont {I.}~\bibnamefont {Bloch}},\ and\ \bibinfo {author}
  {\bibfnamefont {M.}~\bibnamefont {Aidelsburger}},\ }\bibfield  {title}
  {\bibinfo {title} {Observing non-ergodicity due to kinetic constraints in
  tilted fermi-hubbard chains},\ }\href@noop {} {\bibfield  {journal} {\bibinfo
   {journal} {Nature Communications}\ }\textbf {\bibinfo {volume} {12}},\
  \bibinfo {pages} {1} (\bibinfo {year} {2021})}\BibitemShut {NoStop}%
\bibitem [{\citenamefont {Turner}\ \emph
  {et~al.}(2018{\natexlab{a}})\citenamefont {Turner}, \citenamefont
  {Michailidis}, \citenamefont {Abanin}, \citenamefont {Serbyn},\ and\
  \citenamefont {Papi{\'c}}}]{Turner2018}%
  \BibitemOpen
  \bibfield  {author} {\bibinfo {author} {\bibfnamefont {C.~J.}\ \bibnamefont
  {Turner}}, \bibinfo {author} {\bibfnamefont {A.~A.}\ \bibnamefont
  {Michailidis}}, \bibinfo {author} {\bibfnamefont {D.~A.}\ \bibnamefont
  {Abanin}}, \bibinfo {author} {\bibfnamefont {M.}~\bibnamefont {Serbyn}},\
  and\ \bibinfo {author} {\bibfnamefont {Z.}~\bibnamefont {Papi{\'c}}},\
  }\bibfield  {title} {\bibinfo {title} {Weak ergodicity breaking from quantum
  many-body scars},\ }\href {https://doi.org/10.1038/s41567-018-0137-5}
  {\bibfield  {journal} {\bibinfo  {journal} {Nature Physics}\ }\textbf
  {\bibinfo {volume} {14}},\ \bibinfo {pages} {745} (\bibinfo {year}
  {2018}{\natexlab{a}})}\BibitemShut {NoStop}%
\bibitem [{\citenamefont {Turner}\ \emph
  {et~al.}(2018{\natexlab{b}})\citenamefont {Turner}, \citenamefont
  {Michailidis}, \citenamefont {Abanin}, \citenamefont {Serbyn},\ and\
  \citenamefont {Papi\ifmmode~\acute{c}\else \'{c}\fi{}}}]{Turner2018b}%
  \BibitemOpen
  \bibfield  {author} {\bibinfo {author} {\bibfnamefont {C.~J.}\ \bibnamefont
  {Turner}}, \bibinfo {author} {\bibfnamefont {A.~A.}\ \bibnamefont
  {Michailidis}}, \bibinfo {author} {\bibfnamefont {D.~A.}\ \bibnamefont
  {Abanin}}, \bibinfo {author} {\bibfnamefont {M.}~\bibnamefont {Serbyn}},\
  and\ \bibinfo {author} {\bibfnamefont {Z.}~\bibnamefont
  {Papi\ifmmode~\acute{c}\else \'{c}\fi{}}},\ }\bibfield  {title} {\bibinfo
  {title} {Quantum scarred eigenstates in a rydberg atom chain: Entanglement,
  breakdown of thermalization, and stability to perturbations},\ }\href
  {https://doi.org/10.1103/PhysRevB.98.155134} {\bibfield  {journal} {\bibinfo
  {journal} {Phys. Rev. B}\ }\textbf {\bibinfo {volume} {98}},\ \bibinfo
  {pages} {155134} (\bibinfo {year} {2018}{\natexlab{b}})}\BibitemShut
  {NoStop}%
\bibitem [{\citenamefont {Ho}\ \emph {et~al.}(2019)\citenamefont {Ho},
  \citenamefont {Choi}, \citenamefont {Pichler},\ and\ \citenamefont
  {Lukin}}]{Ho2019}%
  \BibitemOpen
  \bibfield  {author} {\bibinfo {author} {\bibfnamefont {W.~W.}\ \bibnamefont
  {Ho}}, \bibinfo {author} {\bibfnamefont {S.}~\bibnamefont {Choi}}, \bibinfo
  {author} {\bibfnamefont {H.}~\bibnamefont {Pichler}},\ and\ \bibinfo {author}
  {\bibfnamefont {M.~D.}\ \bibnamefont {Lukin}},\ }\bibfield  {title} {\bibinfo
  {title} {Periodic orbits, entanglement, and quantum many-body scars in
  constrained models: Matrix product state approach},\ }\href
  {https://doi.org/10.1103/PhysRevLett.122.040603} {\bibfield  {journal}
  {\bibinfo  {journal} {Phys. Rev. Lett.}\ }\textbf {\bibinfo {volume} {122}},\
  \bibinfo {pages} {040603} (\bibinfo {year} {2019})}\BibitemShut {NoStop}%
\bibitem [{\citenamefont {Ok}\ \emph {et~al.}(2019)\citenamefont {Ok},
  \citenamefont {Choo}, \citenamefont {Mudry}, \citenamefont {Castelnovo},
  \citenamefont {Chamon},\ and\ \citenamefont {Neupert}}]{Ok2019}%
  \BibitemOpen
  \bibfield  {author} {\bibinfo {author} {\bibfnamefont {S.}~\bibnamefont
  {Ok}}, \bibinfo {author} {\bibfnamefont {K.}~\bibnamefont {Choo}}, \bibinfo
  {author} {\bibfnamefont {C.}~\bibnamefont {Mudry}}, \bibinfo {author}
  {\bibfnamefont {C.}~\bibnamefont {Castelnovo}}, \bibinfo {author}
  {\bibfnamefont {C.}~\bibnamefont {Chamon}},\ and\ \bibinfo {author}
  {\bibfnamefont {T.}~\bibnamefont {Neupert}},\ }\bibfield  {title} {\bibinfo
  {title} {Topological many-body scar states in dimensions one, two, and
  three},\ }\href {https://doi.org/10.1103/PhysRevResearch.1.033144} {\bibfield
   {journal} {\bibinfo  {journal} {Phys. Rev. Research}\ }\textbf {\bibinfo
  {volume} {1}},\ \bibinfo {pages} {033144} (\bibinfo {year}
  {2019})}\BibitemShut {NoStop}%
\bibitem [{\citenamefont {Schecter}\ and\ \citenamefont
  {Iadecola}(2019)}]{Schecter2019}%
  \BibitemOpen
  \bibfield  {author} {\bibinfo {author} {\bibfnamefont {M.}~\bibnamefont
  {Schecter}}\ and\ \bibinfo {author} {\bibfnamefont {T.}~\bibnamefont
  {Iadecola}},\ }\bibfield  {title} {\bibinfo {title} {Weak ergodicity breaking
  and quantum many-body scars in spin-1 $xy$ magnets},\ }\href
  {https://doi.org/10.1103/PhysRevLett.123.147201} {\bibfield  {journal}
  {\bibinfo  {journal} {Phys. Rev. Lett.}\ }\textbf {\bibinfo {volume} {123}},\
  \bibinfo {pages} {147201} (\bibinfo {year} {2019})}\BibitemShut {NoStop}%
\bibitem [{\citenamefont {Khemani}\ \emph {et~al.}(2019)\citenamefont
  {Khemani}, \citenamefont {Laumann},\ and\ \citenamefont
  {Chandran}}]{Khemani2019Signatures}%
  \BibitemOpen
  \bibfield  {author} {\bibinfo {author} {\bibfnamefont {V.}~\bibnamefont
  {Khemani}}, \bibinfo {author} {\bibfnamefont {C.~R.}\ \bibnamefont
  {Laumann}},\ and\ \bibinfo {author} {\bibfnamefont {A.}~\bibnamefont
  {Chandran}},\ }\bibfield  {title} {\bibinfo {title} {Signatures of
  integrability in the dynamics of rydberg-blockaded chains},\ }\href
  {https://doi.org/10.1103/PhysRevB.99.161101} {\bibfield  {journal} {\bibinfo
  {journal} {Phys. Rev. B}\ }\textbf {\bibinfo {volume} {99}},\ \bibinfo
  {pages} {161101(R)} (\bibinfo {year} {2019})}\BibitemShut {NoStop}%
\bibitem [{\citenamefont {Hudomal}\ \emph {et~al.}(2020)\citenamefont
  {Hudomal}, \citenamefont {Vasi{\'{c}}}, \citenamefont {Regnault},\ and\
  \citenamefont {Papi{\'{c}}}}]{Hudomal2020}%
  \BibitemOpen
  \bibfield  {author} {\bibinfo {author} {\bibfnamefont {A.}~\bibnamefont
  {Hudomal}}, \bibinfo {author} {\bibfnamefont {I.}~\bibnamefont
  {Vasi{\'{c}}}}, \bibinfo {author} {\bibfnamefont {N.}~\bibnamefont
  {Regnault}},\ and\ \bibinfo {author} {\bibfnamefont {Z.}~\bibnamefont
  {Papi{\'{c}}}},\ }\bibfield  {title} {\bibinfo {title} {Quantum scars of
  bosons with correlated hopping},\ }\href
  {https://doi.org/10.1038/s42005-020-0364-9} {\bibfield  {journal} {\bibinfo
  {journal} {Communications Physics}\ }\textbf {\bibinfo {volume} {3}},\
  \bibinfo {pages} {1} (\bibinfo {year} {2020})}\BibitemShut {NoStop}%
\bibitem [{\citenamefont {Moudgalya}\ \emph {et~al.}(2018)\citenamefont
  {Moudgalya}, \citenamefont {Regnault},\ and\ \citenamefont
  {Bernevig}}]{Moudgalya2018Entanglement}%
  \BibitemOpen
  \bibfield  {author} {\bibinfo {author} {\bibfnamefont {S.}~\bibnamefont
  {Moudgalya}}, \bibinfo {author} {\bibfnamefont {N.}~\bibnamefont
  {Regnault}},\ and\ \bibinfo {author} {\bibfnamefont {B.~A.}\ \bibnamefont
  {Bernevig}},\ }\bibfield  {title} {\bibinfo {title} {Entanglement of exact
  excited states of affleck-kennedy-lieb-tasaki models: Exact results,
  many-body scars, and violation of the strong eigenstate thermalization
  hypothesis},\ }\href {https://doi.org/10.1103/PhysRevB.98.235156} {\bibfield
  {journal} {\bibinfo  {journal} {Phys. Rev. B}\ }\textbf {\bibinfo {volume}
  {98}},\ \bibinfo {pages} {235156} (\bibinfo {year} {2018})}\BibitemShut
  {NoStop}%
\bibitem [{\citenamefont {Feldmeier}\ \emph {et~al.}(2020)\citenamefont
  {Feldmeier}, \citenamefont {Sala}, \citenamefont {De~Tomasi}, \citenamefont
  {Pollmann},\ and\ \citenamefont {Knap}}]{feldmeier2020anomalous}%
  \BibitemOpen
  \bibfield  {author} {\bibinfo {author} {\bibfnamefont {J.}~\bibnamefont
  {Feldmeier}}, \bibinfo {author} {\bibfnamefont {P.}~\bibnamefont {Sala}},
  \bibinfo {author} {\bibfnamefont {G.}~\bibnamefont {De~Tomasi}}, \bibinfo
  {author} {\bibfnamefont {F.}~\bibnamefont {Pollmann}},\ and\ \bibinfo
  {author} {\bibfnamefont {M.}~\bibnamefont {Knap}},\ }\bibfield  {title}
  {\bibinfo {title} {Anomalous diffusion in dipole-and higher-moment-conserving
  systems},\ }\href@noop {} {\bibfield  {journal} {\bibinfo  {journal}
  {Physical Review Letters}\ }\textbf {\bibinfo {volume} {125}},\ \bibinfo
  {pages} {245303} (\bibinfo {year} {2020})}\BibitemShut {NoStop}%
\bibitem [{\citenamefont {Serbyn}\ \emph {et~al.}(2021)\citenamefont {Serbyn},
  \citenamefont {Abanin},\ and\ \citenamefont {Papi{\'c}}}]{serbyn2021quantum}%
  \BibitemOpen
  \bibfield  {author} {\bibinfo {author} {\bibfnamefont {M.}~\bibnamefont
  {Serbyn}}, \bibinfo {author} {\bibfnamefont {D.~A.}\ \bibnamefont {Abanin}},\
  and\ \bibinfo {author} {\bibfnamefont {Z.}~\bibnamefont {Papi{\'c}}},\
  }\bibfield  {title} {\bibinfo {title} {Quantum many-body scars and weak
  breaking of ergodicity},\ }\href@noop {} {\bibfield  {journal} {\bibinfo
  {journal} {Nature Physics}\ }\textbf {\bibinfo {volume} {17}},\ \bibinfo
  {pages} {675} (\bibinfo {year} {2021})}\BibitemShut {NoStop}%
\bibitem [{\citenamefont {Desaules}\ \emph {et~al.}(2021)\citenamefont
  {Desaules}, \citenamefont {Hudomal}, \citenamefont {Turner},\ and\
  \citenamefont {Papi\ifmmode~\acute{c}\else
  \'{c}\fi{}}}]{PhysRevLett.126.210601}%
  \BibitemOpen
  \bibfield  {author} {\bibinfo {author} {\bibfnamefont {J.-Y.}\ \bibnamefont
  {Desaules}}, \bibinfo {author} {\bibfnamefont {A.}~\bibnamefont {Hudomal}},
  \bibinfo {author} {\bibfnamefont {C.~J.}\ \bibnamefont {Turner}},\ and\
  \bibinfo {author} {\bibfnamefont {Z.}~\bibnamefont
  {Papi\ifmmode~\acute{c}\else \'{c}\fi{}}},\ }\bibfield  {title} {\bibinfo
  {title} {Proposal for realizing quantum scars in the tilted 1d fermi-hubbard
  model},\ }\href {https://doi.org/10.1103/PhysRevLett.126.210601} {\bibfield
  {journal} {\bibinfo  {journal} {Phys. Rev. Lett.}\ }\textbf {\bibinfo
  {volume} {126}},\ \bibinfo {pages} {210601} (\bibinfo {year}
  {2021})}\BibitemShut {NoStop}%
\bibitem [{\citenamefont {Turner}\ \emph {et~al.}(2021)\citenamefont {Turner},
  \citenamefont {Desaules}, \citenamefont {Bull},\ and\ \citenamefont
  {Papi\ifmmode~\acute{c}\else \'{c}\fi{}}}]{PhysRevX.11.021021}%
  \BibitemOpen
  \bibfield  {author} {\bibinfo {author} {\bibfnamefont {C.~J.}\ \bibnamefont
  {Turner}}, \bibinfo {author} {\bibfnamefont {J.-Y.}\ \bibnamefont
  {Desaules}}, \bibinfo {author} {\bibfnamefont {K.}~\bibnamefont {Bull}},\
  and\ \bibinfo {author} {\bibfnamefont {Z.}~\bibnamefont
  {Papi\ifmmode~\acute{c}\else \'{c}\fi{}}},\ }\bibfield  {title} {\bibinfo
  {title} {Correspondence principle for many-body scars in ultracold rydberg
  atoms},\ }\href {https://doi.org/10.1103/PhysRevX.11.021021} {\bibfield
  {journal} {\bibinfo  {journal} {Phys. Rev. X}\ }\textbf {\bibinfo {volume}
  {11}},\ \bibinfo {pages} {021021} (\bibinfo {year} {2021})}\BibitemShut
  {NoStop}%
\bibitem [{\citenamefont {Magoni}\ \emph {et~al.}(2021)\citenamefont {Magoni},
  \citenamefont {Mazza},\ and\ \citenamefont
  {Lesanovsky}}]{PhysRevLett.126.103002}%
  \BibitemOpen
  \bibfield  {author} {\bibinfo {author} {\bibfnamefont {M.}~\bibnamefont
  {Magoni}}, \bibinfo {author} {\bibfnamefont {P.~P.}\ \bibnamefont {Mazza}},\
  and\ \bibinfo {author} {\bibfnamefont {I.}~\bibnamefont {Lesanovsky}},\
  }\bibfield  {title} {\bibinfo {title} {Emergent bloch oscillations in a
  kinetically constrained rydberg spin lattice},\ }\href
  {https://doi.org/10.1103/PhysRevLett.126.103002} {\bibfield  {journal}
  {\bibinfo  {journal} {Phys. Rev. Lett.}\ }\textbf {\bibinfo {volume} {126}},\
  \bibinfo {pages} {103002} (\bibinfo {year} {2021})}\BibitemShut {NoStop}%
\bibitem [{\citenamefont {Zhao}\ \emph {et~al.}(2021)\citenamefont {Zhao},
  \citenamefont {Smith}, \citenamefont {Mintert},\ and\ \citenamefont
  {Knolle}}]{zhao2021orthogonal}%
  \BibitemOpen
  \bibfield  {author} {\bibinfo {author} {\bibfnamefont {H.}~\bibnamefont
  {Zhao}}, \bibinfo {author} {\bibfnamefont {A.}~\bibnamefont {Smith}},
  \bibinfo {author} {\bibfnamefont {F.}~\bibnamefont {Mintert}},\ and\ \bibinfo
  {author} {\bibfnamefont {J.}~\bibnamefont {Knolle}},\ }\bibfield  {title}
  {\bibinfo {title} {Orthogonal quantum many-body scars},\ }\href@noop {}
  {\bibfield  {journal} {\bibinfo  {journal} {arXiv preprint arXiv:2102.07672}\
  } (\bibinfo {year} {2021})}\BibitemShut {NoStop}%
\bibitem [{\citenamefont {Garrahan}\ \emph {et~al.}(2009)\citenamefont
  {Garrahan}, \citenamefont {Jack}, \citenamefont {Lecomte}, \citenamefont
  {Pitard}, \citenamefont {van Duijvendijk},\ and\ \citenamefont {van
  Wijland}}]{Garrahan2009}%
  \BibitemOpen
  \bibfield  {author} {\bibinfo {author} {\bibfnamefont {J.~P.}\ \bibnamefont
  {Garrahan}}, \bibinfo {author} {\bibfnamefont {R.~L.}\ \bibnamefont {Jack}},
  \bibinfo {author} {\bibfnamefont {V.}~\bibnamefont {Lecomte}}, \bibinfo
  {author} {\bibfnamefont {E.}~\bibnamefont {Pitard}}, \bibinfo {author}
  {\bibfnamefont {K.}~\bibnamefont {van Duijvendijk}},\ and\ \bibinfo {author}
  {\bibfnamefont {F.}~\bibnamefont {van Wijland}},\ }\bibfield  {title}
  {\bibinfo {title} {First-order dynamical phase transition in models of
  glasses: an approach based on ensembles of histories},\ }\href
  {https://doi.org/10.1088/1751-8113/42/7/075007} {\bibfield  {journal}
  {\bibinfo  {journal} {Journal of Physics A: Mathematical and Theoretical}\
  }\textbf {\bibinfo {volume} {42}},\ \bibinfo {pages} {075007} (\bibinfo
  {year} {2009})}\BibitemShut {NoStop}%
\bibitem [{\citenamefont {Chleboun}\ \emph {et~al.}(2013)\citenamefont
  {Chleboun}, \citenamefont {Faggionato},\ and\ \citenamefont
  {Martinelli}}]{Chleboun2013}%
  \BibitemOpen
  \bibfield  {author} {\bibinfo {author} {\bibfnamefont {P.}~\bibnamefont
  {Chleboun}}, \bibinfo {author} {\bibfnamefont {A.}~\bibnamefont
  {Faggionato}},\ and\ \bibinfo {author} {\bibfnamefont {F.}~\bibnamefont
  {Martinelli}},\ }\bibfield  {title} {\bibinfo {title} {Time scale separation
  in the low temperature east model: rigorous results},\ }\href
  {https://doi.org/10.1088/1742-5468/2013/04/l04001} {\bibfield  {journal}
  {\bibinfo  {journal} {J. Stat. Mech.}\ }\textbf {\bibinfo {volume} {2013}},\
  \bibinfo {pages} {L04001} (\bibinfo {year} {2013})}\BibitemShut {NoStop}%
\bibitem [{\citenamefont {Kim}\ \emph {et~al.}(2015)\citenamefont {Kim},
  \citenamefont {Ba\~nuls}, \citenamefont {Cirac}, \citenamefont {Hastings},\
  and\ \citenamefont {Huse}}]{Kim2015}%
  \BibitemOpen
  \bibfield  {author} {\bibinfo {author} {\bibfnamefont {H.}~\bibnamefont
  {Kim}}, \bibinfo {author} {\bibfnamefont {M.~C.}\ \bibnamefont {Ba\~nuls}},
  \bibinfo {author} {\bibfnamefont {J.~I.}\ \bibnamefont {Cirac}}, \bibinfo
  {author} {\bibfnamefont {M.~B.}\ \bibnamefont {Hastings}},\ and\ \bibinfo
  {author} {\bibfnamefont {D.~A.}\ \bibnamefont {Huse}},\ }\bibfield  {title}
  {\bibinfo {title} {Slowest local operators in quantum spin chains},\ }\href
  {https://doi.org/10.1103/PhysRevE.92.012128} {\bibfield  {journal} {\bibinfo
  {journal} {Phys. Rev. E}\ }\textbf {\bibinfo {volume} {92}},\ \bibinfo
  {pages} {012128} (\bibinfo {year} {2015})}\BibitemShut {NoStop}%
\bibitem [{\citenamefont {Gopalakrishnan}\ \emph {et~al.}(2018)\citenamefont
  {Gopalakrishnan}, \citenamefont {Huse}, \citenamefont {Khemani},\ and\
  \citenamefont {Vasseur}}]{gopalakrishnan2018hydrodynamics}%
  \BibitemOpen
  \bibfield  {author} {\bibinfo {author} {\bibfnamefont {S.}~\bibnamefont
  {Gopalakrishnan}}, \bibinfo {author} {\bibfnamefont {D.~A.}\ \bibnamefont
  {Huse}}, \bibinfo {author} {\bibfnamefont {V.}~\bibnamefont {Khemani}},\ and\
  \bibinfo {author} {\bibfnamefont {R.}~\bibnamefont {Vasseur}},\ }\bibfield
  {title} {\bibinfo {title} {Hydrodynamics of operator spreading and
  quasiparticle diffusion in interacting integrable systems},\ }\href@noop {}
  {\bibfield  {journal} {\bibinfo  {journal} {Physical Review B}\ }\textbf
  {\bibinfo {volume} {98}},\ \bibinfo {pages} {220303(R)} (\bibinfo {year}
  {2018})}\BibitemShut {NoStop}%
\bibitem [{\citenamefont {Gopalakrishnan}(2018)}]{gopalakrishnan2018operator}%
  \BibitemOpen
  \bibfield  {author} {\bibinfo {author} {\bibfnamefont {S.}~\bibnamefont
  {Gopalakrishnan}},\ }\bibfield  {title} {\bibinfo {title} {Operator growth
  and eigenstate entanglement in an interacting integrable floquet system},\
  }\href@noop {} {\bibfield  {journal} {\bibinfo  {journal} {Physical Review
  B}\ }\textbf {\bibinfo {volume} {98}},\ \bibinfo {pages} {060302(R)}
  (\bibinfo {year} {2018})}\BibitemShut {NoStop}%
\bibitem [{\citenamefont {Ba\~nuls}\ and\ \citenamefont
  {Garrahan}(2019)}]{Banuls2019}%
  \BibitemOpen
  \bibfield  {author} {\bibinfo {author} {\bibfnamefont {M.~C.}\ \bibnamefont
  {Ba\~nuls}}\ and\ \bibinfo {author} {\bibfnamefont {J.~P.}\ \bibnamefont
  {Garrahan}},\ }\bibfield  {title} {\bibinfo {title} {Using matrix product
  states to study the dynamical large deviations of kinetically constrained
  models},\ }\href {https://doi.org/10.1103/PhysRevLett.123.200601} {\bibfield
  {journal} {\bibinfo  {journal} {Phys. Rev. Lett.}\ }\textbf {\bibinfo
  {volume} {123}},\ \bibinfo {pages} {200601} (\bibinfo {year}
  {2019})}\BibitemShut {NoStop}%
\bibitem [{\citenamefont {Causer}\ \emph {et~al.}(2020)\citenamefont {Causer},
  \citenamefont {Lesanovsky}, \citenamefont {Ba\~nuls},\ and\ \citenamefont
  {Garrahan}}]{PhysRevE.102.052132}%
  \BibitemOpen
  \bibfield  {author} {\bibinfo {author} {\bibfnamefont {L.}~\bibnamefont
  {Causer}}, \bibinfo {author} {\bibfnamefont {I.}~\bibnamefont {Lesanovsky}},
  \bibinfo {author} {\bibfnamefont {M.~C.}\ \bibnamefont {Ba\~nuls}},\ and\
  \bibinfo {author} {\bibfnamefont {J.~P.}\ \bibnamefont {Garrahan}},\
  }\bibfield  {title} {\bibinfo {title} {Dynamics and large deviation
  transitions of the xor-fredrickson-andersen kinetically constrained model},\
  }\href {https://doi.org/10.1103/PhysRevE.102.052132} {\bibfield  {journal}
  {\bibinfo  {journal} {Phys. Rev. E}\ }\textbf {\bibinfo {volume} {102}},\
  \bibinfo {pages} {052132} (\bibinfo {year} {2020})}\BibitemShut {NoStop}%
\bibitem [{\citenamefont {Pancotti}\ \emph {et~al.}(2020)\citenamefont
  {Pancotti}, \citenamefont {Giudice}, \citenamefont {Cirac}, \citenamefont
  {Garrahan},\ and\ \citenamefont {Ba\~nuls}}]{PhysRevX.10.021051}%
  \BibitemOpen
  \bibfield  {author} {\bibinfo {author} {\bibfnamefont {N.}~\bibnamefont
  {Pancotti}}, \bibinfo {author} {\bibfnamefont {G.}~\bibnamefont {Giudice}},
  \bibinfo {author} {\bibfnamefont {J.~I.}\ \bibnamefont {Cirac}}, \bibinfo
  {author} {\bibfnamefont {J.~P.}\ \bibnamefont {Garrahan}},\ and\ \bibinfo
  {author} {\bibfnamefont {M.~C.}\ \bibnamefont {Ba\~nuls}},\ }\bibfield
  {title} {\bibinfo {title} {Quantum east model: Localization, nonthermal
  eigenstates, and slow dynamics},\ }\href
  {https://doi.org/10.1103/PhysRevX.10.021051} {\bibfield  {journal} {\bibinfo
  {journal} {Phys. Rev. X}\ }\textbf {\bibinfo {volume} {10}},\ \bibinfo
  {pages} {021051} (\bibinfo {year} {2020})}\BibitemShut {NoStop}%
\bibitem [{\citenamefont {Walls}\ and\ \citenamefont
  {Milburn}(2007)}]{walls2007quantum}%
  \BibitemOpen
  \bibfield  {author} {\bibinfo {author} {\bibfnamefont {D.~F.}\ \bibnamefont
  {Walls}}\ and\ \bibinfo {author} {\bibfnamefont {G.~J.}\ \bibnamefont
  {Milburn}},\ }\href@noop {} {\emph {\bibinfo {title} {Quantum optics}}}\
  (\bibinfo  {publisher} {Springer Science \& Business Media},\ \bibinfo
  {address} {Berlin, Germany},\ \bibinfo {year} {2007})\BibitemShut {NoStop}%
\bibitem [{\citenamefont {Chandran}\ \emph {et~al.}(2015)\citenamefont
  {Chandran}, \citenamefont {Kim}, \citenamefont {Vidal},\ and\ \citenamefont
  {Abanin}}]{chandran2015constructing}%
  \BibitemOpen
  \bibfield  {author} {\bibinfo {author} {\bibfnamefont {A.}~\bibnamefont
  {Chandran}}, \bibinfo {author} {\bibfnamefont {I.~H.}\ \bibnamefont {Kim}},
  \bibinfo {author} {\bibfnamefont {G.}~\bibnamefont {Vidal}},\ and\ \bibinfo
  {author} {\bibfnamefont {D.~A.}\ \bibnamefont {Abanin}},\ }\bibfield  {title}
  {\bibinfo {title} {Constructing local integrals of motion in the many-body
  localized phase},\ }\href@noop {} {\bibfield  {journal} {\bibinfo  {journal}
  {Physical Review B}\ }\textbf {\bibinfo {volume} {91}},\ \bibinfo {pages}
  {085425} (\bibinfo {year} {2015})}\BibitemShut {NoStop}%
\bibitem [{\citenamefont {Ros}\ \emph {et~al.}(2015)\citenamefont {Ros},
  \citenamefont {M{\"u}ller},\ and\ \citenamefont
  {Scardicchio}}]{ros2015integrals}%
  \BibitemOpen
  \bibfield  {author} {\bibinfo {author} {\bibfnamefont {V.}~\bibnamefont
  {Ros}}, \bibinfo {author} {\bibfnamefont {M.}~\bibnamefont {M{\"u}ller}},\
  and\ \bibinfo {author} {\bibfnamefont {A.}~\bibnamefont {Scardicchio}},\
  }\bibfield  {title} {\bibinfo {title} {Integrals of motion in the many-body
  localized phase},\ }\href@noop {} {\bibfield  {journal} {\bibinfo  {journal}
  {Nuclear Physics B}\ }\textbf {\bibinfo {volume} {891}},\ \bibinfo {pages}
  {420} (\bibinfo {year} {2015})}\BibitemShut {NoStop}%
\bibitem [{\citenamefont {Imbrie}\ \emph {et~al.}(2017)\citenamefont {Imbrie},
  \citenamefont {Ros},\ and\ \citenamefont {Scardicchio}}]{imbrie2017local}%
  \BibitemOpen
  \bibfield  {author} {\bibinfo {author} {\bibfnamefont {J.~Z.}\ \bibnamefont
  {Imbrie}}, \bibinfo {author} {\bibfnamefont {V.}~\bibnamefont {Ros}},\ and\
  \bibinfo {author} {\bibfnamefont {A.}~\bibnamefont {Scardicchio}},\
  }\bibfield  {title} {\bibinfo {title} {Local integrals of motion in many-body
  localized systems},\ }\href@noop {} {\bibfield  {journal} {\bibinfo
  {journal} {Annalen der Physik}\ }\textbf {\bibinfo {volume} {529}},\ \bibinfo
  {pages} {1600278} (\bibinfo {year} {2017})}\BibitemShut {NoStop}%
\bibitem [{\citenamefont {Huse}\ \emph {et~al.}(2014)\citenamefont {Huse},
  \citenamefont {Nandkishore},\ and\ \citenamefont
  {Oganesyan}}]{huse2014phenomenology}%
  \BibitemOpen
  \bibfield  {author} {\bibinfo {author} {\bibfnamefont {D.~A.}\ \bibnamefont
  {Huse}}, \bibinfo {author} {\bibfnamefont {R.}~\bibnamefont {Nandkishore}},\
  and\ \bibinfo {author} {\bibfnamefont {V.}~\bibnamefont {Oganesyan}},\
  }\bibfield  {title} {\bibinfo {title} {Phenomenology of fully
  many-body-localized systems},\ }\href@noop {} {\bibfield  {journal} {\bibinfo
   {journal} {Physical Review B}\ }\textbf {\bibinfo {volume} {90}},\ \bibinfo
  {pages} {174202} (\bibinfo {year} {2014})}\BibitemShut {NoStop}%
\bibitem [{\citenamefont {L\"uschen}\ \emph {et~al.}(2017)\citenamefont
  {L\"uschen}, \citenamefont {Bordia}, \citenamefont {Hodgman}, \citenamefont
  {Schreiber}, \citenamefont {Sarkar}, \citenamefont {Daley}, \citenamefont
  {Fischer}, \citenamefont {Altman}, \citenamefont {Bloch},\ and\ \citenamefont
  {Schneider}}]{PhysRevX.7.011034}%
  \BibitemOpen
  \bibfield  {author} {\bibinfo {author} {\bibfnamefont {H.~P.}\ \bibnamefont
  {L\"uschen}}, \bibinfo {author} {\bibfnamefont {P.}~\bibnamefont {Bordia}},
  \bibinfo {author} {\bibfnamefont {S.~S.}\ \bibnamefont {Hodgman}}, \bibinfo
  {author} {\bibfnamefont {M.}~\bibnamefont {Schreiber}}, \bibinfo {author}
  {\bibfnamefont {S.}~\bibnamefont {Sarkar}}, \bibinfo {author} {\bibfnamefont
  {A.~J.}\ \bibnamefont {Daley}}, \bibinfo {author} {\bibfnamefont {M.~H.}\
  \bibnamefont {Fischer}}, \bibinfo {author} {\bibfnamefont {E.}~\bibnamefont
  {Altman}}, \bibinfo {author} {\bibfnamefont {I.}~\bibnamefont {Bloch}},\ and\
  \bibinfo {author} {\bibfnamefont {U.}~\bibnamefont {Schneider}},\ }\bibfield
  {title} {\bibinfo {title} {Signatures of many-body localization in a
  controlled open quantum system},\ }\href
  {https://doi.org/10.1103/PhysRevX.7.011034} {\bibfield  {journal} {\bibinfo
  {journal} {Phys. Rev. X}\ }\textbf {\bibinfo {volume} {7}},\ \bibinfo {pages}
  {011034} (\bibinfo {year} {2017})}\BibitemShut {NoStop}%
\bibitem [{\citenamefont {Lenar\ifmmode \check{c}\else
  \v{c}\fi{}i\ifmmode~\check{c}\else \v{c}\fi{}}\ \emph
  {et~al.}(2020)\citenamefont {Lenar\ifmmode \check{c}\else
  \v{c}\fi{}i\ifmmode~\check{c}\else \v{c}\fi{}}, \citenamefont {Alberton},
  \citenamefont {Rosch},\ and\ \citenamefont
  {Altman}}]{PhysRevLett.125.116601}%
  \BibitemOpen
  \bibfield  {author} {\bibinfo {author} {\bibfnamefont {Z.}~\bibnamefont
  {Lenar\ifmmode \check{c}\else \v{c}\fi{}i\ifmmode~\check{c}\else
  \v{c}\fi{}}}, \bibinfo {author} {\bibfnamefont {O.}~\bibnamefont {Alberton}},
  \bibinfo {author} {\bibfnamefont {A.}~\bibnamefont {Rosch}},\ and\ \bibinfo
  {author} {\bibfnamefont {E.}~\bibnamefont {Altman}},\ }\bibfield  {title}
  {\bibinfo {title} {Critical behavior near the many-body localization
  transition in driven open systems},\ }\href
  {https://doi.org/10.1103/PhysRevLett.125.116601} {\bibfield  {journal}
  {\bibinfo  {journal} {Phys. Rev. Lett.}\ }\textbf {\bibinfo {volume} {125}},\
  \bibinfo {pages} {116601} (\bibinfo {year} {2020})}\BibitemShut {NoStop}%
\bibitem [{\citenamefont {Medvedyeva}\ \emph {et~al.}(2016)\citenamefont
  {Medvedyeva}, \citenamefont {Prosen},\ and\ \citenamefont {\ifmmode
  \check{Z}\else \v{Z}\fi{}nidari\ifmmode~\check{c}\else
  \v{c}\fi{}}}]{PhysRevB.93.094205}%
  \BibitemOpen
  \bibfield  {author} {\bibinfo {author} {\bibfnamefont {M.~V.}\ \bibnamefont
  {Medvedyeva}}, \bibinfo {author} {\bibfnamefont {T.~c.~v.}\ \bibnamefont
  {Prosen}},\ and\ \bibinfo {author} {\bibfnamefont {M.}~\bibnamefont {\ifmmode
  \check{Z}\else \v{Z}\fi{}nidari\ifmmode~\check{c}\else \v{c}\fi{}}},\
  }\bibfield  {title} {\bibinfo {title} {Influence of dephasing on many-body
  localization},\ }\href {https://doi.org/10.1103/PhysRevB.93.094205}
  {\bibfield  {journal} {\bibinfo  {journal} {Phys. Rev. B}\ }\textbf {\bibinfo
  {volume} {93}},\ \bibinfo {pages} {094205} (\bibinfo {year}
  {2016})}\BibitemShut {NoStop}%
\bibitem [{\citenamefont {Nandkishore}\ \emph {et~al.}(2014)\citenamefont
  {Nandkishore}, \citenamefont {Gopalakrishnan},\ and\ \citenamefont
  {Huse}}]{PhysRevB.90.064203}%
  \BibitemOpen
  \bibfield  {author} {\bibinfo {author} {\bibfnamefont {R.}~\bibnamefont
  {Nandkishore}}, \bibinfo {author} {\bibfnamefont {S.}~\bibnamefont
  {Gopalakrishnan}},\ and\ \bibinfo {author} {\bibfnamefont {D.~A.}\
  \bibnamefont {Huse}},\ }\bibfield  {title} {\bibinfo {title} {Spectral
  features of a many-body-localized system weakly coupled to a bath},\ }\href
  {https://doi.org/10.1103/PhysRevB.90.064203} {\bibfield  {journal} {\bibinfo
  {journal} {Phys. Rev. B}\ }\textbf {\bibinfo {volume} {90}},\ \bibinfo
  {pages} {064203} (\bibinfo {year} {2014})}\BibitemShut {NoStop}%
\bibitem [{\citenamefont {Nandkishore}\ and\ \citenamefont
  {Gopalakrishnan}(2017)}]{nandkishore2017many}%
  \BibitemOpen
  \bibfield  {author} {\bibinfo {author} {\bibfnamefont {R.}~\bibnamefont
  {Nandkishore}}\ and\ \bibinfo {author} {\bibfnamefont {S.}~\bibnamefont
  {Gopalakrishnan}},\ }\bibfield  {title} {\bibinfo {title} {Many body
  localized systems weakly coupled to baths},\ }\href@noop {} {\bibfield
  {journal} {\bibinfo  {journal} {Annalen der Physik}\ }\textbf {\bibinfo
  {volume} {529}},\ \bibinfo {pages} {1600181} (\bibinfo {year}
  {2017})}\BibitemShut {NoStop}%
\bibitem [{\citenamefont {Fischer}\ \emph {et~al.}(2016)\citenamefont
  {Fischer}, \citenamefont {Maksymenko},\ and\ \citenamefont
  {Altman}}]{fischer2016dynamics}%
  \BibitemOpen
  \bibfield  {author} {\bibinfo {author} {\bibfnamefont {M.~H.}\ \bibnamefont
  {Fischer}}, \bibinfo {author} {\bibfnamefont {M.}~\bibnamefont
  {Maksymenko}},\ and\ \bibinfo {author} {\bibfnamefont {E.}~\bibnamefont
  {Altman}},\ }\bibfield  {title} {\bibinfo {title} {Dynamics of a
  many-body-localized system coupled to a bath},\ }\href@noop {} {\bibfield
  {journal} {\bibinfo  {journal} {Physical review letters}\ }\textbf {\bibinfo
  {volume} {116}},\ \bibinfo {pages} {160401} (\bibinfo {year}
  {2016})}\BibitemShut {NoStop}%
\bibitem [{\citenamefont {Blais}\ \emph {et~al.}(2021)\citenamefont {Blais},
  \citenamefont {Grimsmo}, \citenamefont {Girvin},\ and\ \citenamefont
  {Wallraff}}]{RevModPhys.93.025005}%
  \BibitemOpen
  \bibfield  {author} {\bibinfo {author} {\bibfnamefont {A.}~\bibnamefont
  {Blais}}, \bibinfo {author} {\bibfnamefont {A.~L.}\ \bibnamefont {Grimsmo}},
  \bibinfo {author} {\bibfnamefont {S.~M.}\ \bibnamefont {Girvin}},\ and\
  \bibinfo {author} {\bibfnamefont {A.}~\bibnamefont {Wallraff}},\ }\bibfield
  {title} {\bibinfo {title} {Circuit quantum electrodynamics},\ }\href
  {https://doi.org/10.1103/RevModPhys.93.025005} {\bibfield  {journal}
  {\bibinfo  {journal} {Rev. Mod. Phys.}\ }\textbf {\bibinfo {volume} {93}},\
  \bibinfo {pages} {025005} (\bibinfo {year} {2021})}\BibitemShut {NoStop}%
\bibitem [{\citenamefont {Blais}\ \emph {et~al.}(2020)\citenamefont {Blais},
  \citenamefont {Girvin},\ and\ \citenamefont {Oliver}}]{Blais2020}%
  \BibitemOpen
  \bibfield  {author} {\bibinfo {author} {\bibfnamefont {A.}~\bibnamefont
  {Blais}}, \bibinfo {author} {\bibfnamefont {S.~M.}\ \bibnamefont {Girvin}},\
  and\ \bibinfo {author} {\bibfnamefont {W.~D.}\ \bibnamefont {Oliver}},\
  }\bibfield  {title} {\bibinfo {title} {Quantum information processing and
  quantum optics with circuit quantum electrodynamics},\ }\href
  {https://doi.org/10.1038/s41567-020-0806-z} {\bibfield  {journal} {\bibinfo
  {journal} {Nature Physics}\ }\textbf {\bibinfo {volume} {16}},\ \bibinfo
  {pages} {247} (\bibinfo {year} {2020})}\BibitemShut {NoStop}%
\bibitem [{\citenamefont {Joshi}\ \emph {et~al.}(2021)\citenamefont {Joshi},
  \citenamefont {Noh},\ and\ \citenamefont {Gao}}]{Joshi_2021}%
  \BibitemOpen
  \bibfield  {author} {\bibinfo {author} {\bibfnamefont {A.}~\bibnamefont
  {Joshi}}, \bibinfo {author} {\bibfnamefont {K.}~\bibnamefont {Noh}},\ and\
  \bibinfo {author} {\bibfnamefont {Y.~Y.}\ \bibnamefont {Gao}},\ }\bibfield
  {title} {\bibinfo {title} {Quantum information processing with bosonic qubits
  in circuit {QED}},\ }\href {https://doi.org/10.1088/2058-9565/abe989}
  {\bibfield  {journal} {\bibinfo  {journal} {Quantum Science and Technology}\
  }\textbf {\bibinfo {volume} {6}},\ \bibinfo {pages} {033001} (\bibinfo {year}
  {2021})}\BibitemShut {NoStop}%
\bibitem [{\citenamefont {Eickbusch}\ \emph {et~al.}(2021)\citenamefont
  {Eickbusch}, \citenamefont {Sivak}, \citenamefont {Ding}, \citenamefont
  {Elder}, \citenamefont {Jha}, \citenamefont {Venkatraman}, \citenamefont
  {Royer}, \citenamefont {Girvin}, \citenamefont {Schoelkopf},\ and\
  \citenamefont {Devoret}}]{eickbusch2021fast}%
  \BibitemOpen
  \bibfield  {author} {\bibinfo {author} {\bibfnamefont {A.}~\bibnamefont
  {Eickbusch}}, \bibinfo {author} {\bibfnamefont {V.}~\bibnamefont {Sivak}},
  \bibinfo {author} {\bibfnamefont {A.~Z.}\ \bibnamefont {Ding}}, \bibinfo
  {author} {\bibfnamefont {S.~S.}\ \bibnamefont {Elder}}, \bibinfo {author}
  {\bibfnamefont {S.~R.}\ \bibnamefont {Jha}}, \bibinfo {author} {\bibfnamefont
  {J.}~\bibnamefont {Venkatraman}}, \bibinfo {author} {\bibfnamefont
  {B.}~\bibnamefont {Royer}}, \bibinfo {author} {\bibfnamefont {S.~M.}\
  \bibnamefont {Girvin}}, \bibinfo {author} {\bibfnamefont {R.~J.}\
  \bibnamefont {Schoelkopf}},\ and\ \bibinfo {author} {\bibfnamefont {M.~H.}\
  \bibnamefont {Devoret}},\ }\href@noop {} {\bibinfo {title} {Fast universal
  control of an oscillator with weak dispersive coupling to a qubit}} (\bibinfo
  {year} {2021}),\ \Eprint {https://arxiv.org/abs/2111.06414} {arXiv:2111.06414
  [quant-ph]} \BibitemShut {NoStop}%
\bibitem [{\citenamefont {Ma}\ \emph {et~al.}(2021)\citenamefont {Ma},
  \citenamefont {Puri}, \citenamefont {Schoelkopf}, \citenamefont {Devoret},
  \citenamefont {Girvin},\ and\ \citenamefont {Jiang}}]{Ma2021}%
  \BibitemOpen
  \bibfield  {author} {\bibinfo {author} {\bibfnamefont {W.-L.}\ \bibnamefont
  {Ma}}, \bibinfo {author} {\bibfnamefont {S.}~\bibnamefont {Puri}}, \bibinfo
  {author} {\bibfnamefont {R.~J.}\ \bibnamefont {Schoelkopf}}, \bibinfo
  {author} {\bibfnamefont {M.~H.}\ \bibnamefont {Devoret}}, \bibinfo {author}
  {\bibfnamefont {S.}~\bibnamefont {Girvin}},\ and\ \bibinfo {author}
  {\bibfnamefont {L.}~\bibnamefont {Jiang}},\ }\bibfield  {title} {\bibinfo
  {title} {Quantum control of bosonic modes with superconducting circuits},\
  }\href {https://doi.org/10.1016/j.scib.2021.05.024} {\bibfield  {journal}
  {\bibinfo  {journal} {Science Bulletin}\ }\textbf {\bibinfo {volume} {66}},\
  \bibinfo {pages} {1789} (\bibinfo {year} {2021})}\BibitemShut {NoStop}%
\bibitem [{\citenamefont {Wang}\ \emph {et~al.}(2021)\citenamefont {Wang},
  \citenamefont {Noh}, \citenamefont {Lebreuilly}, \citenamefont {Girvin},\
  and\ \citenamefont {Jiang}}]{Wang2021}%
  \BibitemOpen
  \bibfield  {author} {\bibinfo {author} {\bibfnamefont {C.-H.}\ \bibnamefont
  {Wang}}, \bibinfo {author} {\bibfnamefont {K.}~\bibnamefont {Noh}}, \bibinfo
  {author} {\bibfnamefont {J.}~\bibnamefont {Lebreuilly}}, \bibinfo {author}
  {\bibfnamefont {S.~M.}\ \bibnamefont {Girvin}},\ and\ \bibinfo {author}
  {\bibfnamefont {L.}~\bibnamefont {Jiang}},\ }\bibfield  {title} {\bibinfo
  {title} {Photon-number-dependent hamiltonian engineering for cavities},\
  }\href {https://doi.org/10.1103/physrevapplied.15.044026} {\bibfield
  {journal} {\bibinfo  {journal} {Physical Review Applied}\ }\textbf {\bibinfo
  {volume} {15}},\ \bibinfo {pages} {044026} (\bibinfo {year}
  {2021})}\BibitemShut {NoStop}%
\bibitem [{\citenamefont {Wang}\ \emph {et~al.}(2020)\citenamefont {Wang},
  \citenamefont {Curtis}, \citenamefont {Lester}, \citenamefont {Zhang},
  \citenamefont {Gao}, \citenamefont {Freeze}, \citenamefont {Batista},
  \citenamefont {Vaccaro}, \citenamefont {Chuang}, \citenamefont {Frunzio},
  \citenamefont {Jiang}, \citenamefont {Girvin},\ and\ \citenamefont
  {Schoelkopf}}]{PhysRevX.10.021060}%
  \BibitemOpen
  \bibfield  {author} {\bibinfo {author} {\bibfnamefont {C.~S.}\ \bibnamefont
  {Wang}}, \bibinfo {author} {\bibfnamefont {J.~C.}\ \bibnamefont {Curtis}},
  \bibinfo {author} {\bibfnamefont {B.~J.}\ \bibnamefont {Lester}}, \bibinfo
  {author} {\bibfnamefont {Y.}~\bibnamefont {Zhang}}, \bibinfo {author}
  {\bibfnamefont {Y.~Y.}\ \bibnamefont {Gao}}, \bibinfo {author} {\bibfnamefont
  {J.}~\bibnamefont {Freeze}}, \bibinfo {author} {\bibfnamefont {V.~S.}\
  \bibnamefont {Batista}}, \bibinfo {author} {\bibfnamefont {P.~H.}\
  \bibnamefont {Vaccaro}}, \bibinfo {author} {\bibfnamefont {I.~L.}\
  \bibnamefont {Chuang}}, \bibinfo {author} {\bibfnamefont {L.}~\bibnamefont
  {Frunzio}}, \bibinfo {author} {\bibfnamefont {L.}~\bibnamefont {Jiang}},
  \bibinfo {author} {\bibfnamefont {S.~M.}\ \bibnamefont {Girvin}},\ and\
  \bibinfo {author} {\bibfnamefont {R.~J.}\ \bibnamefont {Schoelkopf}},\
  }\bibfield  {title} {\bibinfo {title} {Efficient multiphoton sampling of
  molecular vibronic spectra on a superconducting bosonic processor},\ }\href
  {https://doi.org/10.1103/PhysRevX.10.021060} {\bibfield  {journal} {\bibinfo
  {journal} {Phys. Rev. X}\ }\textbf {\bibinfo {volume} {10}},\ \bibinfo
  {pages} {021060} (\bibinfo {year} {2020})}\BibitemShut {NoStop}%
\bibitem [{\citenamefont {Wallraff}\ \emph {et~al.}(2004)\citenamefont
  {Wallraff}, \citenamefont {Schuster}, \citenamefont {Blais}, \citenamefont
  {Frunzio}, \citenamefont {Huang}, \citenamefont {Majer}, \citenamefont
  {Kumar}, \citenamefont {Girvin},\ and\ \citenamefont
  {Schoelkopf}}]{Wallraff2004}%
  \BibitemOpen
  \bibfield  {author} {\bibinfo {author} {\bibfnamefont {A.}~\bibnamefont
  {Wallraff}}, \bibinfo {author} {\bibfnamefont {D.~I.}\ \bibnamefont
  {Schuster}}, \bibinfo {author} {\bibfnamefont {A.}~\bibnamefont {Blais}},
  \bibinfo {author} {\bibfnamefont {L.}~\bibnamefont {Frunzio}}, \bibinfo
  {author} {\bibfnamefont {R.-S.}\ \bibnamefont {Huang}}, \bibinfo {author}
  {\bibfnamefont {J.}~\bibnamefont {Majer}}, \bibinfo {author} {\bibfnamefont
  {S.}~\bibnamefont {Kumar}}, \bibinfo {author} {\bibfnamefont {S.~M.}\
  \bibnamefont {Girvin}},\ and\ \bibinfo {author} {\bibfnamefont {R.~J.}\
  \bibnamefont {Schoelkopf}},\ }\bibfield  {title} {\bibinfo {title} {Strong
  coupling of a single photon to a superconducting qubit using circuit quantum
  electrodynamics},\ }\href {https://doi.org/10.1038/nature02851} {\bibfield
  {journal} {\bibinfo  {journal} {Nature}\ }\textbf {\bibinfo {volume} {431}},\
  \bibinfo {pages} {162} (\bibinfo {year} {2004})}\BibitemShut {NoStop}%
\bibitem [{\citenamefont {Houck}\ \emph {et~al.}(2012)\citenamefont {Houck},
  \citenamefont {T\"{u}reci},\ and\ \citenamefont {Koch}}]{Houck2012}%
  \BibitemOpen
  \bibfield  {author} {\bibinfo {author} {\bibfnamefont {A.~A.}\ \bibnamefont
  {Houck}}, \bibinfo {author} {\bibfnamefont {H.~E.}\ \bibnamefont
  {T\"{u}reci}},\ and\ \bibinfo {author} {\bibfnamefont {J.}~\bibnamefont
  {Koch}},\ }\bibfield  {title} {\bibinfo {title} {On-chip quantum simulation
  with superconducting circuits},\ }\href {https://doi.org/10.1038/nphys2251}
  {\bibfield  {journal} {\bibinfo  {journal} {Nature Physics}\ }\textbf
  {\bibinfo {volume} {8}},\ \bibinfo {pages} {292} (\bibinfo {year}
  {2012})}\BibitemShut {NoStop}%
\bibitem [{\citenamefont {Koch}\ \emph {et~al.}(2007)\citenamefont {Koch},
  \citenamefont {Yu}, \citenamefont {Gambetta}, \citenamefont {Houck},
  \citenamefont {Schuster}, \citenamefont {Majer}, \citenamefont {Blais},
  \citenamefont {Devoret}, \citenamefont {Girvin},\ and\ \citenamefont
  {Schoelkopf}}]{PhysRevA.76.042319}%
  \BibitemOpen
  \bibfield  {author} {\bibinfo {author} {\bibfnamefont {J.}~\bibnamefont
  {Koch}}, \bibinfo {author} {\bibfnamefont {T.~M.}\ \bibnamefont {Yu}},
  \bibinfo {author} {\bibfnamefont {J.}~\bibnamefont {Gambetta}}, \bibinfo
  {author} {\bibfnamefont {A.~A.}\ \bibnamefont {Houck}}, \bibinfo {author}
  {\bibfnamefont {D.~I.}\ \bibnamefont {Schuster}}, \bibinfo {author}
  {\bibfnamefont {J.}~\bibnamefont {Majer}}, \bibinfo {author} {\bibfnamefont
  {A.}~\bibnamefont {Blais}}, \bibinfo {author} {\bibfnamefont {M.~H.}\
  \bibnamefont {Devoret}}, \bibinfo {author} {\bibfnamefont {S.~M.}\
  \bibnamefont {Girvin}},\ and\ \bibinfo {author} {\bibfnamefont {R.~J.}\
  \bibnamefont {Schoelkopf}},\ }\bibfield  {title} {\bibinfo {title}
  {Charge-insensitive qubit design derived from the cooper pair box},\ }\href
  {https://doi.org/10.1103/PhysRevA.76.042319} {\bibfield  {journal} {\bibinfo
  {journal} {Phys. Rev. A}\ }\textbf {\bibinfo {volume} {76}},\ \bibinfo
  {pages} {042319} (\bibinfo {year} {2007})}\BibitemShut {NoStop}%
\bibitem [{\citenamefont {Frattini}\ \emph {et~al.}(2017)\citenamefont
  {Frattini}, \citenamefont {Vool}, \citenamefont {Shankar}, \citenamefont
  {Narla}, \citenamefont {Sliwa},\ and\ \citenamefont
  {Devoret}}]{Frattini2017}%
  \BibitemOpen
  \bibfield  {author} {\bibinfo {author} {\bibfnamefont {N.~E.}\ \bibnamefont
  {Frattini}}, \bibinfo {author} {\bibfnamefont {U.}~\bibnamefont {Vool}},
  \bibinfo {author} {\bibfnamefont {S.}~\bibnamefont {Shankar}}, \bibinfo
  {author} {\bibfnamefont {A.}~\bibnamefont {Narla}}, \bibinfo {author}
  {\bibfnamefont {K.~M.}\ \bibnamefont {Sliwa}},\ and\ \bibinfo {author}
  {\bibfnamefont {M.~H.}\ \bibnamefont {Devoret}},\ }\bibfield  {title}
  {\bibinfo {title} {3-wave mixing josephson dipole element},\ }\href
  {https://doi.org/10.1063/1.4984142} {\bibfield  {journal} {\bibinfo
  {journal} {Applied Physics Letters}\ }\textbf {\bibinfo {volume} {110}},\
  \bibinfo {pages} {222603} (\bibinfo {year} {2017})}\BibitemShut {NoStop}%
\bibitem [{\citenamefont {Carusotto}\ \emph {et~al.}(2020)\citenamefont
  {Carusotto}, \citenamefont {Houck}, \citenamefont {Koll{\'{a}}r},
  \citenamefont {Roushan}, \citenamefont {Schuster},\ and\ \citenamefont
  {Simon}}]{Carusotto2020}%
  \BibitemOpen
  \bibfield  {author} {\bibinfo {author} {\bibfnamefont {I.}~\bibnamefont
  {Carusotto}}, \bibinfo {author} {\bibfnamefont {A.~A.}\ \bibnamefont
  {Houck}}, \bibinfo {author} {\bibfnamefont {A.~J.}\ \bibnamefont
  {Koll{\'{a}}r}}, \bibinfo {author} {\bibfnamefont {P.}~\bibnamefont
  {Roushan}}, \bibinfo {author} {\bibfnamefont {D.~I.}\ \bibnamefont
  {Schuster}},\ and\ \bibinfo {author} {\bibfnamefont {J.}~\bibnamefont
  {Simon}},\ }\bibfield  {title} {\bibinfo {title} {Photonic materials in
  circuit quantum electrodynamics},\ }\href
  {https://doi.org/10.1038/s41567-020-0815-y} {\bibfield  {journal} {\bibinfo
  {journal} {Nature Physics}\ }\textbf {\bibinfo {volume} {16}},\ \bibinfo
  {pages} {268} (\bibinfo {year} {2020})}\BibitemShut {NoStop}%
\bibitem [{\citenamefont {Yanay}\ \emph {et~al.}(2020)\citenamefont {Yanay},
  \citenamefont {Braum\"{u}ller}, \citenamefont {Gustavsson}, \citenamefont
  {Oliver},\ and\ \citenamefont {Tahan}}]{Yanay2020}%
  \BibitemOpen
  \bibfield  {author} {\bibinfo {author} {\bibfnamefont {Y.}~\bibnamefont
  {Yanay}}, \bibinfo {author} {\bibfnamefont {J.}~\bibnamefont
  {Braum\"{u}ller}}, \bibinfo {author} {\bibfnamefont {S.}~\bibnamefont
  {Gustavsson}}, \bibinfo {author} {\bibfnamefont {W.~D.}\ \bibnamefont
  {Oliver}},\ and\ \bibinfo {author} {\bibfnamefont {C.}~\bibnamefont
  {Tahan}},\ }\bibfield  {title} {\bibinfo {title} {Two-dimensional hard-core
  bose{\textendash}hubbard model with superconducting qubits},\ }\href
  {https://doi.org/10.1038/s41534-020-0269-1} {\bibfield  {journal} {\bibinfo
  {journal} {npj Quantum Information}\ }\textbf {\bibinfo {volume} {6}},\
  \bibinfo {pages} {1} (\bibinfo {year} {2020})}\BibitemShut {NoStop}%
\bibitem [{\citenamefont {Mansikkam\"aki}\ \emph {et~al.}(2021)\citenamefont
  {Mansikkam\"aki}, \citenamefont {Laine},\ and\ \citenamefont
  {Silveri}}]{PhysRevB.103.L220202}%
  \BibitemOpen
  \bibfield  {author} {\bibinfo {author} {\bibfnamefont {O.}~\bibnamefont
  {Mansikkam\"aki}}, \bibinfo {author} {\bibfnamefont {S.}~\bibnamefont
  {Laine}},\ and\ \bibinfo {author} {\bibfnamefont {M.}~\bibnamefont
  {Silveri}},\ }\bibfield  {title} {\bibinfo {title} {Phases of the disordered
  bose-hubbard model with attractive interactions},\ }\href
  {https://doi.org/10.1103/PhysRevB.103.L220202} {\bibfield  {journal}
  {\bibinfo  {journal} {Phys. Rev. B}\ }\textbf {\bibinfo {volume} {103}},\
  \bibinfo {pages} {L220202} (\bibinfo {year} {2021})}\BibitemShut {NoStop}%
\bibitem [{\citenamefont {Schmidt}\ and\ \citenamefont
  {Koch}(2013)}]{Schmidt2013}%
  \BibitemOpen
  \bibfield  {author} {\bibinfo {author} {\bibfnamefont {S.}~\bibnamefont
  {Schmidt}}\ and\ \bibinfo {author} {\bibfnamefont {J.}~\bibnamefont {Koch}},\
  }\bibfield  {title} {\bibinfo {title} {Circuit {QED} lattices: Towards
  quantum simulation with superconducting circuits},\ }\href
  {https://doi.org/10.1002/andp.201200261} {\bibfield  {journal} {\bibinfo
  {journal} {Annalen der Physik}\ }\textbf {\bibinfo {volume} {525}},\ \bibinfo
  {pages} {395} (\bibinfo {year} {2013})}\BibitemShut {NoStop}%
\bibitem [{\citenamefont {Devoret}\ and\ \citenamefont
  {Schoelkopf}(2013)}]{Devoret2013}%
  \BibitemOpen
  \bibfield  {author} {\bibinfo {author} {\bibfnamefont {M.~H.}\ \bibnamefont
  {Devoret}}\ and\ \bibinfo {author} {\bibfnamefont {R.~J.}\ \bibnamefont
  {Schoelkopf}},\ }\bibfield  {title} {\bibinfo {title} {Superconducting
  circuits for quantum information: An outlook},\ }\href
  {https://doi.org/10.1126/science.1231930} {\bibfield  {journal} {\bibinfo
  {journal} {Science}\ }\textbf {\bibinfo {volume} {339}},\ \bibinfo {pages}
  {1169} (\bibinfo {year} {2013})}\BibitemShut {NoStop}%
\bibitem [{\citenamefont {Roushan}\ \emph {et~al.}(2017)\citenamefont
  {Roushan}, \citenamefont {Neill}, \citenamefont {Tangpanitanon},
  \citenamefont {Bastidas}, \citenamefont {Megrant}, \citenamefont {Barends},
  \citenamefont {Chen}, \citenamefont {Chen}, \citenamefont {Chiaro},
  \citenamefont {Dunsworth}, \citenamefont {Fowler}, \citenamefont {Foxen},
  \citenamefont {Giustina}, \citenamefont {Jeffrey}, \citenamefont {Kelly},
  \citenamefont {Lucero}, \citenamefont {Mutus}, \citenamefont {Neeley},
  \citenamefont {Quintana}, \citenamefont {Sank}, \citenamefont {Vainsencher},
  \citenamefont {Wenner}, \citenamefont {White}, \citenamefont {Neven},
  \citenamefont {Angelakis},\ and\ \citenamefont {Martinis}}]{Roushan2017}%
  \BibitemOpen
  \bibfield  {author} {\bibinfo {author} {\bibfnamefont {P.}~\bibnamefont
  {Roushan}}, \bibinfo {author} {\bibfnamefont {C.}~\bibnamefont {Neill}},
  \bibinfo {author} {\bibfnamefont {J.}~\bibnamefont {Tangpanitanon}}, \bibinfo
  {author} {\bibfnamefont {V.~M.}\ \bibnamefont {Bastidas}}, \bibinfo {author}
  {\bibfnamefont {A.}~\bibnamefont {Megrant}}, \bibinfo {author} {\bibfnamefont
  {R.}~\bibnamefont {Barends}}, \bibinfo {author} {\bibfnamefont
  {Y.}~\bibnamefont {Chen}}, \bibinfo {author} {\bibfnamefont {Z.}~\bibnamefont
  {Chen}}, \bibinfo {author} {\bibfnamefont {B.}~\bibnamefont {Chiaro}},
  \bibinfo {author} {\bibfnamefont {A.}~\bibnamefont {Dunsworth}}, \bibinfo
  {author} {\bibfnamefont {A.}~\bibnamefont {Fowler}}, \bibinfo {author}
  {\bibfnamefont {B.}~\bibnamefont {Foxen}}, \bibinfo {author} {\bibfnamefont
  {M.}~\bibnamefont {Giustina}}, \bibinfo {author} {\bibfnamefont
  {E.}~\bibnamefont {Jeffrey}}, \bibinfo {author} {\bibfnamefont
  {J.}~\bibnamefont {Kelly}}, \bibinfo {author} {\bibfnamefont
  {E.}~\bibnamefont {Lucero}}, \bibinfo {author} {\bibfnamefont
  {J.}~\bibnamefont {Mutus}}, \bibinfo {author} {\bibfnamefont
  {M.}~\bibnamefont {Neeley}}, \bibinfo {author} {\bibfnamefont
  {C.}~\bibnamefont {Quintana}}, \bibinfo {author} {\bibfnamefont
  {D.}~\bibnamefont {Sank}}, \bibinfo {author} {\bibfnamefont {A.}~\bibnamefont
  {Vainsencher}}, \bibinfo {author} {\bibfnamefont {J.}~\bibnamefont {Wenner}},
  \bibinfo {author} {\bibfnamefont {T.}~\bibnamefont {White}}, \bibinfo
  {author} {\bibfnamefont {H.}~\bibnamefont {Neven}}, \bibinfo {author}
  {\bibfnamefont {D.~G.}\ \bibnamefont {Angelakis}},\ and\ \bibinfo {author}
  {\bibfnamefont {J.}~\bibnamefont {Martinis}},\ }\bibfield  {title} {\bibinfo
  {title} {Spectroscopic signatures of localization with interacting photons in
  superconducting qubits},\ }\href {https://doi.org/10.1126/science.aao1401}
  {\bibfield  {journal} {\bibinfo  {journal} {Science}\ }\textbf {\bibinfo
  {volume} {358}},\ \bibinfo {pages} {1175} (\bibinfo {year}
  {2017})}\BibitemShut {NoStop}%
\bibitem [{\citenamefont {Ye}\ \emph {et~al.}(2019)\citenamefont {Ye},
  \citenamefont {Ge}, \citenamefont {Wu}, \citenamefont {Wang}, \citenamefont
  {Gong}, \citenamefont {Zhang}, \citenamefont {Zhu}, \citenamefont {Yang},
  \citenamefont {Li}, \citenamefont {Liang}, \citenamefont {Lin}, \citenamefont
  {Xu}, \citenamefont {Guo}, \citenamefont {Sun}, \citenamefont {Cheng},
  \citenamefont {Ma}, \citenamefont {Meng}, \citenamefont {Deng}, \citenamefont
  {Rong}, \citenamefont {Lu}, \citenamefont {Peng}, \citenamefont {Fan},
  \citenamefont {Zhu},\ and\ \citenamefont {Pan}}]{PhysRevLett.123.050502}%
  \BibitemOpen
  \bibfield  {author} {\bibinfo {author} {\bibfnamefont {Y.}~\bibnamefont
  {Ye}}, \bibinfo {author} {\bibfnamefont {Z.-Y.}\ \bibnamefont {Ge}}, \bibinfo
  {author} {\bibfnamefont {Y.}~\bibnamefont {Wu}}, \bibinfo {author}
  {\bibfnamefont {S.}~\bibnamefont {Wang}}, \bibinfo {author} {\bibfnamefont
  {M.}~\bibnamefont {Gong}}, \bibinfo {author} {\bibfnamefont {Y.-R.}\
  \bibnamefont {Zhang}}, \bibinfo {author} {\bibfnamefont {Q.}~\bibnamefont
  {Zhu}}, \bibinfo {author} {\bibfnamefont {R.}~\bibnamefont {Yang}}, \bibinfo
  {author} {\bibfnamefont {S.}~\bibnamefont {Li}}, \bibinfo {author}
  {\bibfnamefont {F.}~\bibnamefont {Liang}}, \bibinfo {author} {\bibfnamefont
  {J.}~\bibnamefont {Lin}}, \bibinfo {author} {\bibfnamefont {Y.}~\bibnamefont
  {Xu}}, \bibinfo {author} {\bibfnamefont {C.}~\bibnamefont {Guo}}, \bibinfo
  {author} {\bibfnamefont {L.}~\bibnamefont {Sun}}, \bibinfo {author}
  {\bibfnamefont {C.}~\bibnamefont {Cheng}}, \bibinfo {author} {\bibfnamefont
  {N.}~\bibnamefont {Ma}}, \bibinfo {author} {\bibfnamefont {Z.~Y.}\
  \bibnamefont {Meng}}, \bibinfo {author} {\bibfnamefont {H.}~\bibnamefont
  {Deng}}, \bibinfo {author} {\bibfnamefont {H.}~\bibnamefont {Rong}}, \bibinfo
  {author} {\bibfnamefont {C.-Y.}\ \bibnamefont {Lu}}, \bibinfo {author}
  {\bibfnamefont {C.-Z.}\ \bibnamefont {Peng}}, \bibinfo {author}
  {\bibfnamefont {H.}~\bibnamefont {Fan}}, \bibinfo {author} {\bibfnamefont
  {X.}~\bibnamefont {Zhu}},\ and\ \bibinfo {author} {\bibfnamefont {J.-W.}\
  \bibnamefont {Pan}},\ }\bibfield  {title} {\bibinfo {title} {Propagation and
  localization of collective excitations on a 24-qubit superconducting
  processor},\ }\href {https://doi.org/10.1103/PhysRevLett.123.050502}
  {\bibfield  {journal} {\bibinfo  {journal} {Phys. Rev. Lett.}\ }\textbf
  {\bibinfo {volume} {123}},\ \bibinfo {pages} {050502} (\bibinfo {year}
  {2019})}\BibitemShut {NoStop}%
\bibitem [{\citenamefont {Chiaro}\ \emph {et~al.}(2020)\citenamefont {Chiaro},
  \citenamefont {Neill}, \citenamefont {Bohrdt}, \citenamefont {Filippone},
  \citenamefont {Arute}, \citenamefont {Arya}, \citenamefont {Babbush},
  \citenamefont {Bacon}, \citenamefont {Bardin}, \citenamefont {Barends},
  \citenamefont {Boixo}, \citenamefont {Buell}, \citenamefont {Burkett},
  \citenamefont {Chen}, \citenamefont {Chen}, \citenamefont {Collins},
  \citenamefont {Dunsworth}, \citenamefont {Farhi}, \citenamefont {Fowler},
  \citenamefont {Foxen}, \citenamefont {Gidney}, \citenamefont {Giustina},
  \citenamefont {Harrigan}, \citenamefont {Huang}, \citenamefont {Isakov},
  \citenamefont {Jeffrey}, \citenamefont {Jiang}, \citenamefont {Kafri},
  \citenamefont {Kechedzhi}, \citenamefont {Kelly}, \citenamefont {Klimov},
  \citenamefont {Korotkov}, \citenamefont {Kostritsa}, \citenamefont
  {Landhuis}, \citenamefont {Lucero}, \citenamefont {McClean}, \citenamefont
  {Mi}, \citenamefont {Megrant}, \citenamefont {Mohseni}, \citenamefont
  {Mutus}, \citenamefont {McEwen}, \citenamefont {Naaman}, \citenamefont
  {Neeley}, \citenamefont {Niu}, \citenamefont {Petukhov}, \citenamefont
  {Quintana}, \citenamefont {Rubin}, \citenamefont {Sank}, \citenamefont
  {Satzinger}, \citenamefont {Vainsencher}, \citenamefont {White},
  \citenamefont {Yao}, \citenamefont {Yeh}, \citenamefont {Zalcman},
  \citenamefont {Smelyanskiy}, \citenamefont {Neven}, \citenamefont
  {Gopalakrishnan}, \citenamefont {Abanin}, \citenamefont {Knap}, \citenamefont
  {Martinis},\ and\ \citenamefont {Roushan}}]{chiaro2020direct}%
  \BibitemOpen
  \bibfield  {author} {\bibinfo {author} {\bibfnamefont {B.}~\bibnamefont
  {Chiaro}}, \bibinfo {author} {\bibfnamefont {C.}~\bibnamefont {Neill}},
  \bibinfo {author} {\bibfnamefont {A.}~\bibnamefont {Bohrdt}}, \bibinfo
  {author} {\bibfnamefont {M.}~\bibnamefont {Filippone}}, \bibinfo {author}
  {\bibfnamefont {F.}~\bibnamefont {Arute}}, \bibinfo {author} {\bibfnamefont
  {K.}~\bibnamefont {Arya}}, \bibinfo {author} {\bibfnamefont {R.}~\bibnamefont
  {Babbush}}, \bibinfo {author} {\bibfnamefont {D.}~\bibnamefont {Bacon}},
  \bibinfo {author} {\bibfnamefont {J.}~\bibnamefont {Bardin}}, \bibinfo
  {author} {\bibfnamefont {R.}~\bibnamefont {Barends}}, \bibinfo {author}
  {\bibfnamefont {S.}~\bibnamefont {Boixo}}, \bibinfo {author} {\bibfnamefont
  {D.}~\bibnamefont {Buell}}, \bibinfo {author} {\bibfnamefont
  {B.}~\bibnamefont {Burkett}}, \bibinfo {author} {\bibfnamefont
  {Y.}~\bibnamefont {Chen}}, \bibinfo {author} {\bibfnamefont {Z.}~\bibnamefont
  {Chen}}, \bibinfo {author} {\bibfnamefont {R.}~\bibnamefont {Collins}},
  \bibinfo {author} {\bibfnamefont {A.}~\bibnamefont {Dunsworth}}, \bibinfo
  {author} {\bibfnamefont {E.}~\bibnamefont {Farhi}}, \bibinfo {author}
  {\bibfnamefont {A.}~\bibnamefont {Fowler}}, \bibinfo {author} {\bibfnamefont
  {B.}~\bibnamefont {Foxen}}, \bibinfo {author} {\bibfnamefont
  {C.}~\bibnamefont {Gidney}}, \bibinfo {author} {\bibfnamefont
  {M.}~\bibnamefont {Giustina}}, \bibinfo {author} {\bibfnamefont
  {M.}~\bibnamefont {Harrigan}}, \bibinfo {author} {\bibfnamefont
  {T.}~\bibnamefont {Huang}}, \bibinfo {author} {\bibfnamefont
  {S.}~\bibnamefont {Isakov}}, \bibinfo {author} {\bibfnamefont
  {E.}~\bibnamefont {Jeffrey}}, \bibinfo {author} {\bibfnamefont
  {Z.}~\bibnamefont {Jiang}}, \bibinfo {author} {\bibfnamefont
  {D.}~\bibnamefont {Kafri}}, \bibinfo {author} {\bibfnamefont
  {K.}~\bibnamefont {Kechedzhi}}, \bibinfo {author} {\bibfnamefont
  {J.}~\bibnamefont {Kelly}}, \bibinfo {author} {\bibfnamefont
  {P.}~\bibnamefont {Klimov}}, \bibinfo {author} {\bibfnamefont
  {A.}~\bibnamefont {Korotkov}}, \bibinfo {author} {\bibfnamefont
  {F.}~\bibnamefont {Kostritsa}}, \bibinfo {author} {\bibfnamefont
  {D.}~\bibnamefont {Landhuis}}, \bibinfo {author} {\bibfnamefont
  {E.}~\bibnamefont {Lucero}}, \bibinfo {author} {\bibfnamefont
  {J.}~\bibnamefont {McClean}}, \bibinfo {author} {\bibfnamefont
  {X.}~\bibnamefont {Mi}}, \bibinfo {author} {\bibfnamefont {A.}~\bibnamefont
  {Megrant}}, \bibinfo {author} {\bibfnamefont {M.}~\bibnamefont {Mohseni}},
  \bibinfo {author} {\bibfnamefont {J.}~\bibnamefont {Mutus}}, \bibinfo
  {author} {\bibfnamefont {M.}~\bibnamefont {McEwen}}, \bibinfo {author}
  {\bibfnamefont {O.}~\bibnamefont {Naaman}}, \bibinfo {author} {\bibfnamefont
  {M.}~\bibnamefont {Neeley}}, \bibinfo {author} {\bibfnamefont
  {M.}~\bibnamefont {Niu}}, \bibinfo {author} {\bibfnamefont {A.}~\bibnamefont
  {Petukhov}}, \bibinfo {author} {\bibfnamefont {C.}~\bibnamefont {Quintana}},
  \bibinfo {author} {\bibfnamefont {N.}~\bibnamefont {Rubin}}, \bibinfo
  {author} {\bibfnamefont {D.}~\bibnamefont {Sank}}, \bibinfo {author}
  {\bibfnamefont {K.}~\bibnamefont {Satzinger}}, \bibinfo {author}
  {\bibfnamefont {A.}~\bibnamefont {Vainsencher}}, \bibinfo {author}
  {\bibfnamefont {T.}~\bibnamefont {White}}, \bibinfo {author} {\bibfnamefont
  {Z.}~\bibnamefont {Yao}}, \bibinfo {author} {\bibfnamefont {P.}~\bibnamefont
  {Yeh}}, \bibinfo {author} {\bibfnamefont {A.}~\bibnamefont {Zalcman}},
  \bibinfo {author} {\bibfnamefont {V.}~\bibnamefont {Smelyanskiy}}, \bibinfo
  {author} {\bibfnamefont {H.}~\bibnamefont {Neven}}, \bibinfo {author}
  {\bibfnamefont {S.}~\bibnamefont {Gopalakrishnan}}, \bibinfo {author}
  {\bibfnamefont {D.}~\bibnamefont {Abanin}}, \bibinfo {author} {\bibfnamefont
  {M.}~\bibnamefont {Knap}}, \bibinfo {author} {\bibfnamefont {J.}~\bibnamefont
  {Martinis}},\ and\ \bibinfo {author} {\bibfnamefont {P.}~\bibnamefont
  {Roushan}},\ }\href@noop {} {\bibinfo {title} {Direct measurement of
  non-local interactions in the many-body localized phase}} (\bibinfo {year}
  {2020}),\ \Eprint {https://arxiv.org/abs/1910.06024} {arXiv:1910.06024
  [cond-mat.dis-nn]} \BibitemShut {NoStop}%
\bibitem [{\citenamefont {Schollwöck}(2011)}]{SCHOLLWOCK201196}%
  \BibitemOpen
  \bibfield  {author} {\bibinfo {author} {\bibfnamefont {U.}~\bibnamefont
  {Schollwöck}},\ }\bibfield  {title} {\bibinfo {title} {The density-matrix
  renormalization group in the age of matrix product states},\ }\href
  {https://doi.org/https://doi.org/10.1016/j.aop.2010.09.012} {\bibfield
  {journal} {\bibinfo  {journal} {Annals of Physics}\ }\textbf {\bibinfo
  {volume} {326}},\ \bibinfo {pages} {96} (\bibinfo {year} {2011})},\ \bibinfo
  {note} {january 2011 Special Issue}\BibitemShut {NoStop}%
\bibitem [{\citenamefont {Sakurai}\ and\ \citenamefont
  {Napolitano}(2017)}]{sakurai2017modern}%
  \BibitemOpen
  \bibfield  {author} {\bibinfo {author} {\bibfnamefont {J.}~\bibnamefont
  {Sakurai}}\ and\ \bibinfo {author} {\bibfnamefont {J.}~\bibnamefont
  {Napolitano}},\ }\href {https://books.google.de/books?id=010yDwAAQBAJ} {\emph
  {\bibinfo {title} {Modern Quantum Mechanics}}}\ (\bibinfo  {publisher}
  {Cambridge University Press},\ \bibinfo {address} {Cambridge, England},\
  \bibinfo {year} {2017})\BibitemShut {NoStop}%
\bibitem [{\citenamefont {Messiah}(2014)}]{messiah2014quantum}%
  \BibitemOpen
  \bibfield  {author} {\bibinfo {author} {\bibfnamefont {A.}~\bibnamefont
  {Messiah}},\ }\href {https://books.google.de/books?id=8FvLAgAAQBAJ} {\emph
  {\bibinfo {title} {Quantum Mechanics}}},\ Dover Books on Physics\ (\bibinfo
  {publisher} {Dover Publications},\ \bibinfo {address} {Mineola, NY},\
  \bibinfo {year} {2014})\BibitemShut {NoStop}%
\bibitem [{\citenamefont {Barthel}\ and\ \citenamefont
  {Schollw\"ock}(2008)}]{PhysRevLett.100.100601}%
  \BibitemOpen
  \bibfield  {author} {\bibinfo {author} {\bibfnamefont {T.}~\bibnamefont
  {Barthel}}\ and\ \bibinfo {author} {\bibfnamefont {U.}~\bibnamefont
  {Schollw\"ock}},\ }\bibfield  {title} {\bibinfo {title} {Dephasing and the
  steady state in quantum many-particle systems},\ }\href
  {https://doi.org/10.1103/PhysRevLett.100.100601} {\bibfield  {journal}
  {\bibinfo  {journal} {Phys. Rev. Lett.}\ }\textbf {\bibinfo {volume} {100}},\
  \bibinfo {pages} {100601} (\bibinfo {year} {2008})}\BibitemShut {NoStop}%
\bibitem [{\citenamefont {Yurke}\ and\ \citenamefont
  {Stoler}(1988)}]{yurke1988dynamic}%
  \BibitemOpen
  \bibfield  {author} {\bibinfo {author} {\bibfnamefont {B.}~\bibnamefont
  {Yurke}}\ and\ \bibinfo {author} {\bibfnamefont {D.}~\bibnamefont {Stoler}},\
  }\bibfield  {title} {\bibinfo {title} {The dynamic generation of
  schr{\"o}dinger cats and their detection},\ }\href@noop {} {\bibfield
  {journal} {\bibinfo  {journal} {Physica B+ C}\ }\textbf {\bibinfo {volume}
  {151}},\ \bibinfo {pages} {298} (\bibinfo {year} {1988})}\BibitemShut
  {NoStop}%
\bibitem [{\citenamefont {Kirchmair}\ \emph {et~al.}(2013)\citenamefont
  {Kirchmair}, \citenamefont {Vlastakis}, \citenamefont {Leghtas},
  \citenamefont {Nigg}, \citenamefont {Paik}, \citenamefont {Ginossar},
  \citenamefont {Mirrahimi}, \citenamefont {Frunzio}, \citenamefont {Girvin},\
  and\ \citenamefont {Schoelkopf}}]{2013}%
  \BibitemOpen
  \bibfield  {author} {\bibinfo {author} {\bibfnamefont {G.}~\bibnamefont
  {Kirchmair}}, \bibinfo {author} {\bibfnamefont {B.}~\bibnamefont
  {Vlastakis}}, \bibinfo {author} {\bibfnamefont {Z.}~\bibnamefont {Leghtas}},
  \bibinfo {author} {\bibfnamefont {S.~E.}\ \bibnamefont {Nigg}}, \bibinfo
  {author} {\bibfnamefont {H.}~\bibnamefont {Paik}}, \bibinfo {author}
  {\bibfnamefont {E.}~\bibnamefont {Ginossar}}, \bibinfo {author}
  {\bibfnamefont {M.}~\bibnamefont {Mirrahimi}}, \bibinfo {author}
  {\bibfnamefont {L.}~\bibnamefont {Frunzio}}, \bibinfo {author} {\bibfnamefont
  {S.~M.}\ \bibnamefont {Girvin}},\ and\ \bibinfo {author} {\bibfnamefont
  {R.~J.}\ \bibnamefont {Schoelkopf}},\ }\bibfield  {title} {\bibinfo {title}
  {Observation of quantum state collapse and revival due to the single-photon
  kerr effect},\ }\href {https://doi.org/10.1038/nature11902} {\bibfield
  {journal} {\bibinfo  {journal} {Nature}\ }\textbf {\bibinfo {volume} {495}},\
  \bibinfo {pages} {205–209} (\bibinfo {year} {2013})}\BibitemShut {NoStop}%
\bibitem [{\citenamefont {Jaschke}\ \emph {et~al.}(2018)\citenamefont
  {Jaschke}, \citenamefont {Montangero},\ and\ \citenamefont
  {Carr}}]{Jaschke2018}%
  \BibitemOpen
  \bibfield  {author} {\bibinfo {author} {\bibfnamefont {D.}~\bibnamefont
  {Jaschke}}, \bibinfo {author} {\bibfnamefont {S.}~\bibnamefont
  {Montangero}},\ and\ \bibinfo {author} {\bibfnamefont {L.~D.}\ \bibnamefont
  {Carr}},\ }\bibfield  {title} {\bibinfo {title} {One-dimensional many-body
  entangled open quantum systems with tensor network methods},\ }\href
  {https://doi.org/10.1088/2058-9565/aae724} {\bibfield  {journal} {\bibinfo
  {journal} {Quantum Science and Technology}\ }\textbf {\bibinfo {volume}
  {4}},\ \bibinfo {pages} {013001} (\bibinfo {year} {2018})}\BibitemShut
  {NoStop}%
\bibitem [{\citenamefont {Daley}(2014)}]{Daley2014}%
  \BibitemOpen
  \bibfield  {author} {\bibinfo {author} {\bibfnamefont {A.~J.}\ \bibnamefont
  {Daley}},\ }\bibfield  {title} {\bibinfo {title} {Quantum trajectories and
  open many-body quantum systems},\ }\href
  {https://doi.org/10.1080/00018732.2014.933502} {\bibfield  {journal}
  {\bibinfo  {journal} {Advances in Physics}\ }\textbf {\bibinfo {volume}
  {63}},\ \bibinfo {pages} {77} (\bibinfo {year} {2014})}\BibitemShut {NoStop}%
\bibitem [{\citenamefont {Dolgirev}\ \emph {et~al.}(2020)\citenamefont
  {Dolgirev}, \citenamefont {Marino}, \citenamefont {Sels},\ and\ \citenamefont
  {Demler}}]{dolgirev2020non}%
  \BibitemOpen
  \bibfield  {author} {\bibinfo {author} {\bibfnamefont {P.~E.}\ \bibnamefont
  {Dolgirev}}, \bibinfo {author} {\bibfnamefont {J.}~\bibnamefont {Marino}},
  \bibinfo {author} {\bibfnamefont {D.}~\bibnamefont {Sels}},\ and\ \bibinfo
  {author} {\bibfnamefont {E.}~\bibnamefont {Demler}},\ }\bibfield  {title}
  {\bibinfo {title} {Non-gaussian correlations imprinted by local dephasing in
  fermionic wires},\ }\href@noop {} {\bibfield  {journal} {\bibinfo  {journal}
  {Physical Review B}\ }\textbf {\bibinfo {volume} {102}},\ \bibinfo {pages}
  {100301(R)} (\bibinfo {year} {2020})}\BibitemShut {NoStop}%
\bibitem [{\citenamefont {Josephson}(1962)}]{JOSEPHSON1962251}%
  \BibitemOpen
  \bibfield  {author} {\bibinfo {author} {\bibfnamefont {B.}~\bibnamefont
  {Josephson}},\ }\bibfield  {title} {\bibinfo {title} {Possible new effects in
  superconductive tunnelling},\ }\href
  {https://doi.org/https://doi.org/10.1016/0031-9163(62)91369-0} {\bibfield
  {journal} {\bibinfo  {journal} {Physics Letters}\ }\textbf {\bibinfo {volume}
  {1}},\ \bibinfo {pages} {251} (\bibinfo {year} {1962})}\BibitemShut {NoStop}%
\bibitem [{\citenamefont {Tinkham}(2004)}]{tinkham2004introduction}%
  \BibitemOpen
  \bibfield  {author} {\bibinfo {author} {\bibfnamefont {M.}~\bibnamefont
  {Tinkham}},\ }\href@noop {} {\emph {\bibinfo {title} {Introduction to
  superconductivity}}}\ (\bibinfo  {publisher} {Courier Corporation},\ \bibinfo
  {address} {Mineola, NY},\ \bibinfo {year} {2004})\BibitemShut {NoStop}%
\bibitem [{\citenamefont {Noguchi}\ \emph {et~al.}(2020)\citenamefont
  {Noguchi}, \citenamefont {Osada}, \citenamefont {Masuda}, \citenamefont
  {Kono}, \citenamefont {Heya}, \citenamefont {Wolski}, \citenamefont
  {Takahashi}, \citenamefont {Sugiyama}, \citenamefont {Lachance-Quirion},\
  and\ \citenamefont {Nakamura}}]{PhysRevA.102.062408}%
  \BibitemOpen
  \bibfield  {author} {\bibinfo {author} {\bibfnamefont {A.}~\bibnamefont
  {Noguchi}}, \bibinfo {author} {\bibfnamefont {A.}~\bibnamefont {Osada}},
  \bibinfo {author} {\bibfnamefont {S.}~\bibnamefont {Masuda}}, \bibinfo
  {author} {\bibfnamefont {S.}~\bibnamefont {Kono}}, \bibinfo {author}
  {\bibfnamefont {K.}~\bibnamefont {Heya}}, \bibinfo {author} {\bibfnamefont
  {S.~P.}\ \bibnamefont {Wolski}}, \bibinfo {author} {\bibfnamefont
  {H.}~\bibnamefont {Takahashi}}, \bibinfo {author} {\bibfnamefont
  {T.}~\bibnamefont {Sugiyama}}, \bibinfo {author} {\bibfnamefont
  {D.}~\bibnamefont {Lachance-Quirion}},\ and\ \bibinfo {author} {\bibfnamefont
  {Y.}~\bibnamefont {Nakamura}},\ }\bibfield  {title} {\bibinfo {title} {Fast
  parametric two-qubit gates with suppressed residual interaction using the
  second-order nonlinearity of a cubic transmon},\ }\href
  {https://doi.org/10.1103/PhysRevA.102.062408} {\bibfield  {journal} {\bibinfo
   {journal} {Phys. Rev. A}\ }\textbf {\bibinfo {volume} {102}},\ \bibinfo
  {pages} {062408} (\bibinfo {year} {2020})}\BibitemShut {NoStop}%
\bibitem [{\citenamefont {Yan}\ \emph {et~al.}(2016)\citenamefont {Yan},
  \citenamefont {Gustavsson}, \citenamefont {Kamal}, \citenamefont {Birenbaum},
  \citenamefont {Sears}, \citenamefont {Hover}, \citenamefont {Gudmundsen},
  \citenamefont {Rosenberg}, \citenamefont {Samach}, \citenamefont {Weber},
  \citenamefont {Yoder}, \citenamefont {Orlando}, \citenamefont {Clarke},
  \citenamefont {Kerman},\ and\ \citenamefont {Oliver}}]{Yan2016}%
  \BibitemOpen
  \bibfield  {author} {\bibinfo {author} {\bibfnamefont {F.}~\bibnamefont
  {Yan}}, \bibinfo {author} {\bibfnamefont {S.}~\bibnamefont {Gustavsson}},
  \bibinfo {author} {\bibfnamefont {A.}~\bibnamefont {Kamal}}, \bibinfo
  {author} {\bibfnamefont {J.}~\bibnamefont {Birenbaum}}, \bibinfo {author}
  {\bibfnamefont {A.~P.}\ \bibnamefont {Sears}}, \bibinfo {author}
  {\bibfnamefont {D.}~\bibnamefont {Hover}}, \bibinfo {author} {\bibfnamefont
  {T.~J.}\ \bibnamefont {Gudmundsen}}, \bibinfo {author} {\bibfnamefont
  {D.}~\bibnamefont {Rosenberg}}, \bibinfo {author} {\bibfnamefont
  {G.}~\bibnamefont {Samach}}, \bibinfo {author} {\bibfnamefont
  {S.}~\bibnamefont {Weber}}, \bibinfo {author} {\bibfnamefont {J.~L.}\
  \bibnamefont {Yoder}}, \bibinfo {author} {\bibfnamefont {T.~P.}\ \bibnamefont
  {Orlando}}, \bibinfo {author} {\bibfnamefont {J.}~\bibnamefont {Clarke}},
  \bibinfo {author} {\bibfnamefont {A.~J.}\ \bibnamefont {Kerman}},\ and\
  \bibinfo {author} {\bibfnamefont {W.~D.}\ \bibnamefont {Oliver}},\ }\bibfield
   {title} {\bibinfo {title} {The flux qubit revisited to enhance coherence and
  reproducibility},\ }\href {https://doi.org/10.1038/ncomms12964} {\bibfield
  {journal} {\bibinfo  {journal} {Nature Communications}\ }\textbf {\bibinfo
  {volume} {7}},\ \bibinfo {pages} {1} (\bibinfo {year} {2016})}\BibitemShut
  {NoStop}%
\bibitem [{\citenamefont {DiCarlo}\ \emph {et~al.}(2009)\citenamefont
  {DiCarlo}, \citenamefont {Chow}, \citenamefont {Gambetta}, \citenamefont
  {Bishop}, \citenamefont {Johnson}, \citenamefont {Schuster}, \citenamefont
  {Majer}, \citenamefont {Blais}, \citenamefont {Frunzio}, \citenamefont
  {Girvin},\ and\ \citenamefont {Schoelkopf}}]{DiCarlo2009}%
  \BibitemOpen
  \bibfield  {author} {\bibinfo {author} {\bibfnamefont {L.}~\bibnamefont
  {DiCarlo}}, \bibinfo {author} {\bibfnamefont {J.~M.}\ \bibnamefont {Chow}},
  \bibinfo {author} {\bibfnamefont {J.~M.}\ \bibnamefont {Gambetta}}, \bibinfo
  {author} {\bibfnamefont {L.~S.}\ \bibnamefont {Bishop}}, \bibinfo {author}
  {\bibfnamefont {B.~R.}\ \bibnamefont {Johnson}}, \bibinfo {author}
  {\bibfnamefont {D.~I.}\ \bibnamefont {Schuster}}, \bibinfo {author}
  {\bibfnamefont {J.}~\bibnamefont {Majer}}, \bibinfo {author} {\bibfnamefont
  {A.}~\bibnamefont {Blais}}, \bibinfo {author} {\bibfnamefont
  {L.}~\bibnamefont {Frunzio}}, \bibinfo {author} {\bibfnamefont {S.~M.}\
  \bibnamefont {Girvin}},\ and\ \bibinfo {author} {\bibfnamefont {R.~J.}\
  \bibnamefont {Schoelkopf}},\ }\bibfield  {title} {\bibinfo {title}
  {Demonstration of two-qubit algorithms with a superconducting quantum
  processor},\ }\href {https://doi.org/10.1038/nature08121} {\bibfield
  {journal} {\bibinfo  {journal} {Nature}\ }\textbf {\bibinfo {volume} {460}},\
  \bibinfo {pages} {240} (\bibinfo {year} {2009})}\BibitemShut {NoStop}%
\bibitem [{\citenamefont {Barends}\ \emph {et~al.}(2014)\citenamefont
  {Barends}, \citenamefont {Kelly}, \citenamefont {Megrant}, \citenamefont
  {Veitia}, \citenamefont {Sank}, \citenamefont {Jeffrey}, \citenamefont
  {White}, \citenamefont {Mutus}, \citenamefont {Fowler}, \citenamefont
  {Campbell}, \citenamefont {Chen}, \citenamefont {Chen}, \citenamefont
  {Chiaro}, \citenamefont {Dunsworth}, \citenamefont {Neill}, \citenamefont
  {O'Malley}, \citenamefont {Roushan}, \citenamefont {Vainsencher},
  \citenamefont {Wenner}, \citenamefont {Korotkov}, \citenamefont {Cleland},\
  and\ \citenamefont {Martinis}}]{Barends2014}%
  \BibitemOpen
  \bibfield  {author} {\bibinfo {author} {\bibfnamefont {R.}~\bibnamefont
  {Barends}}, \bibinfo {author} {\bibfnamefont {J.}~\bibnamefont {Kelly}},
  \bibinfo {author} {\bibfnamefont {A.}~\bibnamefont {Megrant}}, \bibinfo
  {author} {\bibfnamefont {A.}~\bibnamefont {Veitia}}, \bibinfo {author}
  {\bibfnamefont {D.}~\bibnamefont {Sank}}, \bibinfo {author} {\bibfnamefont
  {E.}~\bibnamefont {Jeffrey}}, \bibinfo {author} {\bibfnamefont {T.~C.}\
  \bibnamefont {White}}, \bibinfo {author} {\bibfnamefont {J.}~\bibnamefont
  {Mutus}}, \bibinfo {author} {\bibfnamefont {A.~G.}\ \bibnamefont {Fowler}},
  \bibinfo {author} {\bibfnamefont {B.}~\bibnamefont {Campbell}}, \bibinfo
  {author} {\bibfnamefont {Y.}~\bibnamefont {Chen}}, \bibinfo {author}
  {\bibfnamefont {Z.}~\bibnamefont {Chen}}, \bibinfo {author} {\bibfnamefont
  {B.}~\bibnamefont {Chiaro}}, \bibinfo {author} {\bibfnamefont
  {A.}~\bibnamefont {Dunsworth}}, \bibinfo {author} {\bibfnamefont
  {C.}~\bibnamefont {Neill}}, \bibinfo {author} {\bibfnamefont
  {P.}~\bibnamefont {O'Malley}}, \bibinfo {author} {\bibfnamefont
  {P.}~\bibnamefont {Roushan}}, \bibinfo {author} {\bibfnamefont
  {A.}~\bibnamefont {Vainsencher}}, \bibinfo {author} {\bibfnamefont
  {J.}~\bibnamefont {Wenner}}, \bibinfo {author} {\bibfnamefont {A.~N.}\
  \bibnamefont {Korotkov}}, \bibinfo {author} {\bibfnamefont {A.~N.}\
  \bibnamefont {Cleland}},\ and\ \bibinfo {author} {\bibfnamefont {J.~M.}\
  \bibnamefont {Martinis}},\ }\bibfield  {title} {\bibinfo {title}
  {Superconducting quantum circuits at the surface code threshold for fault
  tolerance},\ }\href {https://doi.org/10.1038/nature13171} {\bibfield
  {journal} {\bibinfo  {journal} {Nature}\ }\textbf {\bibinfo {volume} {508}},\
  \bibinfo {pages} {500} (\bibinfo {year} {2014})}\BibitemShut {NoStop}%
\bibitem [{\citenamefont {Auerbach}(2012)}]{auerbach2012interacting}%
  \BibitemOpen
  \bibfield  {author} {\bibinfo {author} {\bibfnamefont {A.}~\bibnamefont
  {Auerbach}},\ }\href {https://books.google.de/books?id=d-sHCAAAQBAJ} {\emph
  {\bibinfo {title} {Interacting Electrons and Quantum Magnetism}}},\ Graduate
  Texts in Contemporary Physics\ (\bibinfo  {publisher} {Springer New York},\
  \bibinfo {address} {New York, NY},\ \bibinfo {year} {2012})\BibitemShut
  {NoStop}%
\bibitem [{\citenamefont {Magesan}\ and\ \citenamefont
  {Gambetta}(2020)}]{PhysRevA.101.052308}%
  \BibitemOpen
  \bibfield  {author} {\bibinfo {author} {\bibfnamefont {E.}~\bibnamefont
  {Magesan}}\ and\ \bibinfo {author} {\bibfnamefont {J.~M.}\ \bibnamefont
  {Gambetta}},\ }\bibfield  {title} {\bibinfo {title} {Effective hamiltonian
  models of the cross-resonance gate},\ }\href
  {https://doi.org/10.1103/PhysRevA.101.052308} {\bibfield  {journal} {\bibinfo
   {journal} {Phys. Rev. A}\ }\textbf {\bibinfo {volume} {101}},\ \bibinfo
  {pages} {052308} (\bibinfo {year} {2020})}\BibitemShut {NoStop}%
\bibitem [{\citenamefont {Sheldon}\ \emph {et~al.}(2016)\citenamefont
  {Sheldon}, \citenamefont {Magesan}, \citenamefont {Chow},\ and\ \citenamefont
  {Gambetta}}]{PhysRevA.93.060302}%
  \BibitemOpen
  \bibfield  {author} {\bibinfo {author} {\bibfnamefont {S.}~\bibnamefont
  {Sheldon}}, \bibinfo {author} {\bibfnamefont {E.}~\bibnamefont {Magesan}},
  \bibinfo {author} {\bibfnamefont {J.~M.}\ \bibnamefont {Chow}},\ and\
  \bibinfo {author} {\bibfnamefont {J.~M.}\ \bibnamefont {Gambetta}},\
  }\bibfield  {title} {\bibinfo {title} {Procedure for systematically tuning up
  cross-talk in the cross-resonance gate},\ }\href
  {https://doi.org/10.1103/PhysRevA.93.060302} {\bibfield  {journal} {\bibinfo
  {journal} {Phys. Rev. A}\ }\textbf {\bibinfo {volume} {93}},\ \bibinfo
  {pages} {060302(R)} (\bibinfo {year} {2016})}\BibitemShut {NoStop}%
\bibitem [{\citenamefont {C\'orcoles}\ \emph {et~al.}(2013)\citenamefont
  {C\'orcoles}, \citenamefont {Gambetta}, \citenamefont {Chow}, \citenamefont
  {Smolin}, \citenamefont {Ware}, \citenamefont {Strand}, \citenamefont
  {Plourde},\ and\ \citenamefont {Steffen}}]{PhysRevA.87.030301}%
  \BibitemOpen
  \bibfield  {author} {\bibinfo {author} {\bibfnamefont {A.~D.}\ \bibnamefont
  {C\'orcoles}}, \bibinfo {author} {\bibfnamefont {J.~M.}\ \bibnamefont
  {Gambetta}}, \bibinfo {author} {\bibfnamefont {J.~M.}\ \bibnamefont {Chow}},
  \bibinfo {author} {\bibfnamefont {J.~A.}\ \bibnamefont {Smolin}}, \bibinfo
  {author} {\bibfnamefont {M.}~\bibnamefont {Ware}}, \bibinfo {author}
  {\bibfnamefont {J.}~\bibnamefont {Strand}}, \bibinfo {author} {\bibfnamefont
  {B.~L.~T.}\ \bibnamefont {Plourde}},\ and\ \bibinfo {author} {\bibfnamefont
  {M.}~\bibnamefont {Steffen}},\ }\bibfield  {title} {\bibinfo {title} {Process
  verification of two-qubit quantum gates by randomized benchmarking},\ }\href
  {https://doi.org/10.1103/PhysRevA.87.030301} {\bibfield  {journal} {\bibinfo
  {journal} {Phys. Rev. A}\ }\textbf {\bibinfo {volume} {87}},\ \bibinfo
  {pages} {030301(R)} (\bibinfo {year} {2013})}\BibitemShut {NoStop}%
\bibitem [{\citenamefont {Blais}\ \emph {et~al.}(2007)\citenamefont {Blais},
  \citenamefont {Gambetta}, \citenamefont {Wallraff}, \citenamefont {Schuster},
  \citenamefont {Girvin}, \citenamefont {Devoret},\ and\ \citenamefont
  {Schoelkopf}}]{PhysRevA.75.032329}%
  \BibitemOpen
  \bibfield  {author} {\bibinfo {author} {\bibfnamefont {A.}~\bibnamefont
  {Blais}}, \bibinfo {author} {\bibfnamefont {J.}~\bibnamefont {Gambetta}},
  \bibinfo {author} {\bibfnamefont {A.}~\bibnamefont {Wallraff}}, \bibinfo
  {author} {\bibfnamefont {D.~I.}\ \bibnamefont {Schuster}}, \bibinfo {author}
  {\bibfnamefont {S.~M.}\ \bibnamefont {Girvin}}, \bibinfo {author}
  {\bibfnamefont {M.~H.}\ \bibnamefont {Devoret}},\ and\ \bibinfo {author}
  {\bibfnamefont {R.~J.}\ \bibnamefont {Schoelkopf}},\ }\bibfield  {title}
  {\bibinfo {title} {Quantum-information processing with circuit quantum
  electrodynamics},\ }\href {https://doi.org/10.1103/PhysRevA.75.032329}
  {\bibfield  {journal} {\bibinfo  {journal} {Phys. Rev. A}\ }\textbf {\bibinfo
  {volume} {75}},\ \bibinfo {pages} {032329} (\bibinfo {year}
  {2007})}\BibitemShut {NoStop}%
\bibitem [{\citenamefont {Haq}\ \emph {et~al.}(2019)\citenamefont {Haq},
  \citenamefont {Bharadwaj},\ and\ \citenamefont {Wani}}]{haq2019explicit}%
  \BibitemOpen
  \bibfield  {author} {\bibinfo {author} {\bibfnamefont {R.~U.}\ \bibnamefont
  {Haq}}, \bibinfo {author} {\bibfnamefont {S.~S.}\ \bibnamefont {Bharadwaj}},\
  and\ \bibinfo {author} {\bibfnamefont {T.~A.}\ \bibnamefont {Wani}},\
  }\href@noop {} {\bibinfo {title} {An explicit method for schrieffer-wolff
  transformation}} (\bibinfo {year} {2019}),\ \Eprint
  {https://arxiv.org/abs/1901.08617} {arXiv:1901.08617 [cond-mat.str-el]}
  \BibitemShut {NoStop}%
\bibitem [{\citenamefont {{\v{Z}}nidari{\v{c}}}\ \emph
  {et~al.}(2016)\citenamefont {{\v{Z}}nidari{\v{c}}}, \citenamefont
  {Scardicchio},\ and\ \citenamefont {Varma}}]{vznidarivc2016diffusive}%
  \BibitemOpen
  \bibfield  {author} {\bibinfo {author} {\bibfnamefont {M.}~\bibnamefont
  {{\v{Z}}nidari{\v{c}}}}, \bibinfo {author} {\bibfnamefont {A.}~\bibnamefont
  {Scardicchio}},\ and\ \bibinfo {author} {\bibfnamefont {V.~K.}\ \bibnamefont
  {Varma}},\ }\bibfield  {title} {\bibinfo {title} {Diffusive and subdiffusive
  spin transport in the ergodic phase of a many-body localizable system},\
  }\href@noop {} {\bibfield  {journal} {\bibinfo  {journal} {Physical review
  letters}\ }\textbf {\bibinfo {volume} {117}},\ \bibinfo {pages} {040601}
  (\bibinfo {year} {2016})}\BibitemShut {NoStop}%
\bibitem [{\citenamefont {Fishman}\ \emph {et~al.}(2020)\citenamefont
  {Fishman}, \citenamefont {White},\ and\ \citenamefont
  {Stoudenmire}}]{fishman2020itensor}%
  \BibitemOpen
  \bibfield  {author} {\bibinfo {author} {\bibfnamefont {M.}~\bibnamefont
  {Fishman}}, \bibinfo {author} {\bibfnamefont {S.~R.}\ \bibnamefont {White}},\
  and\ \bibinfo {author} {\bibfnamefont {E.~M.}\ \bibnamefont {Stoudenmire}},\
  }\href@noop {} {\bibinfo {title} {The itensor software library for tensor
  network calculations}} (\bibinfo {year} {2020}),\ \Eprint
  {https://arxiv.org/abs/2007.14822} {arXiv:2007.14822 [cs.MS]} \BibitemShut
  {NoStop}%
\end{thebibliography}%
\appendix 
\section{Role of on-site density-density interaction \label{appendix_on_site_density_density_interaction}}
\begin{figure}[t!]
\centering
\includegraphics[width=\linewidth]{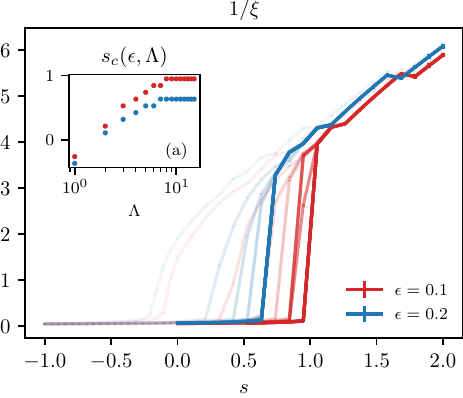}
\caption{\ric{The inverse of the localization length $\xi$ in a system of $L=15$ ``active'' sites in the symmetry sector $n_0=1$, $\beta_{r=0}$. The main plot shows the inverse of the localization length $\xi^{-1}$ as a function of $s$ for different values of $\Lambda \in [1,15]$ and $\epsilon$ at $U=0$. The darker lines correspond to larger values of $\Lambda$. The inset (a) shows the behavior of $s_c(\epsilon,\Lambda)$ as a function of $\Lambda$ for $\epsilon=0.1$ (red) and $\epsilon=0.2$ (blue) at $U=0$.}}
\label{fig_localization_length_scaling_lambda_different_epsilons}
\end{figure}

\ric{In the main text, we focus on a simplified version of the model without   on-site density-density interactions, to keep to a minimum the amount of   technical details in the course of the presentation. In the following, we address the role of on-site density-density interactions, focusing on the localization properties of the ground state and comparing with the statements  in  the main text  resulting from numerics performed at $U>0$ and $\epsilon=0$. }\\
\ric{Starting from the Hamiltonian in Eq.~\eqref{eq_H_specific_symmetry_sector}, we consider $U=0$ and $\epsilon \geq 0$. For $\epsilon=0$, the model does not display localization at finite $s$ in the bosonic limit, as extensively discussed in Sec.~\ref{section_localization}. On the other hand, for $\epsilon>0$, the ground state is localized for $s>s_c$ in the bosonic limit, with $s_c$ being parametrically small in $\epsilon$. We perform the same scaling analysis as a function of the cutoff $\Lambda$ discussed in Sec.~\ref{section_localization}.
In Fig.~\ref{fig_localization_length_scaling_lambda_different_epsilons}, we show the inverse of the localization length $\xi$ swiping $s$ for different values of $\Lambda$ at fixed $\epsilon$. The scaling analysis suggests that the transition point $s_c(\Lambda,\epsilon)$ converges to a finite value independent of $\Lambda$ for $\Lambda\to\infty$. The overall qualitative picture is therefore unaffected if one considers on-site or nearest-neighbor nonlinearities.  }\\
\ric{A nonzero value of $\epsilon$ introduces, however, anharmonic spacings between ground states with different values of   $n_0$. Indeed, we have, for the energy of the ground state, $E(n_0) \approx n_0/2 + \epsilon n_0^2/2$. This additional anharmonicity   has an impact on the adiabatic protocol discussed in Sec.~\ref{section_state_preparation}, since each adiabatically evolved state $\mathcal{U}|n_0\rangle_0 \otimes \bigotimes_{j>0}|0\rangle_j$ in Eq.~\eqref{eq_generic_dressed_state} would acquire a phase with a nonlinear dependence in $n_0$, which  technically complicates state preparation without altering the main physical message. Nonetheless, it is still possible to tame the effect of this nonlinearity by considering a small enough $\epsilon$, at the cost of having a smaller $e^{-s}$ (larger $s$) and therefore working effectively deeper in the localized phase. These types of unnecessary technical complications are at the root of our choice of working throughout the main text with $\epsilon=0$ and $U>0$.}
\section{Properties of the localized ground state upon changing $n_0$ \label{section_appendix_properties_localized_state}}
\begin{figure}
\centering
\includegraphics[width=\linewidth]{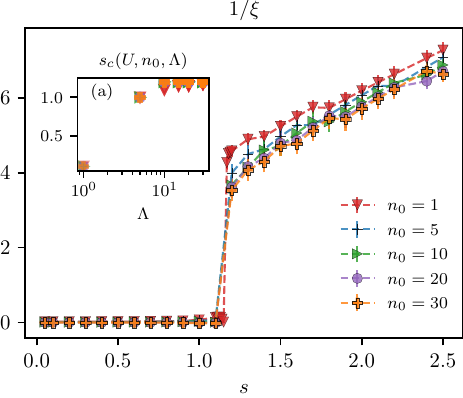}
\caption{
The inverse of localization length $\xi$ in a system of $L=15$ ``active'' sites upon changing $s$ for different values of $n_0=1$. We fix $U=0.1$. The main plot shows the inverse of the localization length $\xi^{-1}$ as a function of $s$ for $\Lambda=30$. The inset (a) shows the behavior of $s_c$ as a function of $\Lambda$ for different values of $n_0$. The circles correspond to  numerically extracted values from DMRG results. The points are indistinguishable upon changing $n_0$ for $\Lambda \gtrsim 10$.}
\label{fig_localization_length_swipe_n0}
\end{figure}
In this appendix, we discuss the properties of the ground state upon changing the symmetry sector specified by the occupation $n_0$ of the first nonempty site. 
We show that the transition point and the exponential decaying tail of the ground state occupation is weakly dependent on $n_0$. We discuss the dependence of the ground state energy on $n_0$, which is relevant in the state preparation via the adiabatic protocol discussed in Sec.~\ref{section_state_preparation}.

We perform the same scaling analysis as a function of the cutoff $\Lambda$ discussed in Sec.~\ref{section_localization} (see Fig.~\ref{fig_localization_length_swipe_n0}).
We extract the transition point $s_c$ for different values of $n_0$ from the inverse of the localization length $\xi$.
The existence of a finite critical point $s_c$ in the $\Lambda \to \infty$ limit turns out to be weakly dependent on the specific symmetry sector $n_0$ at fixed $U$.
\begin{figure}
\centering
\includegraphics[width=\linewidth]{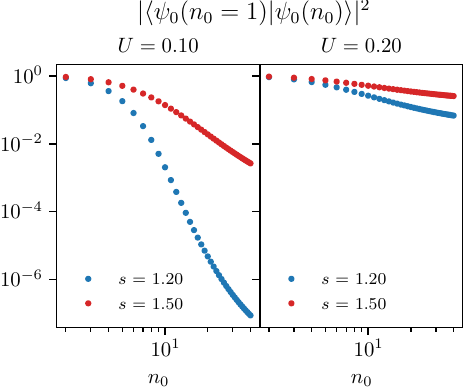}
\caption{The overlap of the exponential tail as a function of $n_0 \in [2,40]$ for two different values of $U=\{0.1,0.2\}$ and $s = \{1.20,1.50\}$. We choose these values of $U$ and $s$ since we are not so deep in the localized phase. The more the system is within the localized phase, the more the localized tails are weakly dependent on $n_0$.}
\label{fig_overlap_vs_n0}
\end{figure}
\begin{figure}
\centering
\includegraphics[width=\linewidth]{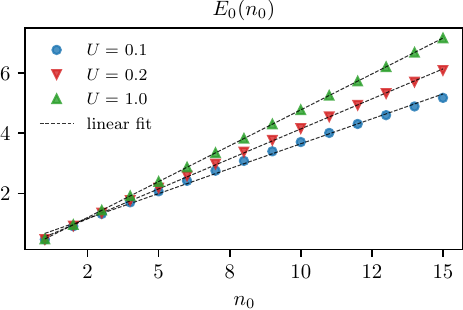}
\caption{The energies of the ground state as a function of $n_0$ for different values of $U$ at fixed $s=1.5>s_c(U)$ and cutoff $\Lambda=15$. The dashed lines are the linear fit. The more we are deep in the localized phase, the more $E(n_0) \propto n_0$.}
\label{fig_energies_vs_n0}
\end{figure}
We investigate the dependence of the localized tail of the ground state $|\psi_0(n_0)\rangle$ as a function of $n_0$ (we exclude the first site, which fixes the symmetry). To this end, we compute $|\langle \psi_0(n_0=1)|\psi_0(n_0)\rangle|^2$, with $n_0 \geq 1$ (see Fig.~\ref{fig_overlap_vs_n0}).
We fix $n_0=1$ as a reference as we want to see whether or not the tail is weakly dependent on $n_0$. All the ground states are computed by fixing $\Lambda=30$. The overlap $|\langle \psi_0(n_0)|\psi_0(n_0=1)\rangle|$ strongly depends on $s$ and $U$. Indeed, the more the system is in the localized phase, the more the exponentially localized tail is weakly dependent on $n_0$. Therefore, deep in the localized phase, $|\psi_0(n_0)\rangle$ is approximately independent on the specific sector $n_0$ and we can write
\begin{equation}
\label{eq_approximated_state}
|\widetilde{n_0}\rangle \equiv |n_0\rangle \otimes |\psi_0(n_0)\rangle \approx |n_0\rangle \otimes |\psi_0\rangle,
\end{equation}
where $|\psi_0\rangle$ is explicitly independent of $n_0$.

The weak dependence of $|\psi_0(n_0)\rangle$ with respect to $n_0$ has consequences on the ground state energy.
Indeed, the expectation value of the Hamiltonian on Eq.~\eqref{eq_approximated_state} is
\begin{equation}
\label{eq_expectation_value}
E_0(n_0) \equiv \langle \widetilde{n_0} | \hat{H} | \widetilde{n_0}\rangle \approx  \frac{1}{2}n_0 + \mathcal{O}(n_0e^{-1/\xi(n_0)}) ,
\end{equation}
where $\langle \hat{n}_j \rangle \sim e^{-j/\xi(n_0)}$ since we are in the localized phase. In Fig.~\ref{fig_energies_vs_n0}, we give a numerical evidence of Eq.~\eqref{eq_expectation_value}.
\section{Scaling analysis in $\Lambda$\label{scaling_analysis_in_lambda}}

\begin{figure}[b]
\centering
\includegraphics[width=\linewidth]{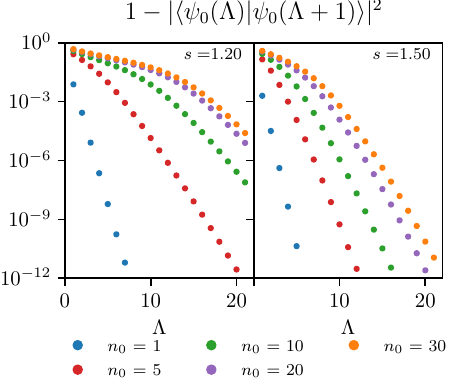}
\caption{The scaling analysis of $1-|\langle \psi_0(\Lambda)|\psi_0(\Lambda+1)\rangle|$ as a function of $\Lambda$ at fixed $U=0.1$ and $s=\{1.2,1.5\}$ for different values of $n_0 \in [1,30]$. The dots and squares refers to the numerical results obtained at $s=1.2$ and $s=1.5$, respectively. The overlap tends exponentially fast to 1 in $\Lambda$. The decay is slower as $n_0$ increases at fixed $s$ and $U$. \label{fig_overlap_scaling_lambda}}
\end{figure}

In the main text, we show that the bosonic system displays a delocalized-localized transition at finite $s$ if $U>0$. Here, we show that the ground state is not only localized but it is weakly dependent on the physical cutoff $\Lambda$. This provides quantitative proof that we can investigate the bosonic system with a finite $\Lambda$ in the localized phase. \\

We fix the symmetry sector $n_0$ and $(s>s_c(U),U>0)$ in the localized phase. We compute $|\psi_0(\Lambda)\rangle$ for different values of $\Lambda$. We calculate $1-|\langle \psi_0(\Lambda)|\psi_0(\Lambda+1)\rangle|^2$ as a function of $\Lambda$ (see Fig.~\ref{fig_overlap_scaling_lambda}). The fidelity $|\langle \psi_0(\Lambda)|\psi_0(\Lambda+1)\rangle|^2$ approaches $1$ exponentially fast in $\Lambda$. The more the system is in the localized phase and $n_0$ is small, the faster is the convergence.
This gives the first evidence that the ground state of the actual bosonic system is well described with small effective cutoffs.

We compute the variance of the Hamiltonian given in Eq.~\eqref{eq_H_full_model} over the ground state $|n_0\rangle_0 \otimes |\psi_0(\Lambda)\rangle$, taking into account the bosonic nature of the original Hamiltonian in Eq.~\eqref{eq_H_full_model}. This quantity is exactly zero if the state $|n_0\rangle_0\otimes|\psi_0(\Lambda)\rangle$ is an eigenstate of $H$. We aim to see how this quantity goes to zero as a function of $\Lambda$. In order to do so, we write the Hamiltonian given in Eq.~\eqref{eq_H_full_model} as the sum of two terms $H=H_-+H_+$. $H_-$ acts on the Hilbert space spanned by states with an occupation number up to $\Lambda$, while $H_+$ acts on the Hilbert space spanned by states with an occupation number greater than $\Lambda$. We label the sectors on which $H_\pm$ acts nontrivially as the $\mathcal{H}_\pm$ sectors, respectively. We apply the same procedure to the number operator and the annihilation(creation) operator:
\begin{equation}
\label{eq_splitting_minus_plus_sector_operators}
\begin{split}
\hat{n}&=\sum_{k=0}^{\Lambda } k \ket{k} \bra{k} + \sum_{k=\Lambda+1}^{\infty} k \ket{k} \bra{k} \\
&= \hat{n}_- + \hat{n}_+ ,\\
\hat{a} &=\sum_{k=0}^{\Lambda} \sqrt{k}|k-1\rangle \langle k |+\sum_{k=\Lambda+1}^{\infty}  \sqrt{k}|k-1\rangle \langle k | \\
&=\hat{a}_- + \hat{a}_+.
\end{split}
\end{equation}
\begin{figure}[t]
\centering
\includegraphics[width=\linewidth]{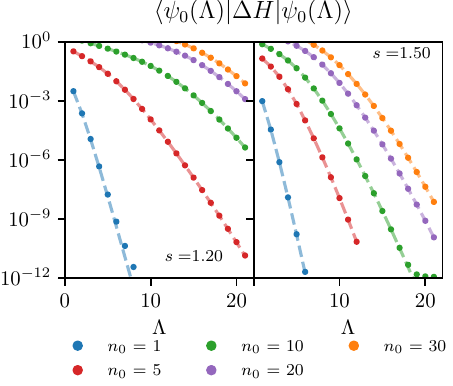}
\caption{The scaling analysis of $\langle \psi_0(\Lambda)|\Delta H|\psi_0(\Lambda)\rangle$ as a function of $\Lambda$ at fixed $U=0.1$ and $s=\{1.2,1.5\}$ for different values of $n_0 \in [1,30]$. The dots refer to the numerical results. The dashed lines are the analytical estimation given by Eq.~\eqref{eq_variance_enlarging_H}. The variance $\Delta H$ decays exponentially fast in $\Lambda$. The decay is slower as $n_0$ increases at fixed $s$ and $U$. \label{fig_variance_H_scaling_lambda}}
\end{figure}
The commutator $[\hat{n}_-,\hat{n}_+]=0$, while $[\hat{a}_-,\hat{a}_+]=\sqrt{\Lambda (\Lambda+1)}|\Lambda-1\rangle \langle \Lambda +1| \neq 0$. This is because the operators $\hat{a}_\pm^{(\dagger)}$ connect the two sectors $\mathcal{H}_\pm$. 
From Eq.~\eqref{eq_splitting_minus_plus_sector_operators}, we straightforwardly obtain the expressions for $H_\pm$:
\begin{equation}
\label{eq_H_plus_minus_expressions}
\begin{split}
H_\pm = -\frac{1}{2}\sum_{i}\hat{n}_{i,\pm} &\Big[e^{-s}\left(\hat{a}_{i+1,\pm}^\dagger + \hat{a}_{i+1,\pm}\right)+\\
&- U \hat{n}_{i+1,\pm}-1\Big].
\end{split}
\end{equation} 
In our numerical scheme we fix a finite cutoff $\Lambda$. Therefore we are computing the ground state $|\psi_0(\Lambda)\rangle$ of $H_-$. Since $\hat{a}_\pm$ are noncommuting operators, the two Hamiltonians $H_-$ and $H_+$ do not commute as well. Therefore, it is not ensured that $|\psi_0(\Lambda)\rangle$ is an eigenstate of the full Hamiltonian $H$. We compute the variance $\Delta H$ over $|\psi_0(\Lambda)\rangle$ of the Hamiltonian $H=H_- + H_+$,
\begin{equation}
\label{eq_deltaH}
\Delta H = \langle  H_+ H_+  \rangle + \langle \{  H_+ ,  H_- \}  \rangle  + \langle H_- H_- \rangle   - \langle H \rangle^2,
\end{equation}
to check whether $|\psi_0(\Lambda)\rangle$ is an eigenstate of $H$. The terms in $H_\pm$ that preserve the sectors $\mathcal{H}_\pm$ give a zero contribution in Eq.~\eqref{eq_deltaH}. Indeed, the ones that keep the system in the $\mathcal{H}_-$ sector give a zero contribution since $|\psi_0(\Lambda)\rangle$ is an eigenstate within this sector by definition. Instead, the ones that keep the system in the $\mathcal{H}_+$ sector trivially give zero since we do not have any occupation larger than $\Lambda$. The only contribution comes from the operators $\hat{a}_\pm^{(\dagger)}$ or, more precisely, the term $\left(\sqrt{\Lambda+1}|\Lambda+1\rangle \langle \Lambda| + h.c.\right)$ which connects the two sectors. Using Eq.~\eqref{eq_H_plus_minus_expressions}, we straightforwardly obtain
\begin{equation}
\label{eq_variance_enlarging_H}
\Delta H= \Lambda\frac{e^{-2s}}{4} \sum_{j=0}^{L-1}  \langle n_{j}^2\rangle \langle \mathcal{P}_{j+1,\Lambda}\rangle,
\end{equation}
where $\mathcal{P}_{j,k} = |k\rangle_j {}_j\langle k|$ is the projector on the Fock state with occupation $k$ on site $j$.
The first term of the sum ($j=0$) encodes the information about the fixed symmetry sector, since $\langle \hat{n}_0^2\rangle = n_0^2$. The variance given in Eq.~\eqref{eq_variance_enlarging_H} depends on the mean occupation number and on the projector over the Fock space on $\Lambda$. In the main text, we show that the system displays a localized phase in the bosonic limit, $\Lambda\to\infty$, if $U>0$. This enables us to estimate Eq.~\eqref{eq_variance_enlarging_H} in the localized phase. In the localized phase, the average occupation number of the ground state is $\langle \hat{n}_j\rangle \sim e^{-j/\xi}$ (cf. Eq.~\eqref{eq_average_occupation_number}). The exponential decay of the occupation number along the chain reflects on the behavior of the expectation value of $\mathcal{P}_{k,j}$, which decays exponentially fast in $k$ (cf. Eq.~\eqref{eq_decay_projector_localized_phase}). Therefore, the series in Eq.~\eqref{eq_variance_enlarging_H} is finite for $\Lambda \to \infty$ and $L\to \infty$, since each term is exponentially suppressed.
\begin{figure}[b]
\centering
\includegraphics[width=\linewidth]{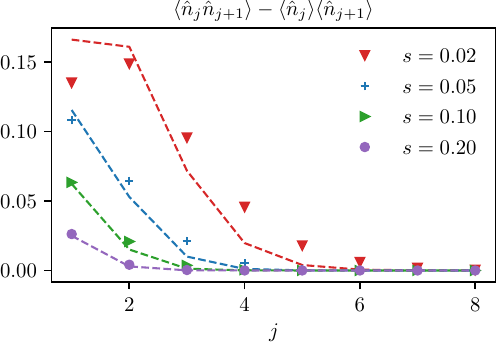}
\caption{The points correspond to the quantity computed on the ground state obtained via the DMRG; the continuous lines are the results obtained assuming that the state is Gaussian. We fix $U=1$, $n_0=1$, and $\Lambda=15$.}
\label{fig_correlation_matrix}
\end{figure}
In Fig.~\ref{fig_variance_H_scaling_lambda}, we numerically compute the variance $\Delta H$ over $|\psi_0(n_0,\Lambda)\rangle$ for different values of $\Lambda$ and $n_0$. Rigorously, the cutoff $\Lambda$ limits the accessible $n_0$, since $\langle \hat{n}_i\rangle \leq \Lambda$. Nevertheless, because $n_0$ appears as a constant in the Hamiltonian, we can also compute the ground state $|\psi_0(n_0,\Lambda)\rangle$ for $n_0>\Lambda$. The numerical results match Eq.~\eqref{eq_variance_enlarging_H} perfectly. The variance goes exponentially fast to zero. Therefore, an eigenstate of $H_-$ is an eigenstate of the fully bosonic system as well, with a reasonably small cutoff $\Lambda$ when $U>0$.

\section{Gaussianity and non-Gaussianity in the ground state \label{appendix_signatures_gaussianity_ground_state}}
In Fig.~\ref{fig_correlation_matrix}, we show the correlator $\Delta_j=\langle \hat{n}_j\hat{n}_{j+1}\rangle - \langle \hat{n}_j\rangle \langle \hat{n}_{j+1}\rangle$ as a function of $j$ for different values of $s$ at fixed $U=1$. We compare $\Delta_j$ computed on the ground state obtained via the DMRG and the one computed assuming that the same state is Gaussian in the operators $\{\hat{a}_j^{(\dagger)}\}_{j=1}^L$, which we call $\Delta^G_j$.

\section{Numerical methods \label{appendix_numerical_methods}}
In this appendix, we provide the details of the \ric{parameters} adopted for simulating \ric{a single stochastic trajectory at the core of the quantum trajectories method, \ricII{while we refer to Ref.~\cite{Jaschke2018,Daley2014} for details of the algorithm}. As stated in the main text, we resort to tensor networks in order to numerically integrate a single trajectory. The deterministic part of the dynamics given by the action of the effective Hamiltonian defined in Eq.~\eqref{eq_effective_Hamiltonian} is performed via the time-evolving block-decimation (TEBD) algorithm with second-order Suzuki-Trotter decomposition. When a jump occurs, the corresponding jump operator is easily applied being a single-site gate.} We fix \ric{a time step} $\delta t = 5 \times 10^{-3}$, a maximal bond dimension $\chi_\text{max}=75$ and we keep the singular values greater than $10^{-10}$. We verify that the results are not affected by the time step $\delta t$ and $\chi_\text{max}$.
All the simulations are performed using the ITensor library~\cite{fishman2020itensor}.
\onecolumngrid
\section{Details about \ric{superconducting circuit} implementation \label{details_cQED_implementation}}
\subsection{Perturbative construction of the generator $S$ of the Schrieffer-Wolff transformation \label{section_SW_transformation}}
We write the Hamiltonian $H_0$ and the perturbation $V$ as a function of the operators $\hat{c}_{\ell,j}^{(\dagger)}$ defined in Eq.~\eqref{eq_definition_ladder_operators}
\begin{equation}
H_0 = \sum_{j=1}^L \sum_{\ell = 0}^\infty \omega_{\ell,j} |\ell,j\rangle \langle \ell,j| \equiv \sum_{j=1}^L \sum_{\ell = 0}^\infty \omega_{\ell,j} p_{\ell,j},\qquad V =  \ric{g\sum_{j=1}^{L-1} \sum_{\ell,s=0}^\infty \left(c_{\ell,j}^\dagger c_{s,j+1} + h.c  \right)},
\end{equation}
where $\omega_{\ell,j} = \left(\omega_j \ric{-} E_C/2\right) j \ric{+} E_C j^2/2$ and we introduce $p_{\ell,j} \equiv |\ell,j\rangle \langle \ell,j|$ for convenience. We compute the generator $\eta = [H_0,V]$
\begin{equation}
\eta =\sum_{j=1}^L \sum_{\ell,s=0}^\infty \:g\left(\tilde{\Delta}_{\ell,j+1} -\tilde{\Delta}_{s,j}\right) \left(c_{s,j}^\dagger c_{\ell,j+1} - c_{s,j} c_{\ell,j+1}^\dagger\right),
\end{equation}
where $\tilde{\Delta}_{\ell,j} = \omega_{\ell+1,j}-\omega_{\ell,j}=(\omega_j \ric{+} E_C \ell)$.
Following Ref.~\cite{haq2019explicit}, the \textit{ansatz} for the generator $S$ of the \ric{SW transformation} is $S= \sum_{j=1}^L \sum_{\ell,s} A_{j,\ell,s} \left(c_{s,j} c_{\ell,j+1}^\dagger \ric{-} c_{s,j}^\dagger c_{\ell,j+1}\right)$.
We compute $[S,H_0]$ and we impose $[S,H_0]=-V$. This condition is satisfied if $A_{j,\ell,s} = g/\left(\tilde{\Delta}_{\ell,\ric{j+1}} -\tilde{\Delta}_{s,j}\right)$.
Therefore 
\begin{equation}
S=\sum_{j=1}^{L-1}S_{j,j+1},\qquad S_{j,j+1} \equiv \sum_{\ell,s=0}^\infty \frac{g}{\tilde{\Delta}_{\ell,j+1} -\tilde{\Delta}_{s,j}} \left(c_{s,j} c_{\ell,j+1}^\dagger \ric{-} c_{s,j}^\dagger c_{\ell,j+1}\right).
\end{equation}
\subsection{Commutator of the Hamiltonian with the generator $S$ of the Schrieffer-Wolff transformation \label{section_after_S}}
We write the perturbation $V = \sum_{j=1}^{L-1}V_{j,j+1}$, where $V_{j,j+1} = g \sum_{\ell,s=0}^\infty \left(c_{\ell,j}^\dagger c_{s,j+1} + h.c  \right)$. We compute the commutators $[S_{j-1,j},V_{j,j+1}]$, $[S_{j,j+1},V_{j,j+1}]$ and $[S_{j-1,j},V_{j,j+1}]$
\begin{equation}
\label{eq_supplemental_V_dressed}
\begin{split}
[S_{j,j+1},V_{j,j+1}]
&=\sum_{\ell,s} \frac{2g^2E_C}{\left(\tilde{\Delta}_{\ell+1,j+1}-\tilde{\Delta}_{s,j}\right) \left( \tilde{\Delta}_{\ell,j+1}-\tilde{\Delta}_{s+1,j}\right)} c_{s,j}c_{s+1,j}c_{\ell+1,j+1}^\dagger c_{\ell,j+1}^\dagger \ric{+} \sum_{\ell,s}\frac{g^2}{\tilde{\Delta}_{\ell-1,j+1} -\tilde{\Delta}_{s,j}} \ell p_{s,j}p_{\ell,j+1}+\\
&\ric{+}\sum_{\ell,s}  \frac{2g^2 E_C}{\left(\tilde{\Delta}_{\ell-1,j+1}-\tilde{\Delta}_{s,j}\right) \left( \tilde{\Delta}_{\ell,j+1}-\tilde{\Delta}_{s-1,j}\right)}s \ell p_{s,j}p_{\ell,\ric{j+1}} \ric{-}\sum_{\ell,s}\frac{g^2}{\tilde{\Delta}_{\ell,j+1} -\tilde{\Delta}_{s-1,j}} s p_{s,j}p_{\ell,j+1} + h.c.\:,\\
[S_{j,j+1},V_{j-1,j}]
&= \sum_{\ell,s,q}
\frac{g^2 E_C}{\left(\tilde{\Delta}_{\ell,j+1}-\tilde{\Delta}_{s,j}\right) \left( \tilde{\Delta}_{\ell,j+1}-\tilde{\Delta}_{s-1,j}\right)}s c_{q,j-1}p_{s,j}c_{\ell,j+1}^\dagger + \sum_{\ell,s,q} \frac{g^2}{\tilde{\Delta}_{\ell,j+1}-\tilde{\Delta}_{s,j}}c_{q,j-1} p_{s,j}  c_{\ell,j+1}^\dagger +\\
&-\sum_{\ell,s,q}\frac{g^2 E_C}{\left(\tilde{\Delta}_{\ell,j+1}-\tilde{\Delta}_{s,j}\right) \left( \tilde{\Delta}_{\ell,j+1}-\tilde{\Delta}_{s+1,j}\right)}c_{q,j-1}^\dagger c_{s,j}c_{s+1,j} c_{\ell,j+1}^\dagger  + h.c.\:,\\
[S_{j-1,j},V_{j,j+1}]&= \sum_{\ell,s,k}
\frac{g^2E_C }{\left(\tilde{\Delta}_{\ell,j}-\tilde{\Delta}_{s,j}\right) \left( \tilde{\Delta}_{\ell-1,j}-\tilde{\Delta}_{s,j-1}\right)} \ell c_{s,j-1} p_{\ell,j}c_{k,j+1}^\dagger-\sum_{\ell,s,k}\frac{g^2}{\tilde{\Delta}_{\ell,j}-\tilde{\Delta}_{s,j-1}} c_{s,j-1} p_{\ell,j} c_{k,j+1}^\dagger +\\
&\ric{-} \sum_{\ell,s,k}
\frac{g^2E_C }{\left(\tilde{\Delta}_{\ell,j}-\tilde{\Delta}_{s,j-1}\right) \left( \tilde{\Delta}_{\ell+1,j}-\tilde{\Delta}_{s,j-1}\right)} c_{s,j-1}  c_{\ell+1,j}^\dagger c_{\ell,j}^\dagger c_{k,j+1} + h.c.\:,
\end{split}
\end{equation}
which constitute the building blocks for computing $[S,V]$. We consider a drive field acting on site $j$, $H_{\text{drive},j}=\Omega_j \left(e^{i\alpha_j t}a_j + h.c\right)$. We compute the commutator $[S,H_{\text{drive},j}]=[S_{j-1,j},H_{\text{drive},j}]+[S_{j,j+1},H_{\text{drive},j}]$:
\begin{equation}
\label{eq_supplemental_H_drive_dressed}
\begin{split}
[S_{j-1,j},H_{\text{drive},j}]&=
\sum_{\ell,s}\frac{\ric{g\Omega_j}E_C }{\left(
\tilde{\Delta}_{\ell-1,j}-\tilde{\Delta}_{s,j-1}\right)\left(\tilde{\Delta}_{\ell,j}-\tilde{\Delta}_{s,j-1}\right)} e^{i\alpha_j t} \ell c_{s,j-1}p_{\ell,j}\ric{-}
\sum_{\ell,s}\frac{\ric{g\Omega_j}}{
\tilde{\Delta}_{\ell,j}-\tilde{\Delta}_{s,j-1}} e^{i\alpha_j t}c_{s,j-1} p_{\ell,j} +\\
& -\sum_{\ell,s}\frac{\ric{ g\Omega_j} E_C }{   
\left(
\tilde{\Delta}_{\ell+1,j}-\tilde{\Delta}_{s,j-1}\right)\left(\tilde{\Delta}_{\ell,j}-\tilde{\Delta}_{s,j-1}\right)
 } e^{-i\alpha_j t}c_{s,j-1} c_{\ell+1,j}^\dagger c_{\ell,j}^\dagger +h.c.\:,\\
 [S_{j,j+1},H_{\text{drive},j}]&=
 \sum_{\ell,s}\frac{\ric{g\Omega_j} E_C}{
\left(\tilde{\Delta}_{\ell,j+1}-\tilde{\Delta}_{s,j}\right)\left(\tilde{\Delta}_{\ell,j+1}-\tilde{\Delta}_{s+1,j}\right)}e^{-i\alpha_j t} s p_{s,j}c_{\ell,j+1}^\dagger
\ric{+}\sum_{\ell,s}\frac{\ric{g\Omega_j}}{\tilde{\Delta}_{\ell,j+1}-\tilde{\Delta}_{s,j}}  e^{-i\alpha_j t}p_{s,j}c_{\ell,j+1}^\dagger+h.c.\\
&\ric{-}\sum_{\ell,s}\frac{\ric{\Omega_j}g E_C}{
\left(
\tilde{\Delta}_{\ell,j+1}-\tilde{\Delta}_{s,j}\right)\left(\tilde{\Delta}_{\ell,j+1}-\tilde{\Delta}_{s+1,j}\right)} e^{i\alpha_j t}c_{s,j} c_{s+1,j}c_{\ell,j+1}^\dagger +h.c.
\end{split}
\end{equation}
\subsection{Low-anharmonicity limit \label{sez_small_anharmonicity_limit}}
In the following, we explicitly consider the results with $L=4$ \ric{superconducting \ricII{qubits}} for clarity. The generalization to a larger number of \ric{superconducting \ricII{qubits}} is straightforward. We work in the limit $E_C \ll \Delta_{ij}$, such that $\tilde{\Delta}_{\ell,j+1}-\tilde{\Delta}_{s,j}\ric{\approx}\Delta_{j+1,j}=\omega_{j+1}-\omega_j$. We neglect the contributions coming from the the commutators of the drive fields controlled by $\{\epsilon_j\}$, since, as we show, they give subleading corrections. From Eqs.~\eqref{eq_supplemental_V_dressed} and \eqref{eq_supplemental_H_drive_dressed} and using the identities $\sum_{\ell=0}^\infty c_{\ell,j}=a_j$, $\sum_{\ell=0}^\infty \ell p_{\ell,j} = n_j$ and $\sum_{\ell=0}^\infty p_{\ell,j}=1$, we obtain
\begin{equation}
\begin{split}
[S,V]\approx & \ric{+}{\frac{2g^2E_C}{\Delta_{12}^2}a_1 a_1 a_2^\dagger a_2^\dagger } + \frac{2g^2 E_C}{\Delta_{12}^2} n_1 n_2 \ric{+} \frac{g^2}{\Delta_{12}} n_1 \ric{-} \frac{g^2}{\Delta_{12}}n_2\ric{+} {\frac{g^2E_C}{\Delta_{23}^2} a_1 n_2 a_3^\dagger - \frac{g^2}{\Delta_{23}} a_1 a_3^\dagger \ric{-}\frac{g^2E_C}{\Delta_{23}^2}a_1^\dagger a_2 a_2 a_3^\dagger} +\\
&\ric{+}{\frac{g^2 E_C}{\Delta_{12}^2} a_1 n_2 a_3^\dagger + \frac{g^2}{\Delta_{12}} a_1 a_3^\dagger \ric{-} \frac{g^2 E_C}{\Delta_{12}^2}a_1^\dagger a_2 a_2 a_3^\dagger}+{\frac{2g^2E_C}{\Delta_{23}^2}a_2 a_2 a_3^\dagger a_3^\dagger} + \frac{2g^2 E_C}{\Delta_{23}^2} n_2 n_3 \ric{+} \frac{g^2}{\Delta_{23}} n_2 \ric{-} \frac{g^2}{\Delta_{23}}n_3+h.c.\:,\\
&\ric{\ric{+} {\frac{g^2E_C}{\Delta_{34}^2} a_2 n_3 a_4^\dagger - \frac{g^2}{\Delta_{34}} a_2a_4^\dagger -\frac{g^2E_C}{\Delta_{34}^2}a_2^\dagger a_3 a_3 a_4^\dagger}} +\\
&\ric{\ric{+}{\frac{g^2 E_C}{\Delta_{23}^2} a_2 n_3 a_4^\dagger + \frac{g^2}{\Delta_{23}} a_2 a_4^\dagger \ric{-} \frac{g^2 E_C}{\Delta_{23}^2}a_2^\dagger a_3 a_3 a_4^\dagger}+{\frac{2g^2E_C}{\Delta_{34}^2}a_3 a_3 a_4^\dagger a_4^\dagger} + \frac{2g^2 E_C}{\Delta_{34}^2} n_3 n_4 \ric{+} \frac{g^2}{\Delta_{34}} n_3 \ric{-} \frac{g^2}{\Delta_{34}}n_4+h.c.}\:,\\
[S,H_\text{drive}]\approx & \ric{+}\frac{gE_C\Omega_1}{\Delta_{12}^2} \left\{n_1 \left(e^{-i\alpha_1 t} a_2^\dagger + h.c.\right) - {\left(  e^{i\alpha_1 t} a_1 a_1 a_2^\dagger + h.c. \right)} \right\}-{ \frac{g\Omega_1}{\Delta_{12}} \left(e^{-i\alpha_1 t} a_2^\dagger + h.c.\right)} +\\
&+\frac{gE_C\Omega_2}{\Delta_{12}^2} \left\{ {n_2 \left(e^{-i\alpha_2 t} a_1^\dagger + h.c.\right) \ric{-}\left( e^{i\alpha_2 t} a_2 a_2 a_1^\dagger + h.c.\right)}\right\}\ric{+}{\frac{g\Omega_2}{\Delta_{12}} \left(e^{-i\alpha_2 t} a_1^\dagger + h.c.\right)} +\\
&\ric{+}\frac{gE_C\Omega_2}{\Delta_{23}^2} \left\{n_2 \left(  e^{-i\alpha_2 t} a_3^\dagger + h.c.\right) -{ \left( e^{i\alpha_2 t} a_2 a_2 a_3^\dagger + h.c. \right)} \right\}- {\frac{g\Omega_2}{\Delta_{23}} \left(e^{-i\alpha_2 t} a_3^\dagger + h.c.\right) }+\\
&+\frac{gE_C\Omega_3}{\Delta_{23}^2} \left\{ {n_3 \left(e^{-i\alpha_3 t} a_2^\dagger + h.c.\right) \ric{-}\left( e^{i\alpha_3 t} a_3 a_3 a_2^\dagger + h.c.\right)}\right\}\ric{+}{\frac{g\Omega_3}{\Delta_{23}} \left(e^{-i\alpha_3 t} a_2^\dagger + h.c.\right)} +\\
&\ric{+}\frac{gE_C\Omega_3}{\Delta_{34}^2} \left\{n_3 \left(  e^{-i\alpha_3 t} a_4^\dagger + h.c.\right) -{ \left( e^{i\alpha_3 t} a_3 a_3 a_4^\dagger + h.c. \right)} \right\}- {\frac{g\Omega_3}{\Delta_{34}} \left(e^{-i\alpha_3 t} a_4^\dagger + h.c.\right) }.
\end{split}
\end{equation}
\subsection{Rotating frame of reference \label{section_appendix_RWA}}
We focus again on the four \ric{superconducting \ricII{qubits}} system (cf. Appendix~\ref{sez_small_anharmonicity_limit}). We change the frame of reference via the unitary transformation $U= \exp(it(\alpha_1 n_2 + \alpha_2 n_3 + \alpha_3 n_4))$, from which 
\begin{equation}
\label{eq_H_rotating_frame_no_approx}
\begin{split}
UH_\text{drive}U^\dagger &= \Omega_1(e^{i\alpha_1 t} a_1 + h.c.) + \Omega_2(e^{i(\alpha_2-\alpha_1)t}a_2+h.c.) + \Omega_3(e^{i(\alpha_3-\alpha_2)t}a_3+h.c.)  +\\
&+\epsilon_2 (a_2+ h.c.) + \epsilon_3 (a_3+h.c.) +\ric{\epsilon_4 (a_4+h.c.)}\\
U[S,V]U^\dagger&\approx {\ric{+}\frac{2g^2E_C}{\Delta_{12}^2}e^{2i\ric{\alpha_1} t}a_1 a_1 a_2^\dagger a_2^\dagger } + \frac{2g^2 E_C}{\Delta_{12}^2} n_1 n_2 \ric{+} \frac{g^2}{\Delta_{12}} n_1 \ric{-} \frac{g^2}{\Delta_{12}}n_2+\\
&\ric{+} {\frac{g^2 E_C}{\Delta_{23}^2} e^{i\alpha_2 t}a_1 n_2 a_3^\dagger - \frac{g^2}{\Delta_{23}}  e^{i\alpha_2 t} a_1 a_3^\dagger \ric{-} \frac{g^2 E_C}{\Delta_{23}^2} e^{-i(2\alpha_1 - \alpha_2)t} a_1^\dagger a_2 a_2 a_3^\dagger} +\\
&\ric{+}{\frac{g^2E_C}{\Delta_{12}^2} e^{i\alpha_2 t} a_1 n_2 a_3^\dagger + \frac{g^2}{\Delta_{12}}  e^{i\alpha_2 t}a_1 a_3^\dagger - \frac{g^2 E_C}{\Delta_{12}^2} e^{-i(2\alpha_1-\alpha_2) t} a_1^\dagger a_2 a_2 a_3^\dagger}+\\
&+{\frac{2g^2E_C}{\Delta_{23}^2}e^{-2i(\alpha_1-\alpha_2) t} a_2 a_2 a_3^\dagger a_3^\dagger} + \frac{2g^2 E_C}{\Delta_{23}^2} n_2 n_3 \ric{+} \frac{g^2}{\Delta_{23}} n_2 \ric{-} \frac{g^2}{\Delta_{23}}n_3 +\\
&\ric{+} {\frac{g^2E_C}{\Delta_{34}^2} e^{i(\alpha_3-\alpha_1)t}a_2 n_3 a_4^\dagger - \frac{g^2}{\Delta_{34}} e^{i(\alpha_3-\alpha_1)t}a_2a_4^\dagger -\frac{g^2E_C}{\Delta_{34}^2}e^{i(\alpha_1 - 2 \alpha_2 + \alpha_3)t}a_2^\dagger a_3 a_3 a_4^\dagger} +\\
&\ric{+}{\frac{g^2 E_C}{\Delta_{23}^2} e^{-i(\alpha_1 - \alpha_3)t}a_2 n_3 a_4^\dagger \ric{-} \frac{g^2}{\Delta_{23}} e^{-i(\alpha_1-\alpha_3)t}a_2 a_4^\dagger + \frac{g^2 E_C}{\Delta_{23}^2}e^{i(\alpha_1-2\alpha_2 + \alpha_3)t}a_2^\dagger a_3 a_3 a_4^\dagger}+\\
&+{\frac{2g^2E_C}{\Delta_{34}^2}e^{-2i(\alpha_2 - \alpha_3)t}a_3 a_3 a_4^\dagger a_4^\dagger} + \frac{2g^2 E_C}{\Delta_{34}^2} n_3 n_4 - \frac{g^2}{\Delta_{34}} n_3 + \frac{g^2}{\Delta_{34}}n_4+h.c.
\end{split}
\end{equation}
\begin{equation}
\label{eq_H_rotating_frame_no_approx_2}
\begin{split}
U[S,H_\text{drive}]U^\dagger &\approx  \ric{+}\frac{g\Omega_1 E_C}{\Delta_{12}^2} \left\{ n_1 \left(a_2^\dagger + h.c.\right) - {\left( e^{\ric{2}i\alpha_1 t} a_1 a_1 a_2^\dagger + h.c. \right)} \right\}-{ \frac{g\Omega_1}{\Delta_{12}} \left(a_2^\dagger + h.c.\right)} +\\
&+\frac{g\Omega_2 E_C}{\Delta_{12}^2} \left\{ {n_2 \left(e^{-i\alpha_2 t} a_1^\dagger + h.c.\right) \ric{-}\left(e^{i(\alpha_2-2\alpha_1) t} a_2 a_2 a_1^\dagger + h.c.\right)}\right\}\ric{+} {\frac{g\Omega_2}{\Delta_{12}} \left(e^{-i\alpha_2 t} a_1^\dagger + h.c.\right)} +\\
&\ric{+}\frac{g\Omega_2 E_C}{\Delta_{23}^2} \left\{ n_2 \left( a_3^\dagger + h.c.\right) - \left(\ric{e^{2i(\alpha_2-\alpha_1)t}} a_2 a_2 a_3^\dagger + h.c. \right) \right\}- {\frac{g\Omega_2}{\Delta_{23}} \left( a_3^\dagger + h.c.\right) }+\\
&+\frac{gE_C\Omega_3}{\Delta_{23}^2} \left\{ {n_3 \left(e^{-i(\alpha_3-\alpha_1) t} a_2^\dagger + h.c.\right) \ric{-}\left( e^{i(\alpha_3-2\alpha_2+\alpha_1) t} a_3 a_3 a_2^\dagger + h.c.\right)}\right\}\ric{+}{\frac{g\Omega_3}{\Delta_{23}} \left(e^{-i(\alpha_3-\alpha_1) t} a_2^\dagger + h.c.\right)} +\\
&\ric{+}\frac{gE_C\Omega_3}{\Delta_{34}^2} \left\{n_3 \left( a_4^\dagger + h.c.\right) -{ \left( e^{2i(\alpha_3 -\alpha_2)t} a_3 a_3 a_4^\dagger + h.c. \right)} \right\}- {\frac{g\Omega_3}{\Delta_{34}} \left( a_4^\dagger + h.c.\right) }.
\end{split}
\end{equation}
We discard all the oscillating terms employing the RWA in the limits,
\begin{align}
&\alpha_1 \gg \max \left(\Omega_1\ric{,\frac{g^2E_C}{\Delta_{12}^2},\frac{g\Omega_1E_C}{\Delta_{12}^2}}\right)\:,\\
&\alpha_2 \gg \max \left(\frac{g^2}{\Delta_{12}},\frac{g^2}{\Delta_{23}},\frac{g^2 E_C}{\Delta_{12}^2},\frac{g^2E_C}{\Delta_{23}^2},\frac{g\Omega_2}{\Delta_{12}}\right)\:,\\
&|\alpha_1-\alpha_2| \gg \max \left(\Omega_2,\ric{\frac{g\Omega_2 E_C}{\Delta_{23}^2},}\frac{g^2E_C}{\Delta_{23}^2}\right)\:,\\
&|\alpha_2-\alpha_3| \gg \max \left(\Omega_3,\ric{\frac{g\Omega_3 E_C}{\Delta_{34}^2},}\frac{g^2E_C}{\Delta_{34}^2}\right)\:,\\
&|2\alpha_1-\alpha_2| \gg \max\left(\frac{g^2E_C}{\Delta_{23}^2},\frac{g^2E_C}{\Delta_{12}^2},\frac{g\Omega_2E_C}{\Delta_{12}^2}\right),\\
&|\alpha_3-\alpha_1| \gg \max\left(\frac{g^2E_C}{\Delta_{34}^2}, \frac{g^2}{\Delta_{34}}, \frac{g^2E_C}{\Delta_{23}^2},\frac{g^2}{\Delta_{23}},\frac{gE_C\Omega_3}{\Delta_{23}^2},\frac{g\Omega_3}{\Delta_{23}}\right),\\
&|\alpha_1 - 2\alpha_2 + \alpha_3| \gg \max\left(\frac{g^2E_C}{\Delta_{34}^2},\frac{g^2E_C}{\Delta_{23}^2},\frac{gE_C\Omega_3}{\Delta_{34}^2}\right).
\end{align}
\ric{which are satisfied in the dispersive regime and at low-anharmonicity $E_C$ limit via a staggered configuration of the drive field frequencies with a little dishomogeneity, as discussed in Sec.~\ref{section_cQED_implementation}.}
Discarding the oscillating terms in Eq.~\eqref{eq_H_rotating_frame_no_approx} \ric{and Eq.~\eqref{eq_H_rotating_frame_no_approx_2}} we obtain
\begin{equation}
\ric{\tilde{H}= \sum_{j=1}^4 \left(\tilde{\omega}_j \hat{n}_j \ric{+} \frac{E_C}{2}\hat{a}_j^\dagger\hat{a}_j^\dagger\hat{a}_j \hat{a}_j\right) + \sum_{j=1}^3 \left(\frac{2g^2E_C}{\Delta_{j,j+1}^2}\hat{n}_j\hat{n}_{j+1} \ric{+}\frac{g\Omega_jE_C}{\Delta_{j,j+1}^2}\hat{n}_j \left(\hat{a}_{j+1} + \hat{a}_{j+1}^\dagger\right)\right) + \sum_{j=2}^4 \left(\epsilon_j-\frac{g\Omega_{j-1}}{\Delta_{j-1,j}}\right) \left(a_j+a_j^\dagger\right).}
\end{equation}
Since we do not want local fields $\propto (a_j + a_j^\dagger)$ we fix the condition $\epsilon_j = g\Omega_{j-1}/\Delta_{j-1,j}$ with $j=2,3\ric{,4}$. We obtain
\begin{equation}
\ric{\tilde{H}= \sum_{j=1}^4 \left(\tilde{\omega}_j \hat{n}_j \ric{+} \frac{E_C}{2}\hat{a}_j^\dagger\hat{a}_j^\dagger\hat{a}_j \hat{a}_j\right) +\sum_{j=1}^3\left(\frac{2g^2E_C}{\Delta_{j,j+1}^2}\hat{n}_j\hat{n}_{j+1}\ric{+} \frac{g\Omega_jE_C}{\Delta_{j,j+1}^2}\hat{n}_j \left(\hat{a}_{j+1} + \hat{a}_{j+1}^\dagger\right)\right).}
\end{equation}
In the dispersive regime, \ric{the drive fields amplitudes} $\{\epsilon_j\}$ are very small compared to the drive fields controlled by $\{\Omega_j\}$. Therefore, it is appropriate to neglect the contributions coming from their commutators \ric{with $S$}.
The above calculations can be straightforwardly generalized to the multisite case, since the \ric{superconducting \ricII{circuits}} in the bulk will behave analogously to the second one in the case treated explicitly above.

\end{document}